 \documentclass[final,5p,times,twocolumn,authoryear]{elsarticle}

\usepackage{amssymb}
\usepackage{lipsum}
\usepackage{amsmath}
\setcitestyle{square,numbers,sort&compress}
\setcitestyle{citesep={,}}
\biboptions{numbers,sort&compress}

\makeatletter
\def\ps@pprintTitle{%
	\let\@oddhead\@empty
	\let\@evenhead\@empty
	\let\@oddfoot\@empty
	\let\@evenfoot\@oddfoot
}
\makeatother

\usepackage{caption}
\usepackage{float}
\usepackage{graphicx}
\usepackage{subcaption}

\begin{document}

\begin{frontmatter}



\title{Holographic Einstein Ring of Quantum Corrected
AdS-Reissner-Nordstrom Black Holes in Kiselev Spacetime}


 \author[first]{Jin-Yu Gui\fnref{1}} 
\fntext[1]{Email: guijinyu0619@qq.com}

\affiliation[first]{organization={Department of Mechanics, Chongqing Jiaotong University},
            city={Chongqing},
            postcode={400074}, 
            country={China}}           

\author[first]{Ke-Jian He\fnref{2}}
\fntext[2]{Email: kjhe94@163.com }

\author[second]{Xiao-Xiong Zeng\fnref{3}}
\fntext[3]{Email: xxzengphysics@163.com (Corresponding author)}
\affiliation[second]{organization={College of Physics and Electronic Engineering, Chongqing Normal University},
            city={Chongqing},
            postcode={401331}, 
            country={China}}

\begin{abstract}
This study, grounded in AdS/CFT correspondence, utilizes wave optics theory to explore the Einstein ring of a quantum-corrected AdS-Reissner-Nordström black hole (BH) in Kiselev spacetime. By fixing the wave source on the AdS boundary, the corresponding response function generated on the antipodal side of the boundary is successfully obtained. Using a virtual optical system with a convex lens, the holographic image of the Einstein ring of the BH is captured on a screen. The study also investigates the impact of various physical parameters and the observer's position on the characteristics of the Einstein ring. The results indicate that changes in the observer's position cause the image to transition from an axisymmetric ring to an arc, ultimately converging to a single luminous point. Additionally, the Einstein ring radius decreases with increasing values of the quantum correction parameter $a$, the equation of state parameter $\Omega$, temperature $T$, and chemical potential $\mu$ , respectively. In contrast, the ring radius increases as the cosmological fluid parameter $c$ increases. Furthermore, the ring radius becomes more distinct as the wave source frequency $\omega$ increases. From the perspective of geometric optics, the photon ring of the quantum-corrected AdS-Reissner-Nordström BH in Kiselev spacetime is further studied. Numerical results suggest that the incident angle of the photon ring aligns with that of the Einstein ring. 
\end{abstract}




\end{frontmatter}




\section{Introduction}
\label{introduction}

Since the advent of the AdS/CFT correspondence theory\cite{Maldacena:1997re}, its theoretical foundation and applications have sparked extensive and in-depth discussions within the physics community. The AdS/CFT correspondence not only validates the internal consistency of the theory but also provides a novel framework for studying strongly coupled field theories. The complexity of these systems makes it difficult for traditional methods to directly uncover their underlying laws. However, the AdS/CFT correspondence offers an ingenious and effective approach, enabling researchers to explore the intricate nature of quantum field theories using the powerful tools of gravitational theory.

In recent years, the AdS/CFT correspondence has been widely applied to various fields of physics. In condensed matter physics, research on holographic superconductivity\cite{Hartnoll:2008kx,Hartnoll:2008vx,Gubser:2008px} has revealed novel physical mechanisms behind superconducting phenomena. Studies on Fermi liquids and non-Fermi liquids \cite{Cubrovic:2009ye,Liu:2009dm,Faulkner:2009wj} have deepened our understanding of the properties of these materials. In areas such as metal-insulator transitions in quantum phase transitions\cite{Mefford:2014gia,Kiritsis:2015oxa,Donos:2014uba}, the AdS/CFT correspondence has demonstrated its powerful predictive and explanatory capabilities.

Furthermore, significant advancements have been made in BH physics through the AdS/CFT correspondence. The shadow of a BH, as an observable feature, has long been a major topic of research in both theoretical physics and astronomy\cite{Shipley:2016omi,
Bambi:2008jg,Atamurotov:2013dpa,Mishra:2019trb,Nedkova:2013msa,Chen:2024wtw,He:2024yeg,Shaikh:2018lcc,Wang:2025ihg,Yang:2024nin,He:2024amh,He:2024qka,He:2022yse,He:2022aox,He:2021htq,Zeng:2021dlj,He:2025rjq}. The Event Horizon Telescope (EHT) project successfully captured images of the shadows of BHs located at the center of the M87 galaxy and the Milky Way\cite{EventHorizonTelescope:2019dse,EventHorizonTelescope:2022wkp}, marking a milestone in astronomical observations. Recently, Hashimoto and colleagues made notable progress in BH shadow research. By applying the AdS/CFT correspondence and principles of wave optics, they proposed a method to study BH shadows and successfully reconstructed the Einstein ring within the BH shadow\cite{Hashimoto:2019jmw}. Notably, for AdS BHs \cite{Hashimoto:2018okj}, studies have shown that holographic BH images can be constructed through gravitational lensing response functions, with the Einstein ring clearly visible in the images. This method has also been applied to holographic superconductor models\cite{Kaku:2021xqp}, revealing discontinuous changes in the size of the photon ring. Correspondingly, studies on charged BHs\cite{Liu:2022cev} suggest that the chemical potential does not affect the radius of the Einstein ring, while the temperature dependence exhibits unique characteristics. In other modified gravity theories, intriguing results have also been found in the study of holographic Einstein rings of BHs\cite{Zeng:2023zlf,Zeng:2023tjb,Zeng:2023ihy,Hu:2023mai,Li:2024mdd,He:2024bll,Luo:2024wih,Zeng:2024ptv,Zeng:2024ybj,Gui:2024owz,He:2024mal}.

The primary goal of this study is to investigate the Einstein ring of a quantum-corrected AdS-Reissner-Nordström BH in Kiselev spacetime, based on the AdS/CFT correspondence and wave optics principles. Quantum corrections play a crucial role in addressing the singularity problems inherent in classical general relativity. The traditional Schwarzschild solution has a singularity at $r=0$, where the spacetime curvature diverges. However, as suggested in the literature\cite{Kazakov:1993ha}, quantum fluctuations shift this singularity to a finite radius (near the Planck length), and the scalar curvature remains finite. This implies that at quantum scales, spacetime exhibits regularity, free from the singularities seen in classical theories. As a result, quantum corrections have attracted considerable attention from researchers and have advanced numerous related fields. In BH physics, quantum corrections are widely applied to study quasi-normal modes, BH shadows, thermodynamics, and horizons\cite{Liu:2020ola,Konoplya:2019xmn,Bezerra:2019qkx,Shahjalal:2019pqb}. In particular, the study of quantum-corrected AdS-Reissner-Nordström BHs in Kiselev spacetime is of significant importance \cite{MoraisGraca:2021ife,Sadeghi:2024moo}.

The introduction of cosmological fluids is also pivotal due to their profound influence on the large-scale structure and evolution of the universe. In the context of the Kiselev metric\cite{Kiselev:2002dx}, these fluids can be characterized by an equation of state (e.g., $P=\Omega p$, where $\Omega$ denotes the equation of state parameter). Building on previous research, this paper examines the Einstein rings formed by quantum-corrected AdS-Reissner-Nordström BHs in Kiselev spacetime. The objectives of this study are twofold. First, to assess whether the holographic scheme can effectively and accurately reconstruct the image of Einstein rings within the current spacetime framework. Second, to thoroughly investigate how key physical parameters (such as the quantum correction parameter $a$, the cosmological fluid parameter $c$, the equation of state parameter $\Omega$, and chemical potential $\mu$) influence the formation mechanism and characteristics of Einstein rings in this spacetime background. Through these investigations, we aim to provide new insights and deepen our understanding of the structures of the spacetime and the shadows

The structure of this article is as follows: Section 2 provides a brief overview of quantum-corrected AdS-Reissner-Nordström BH in Kiselev spacetime, deriving the lensing response function that describes the diffraction of wave sources by BHs in this specific spacetime. This function is essential for understanding the gravitational lensing effects of these BHs. Section 3 presents the experimental setup, where an optical system equipped with a convex lens is used to observe the Einstein ring formed on a screen. This section also investigates the impact of physical parameters such as the quantum correction parameter $a$, the cosmological fluid parameter $c$, the equation of state parameter $\Omega$, and chemical potential $\mu$, the BH's charge $e$, the equation of state parameter $\Omega$, and the observer's position on the Einstein ring's characteristics. Additionally, we compare the conclusions drawn from wave optics with those from geometric optics, highlighting the similarities and differences between the two approaches. Finally, Section 4 summarizes the findings of this study.

\section{The quantum corrected AdS-Reissner-Nordstrom BHs
in Kiselev Spacetime and its response
function}
The quantum-corrected charged AdS BH, within a Kiselev spacetime, is described by the following spherically symmetric metric: 
\begin{equation}
\mathrm{d}s^{2} =-f(r)\mathrm{d}t^{2}  +\frac{1}{f(r)} \mathrm{d}r^{2} +r^{2} (\mathrm{d}\theta ^{2}+\sin ^{2}\theta \mathrm{d}\varphi ^{2}  ),\label{1}
\end{equation}
where the horizon function is\cite{MoraisGraca:2021ife}.
\begin{equation}
f(r)=-\frac{2M}{r} +\frac{\sqrt{r^{2}-a^{2}}}{r}+\frac{r^{2}}{\ell ^{2}}-\frac{c}{r^{3\Omega+1}}+\frac{Q^{2}}{r^{2}}.\label{2}
\end{equation}
The parameter $M$ represents the mass of the BH. while $\ell$ denotes the length scale associated with asymptotically AdS spacetime. For the purposes of subsequent calculations, the value of $\ell$ is set to 1. The parameter $a$ is associated with quantum correction, and $c$ characterizes the cosmological fluid surrounding the BH. The parameter $\Omega$ represents the equation of state parameter, while $Q$ is the electric charge of the BH. By substituting the metric function with the variable $r=1/y$ and $F(y)=f(1/r)$, Eq.(\ref{1}) and Eq.(\ref{2}) can be rewritten as \begin{equation}
\mathrm{d}s^{2} =\frac{1}{y^{2} }[-F(y)\mathrm{d}t ^{2}+\frac{\mathrm{d}y^{2}}{F(y)}+ \mathrm{d}\theta ^{2}+\sin ^{2}\theta \mathrm{d}\varphi ^{2}  ],\label{3}
\end{equation}
\begin{equation}
F(y)=1-cy^{3+3\Omega}-2My^{3}+y^{2}\sqrt{1-a^{2}y^{2}}+Q^{2}y^{2} .
\end{equation}

At $y=\infty$, a spacetime singularity arises, while at $y = 0$, the boundary of the AdS spacetime is located. The temperature of the BH is defined by the Hawking temperature, which is given by the relation $T=\frac{1}{4\pi} {F}'(y_{h})$, where $y=y_{h}$ represents the event horizon of the BH. It is important to note that when $y_{h}=\infty$, the spacetime becomes pure AdS.

Next, we examine the dynamics of a massless scalar field using the Klein-Gordon equation\cite{Hashimoto:2018okj} \begin{equation}
D_{\alpha}D^{\alpha}\tilde{\Psi}-{\mathcal{M}}^{2}\tilde{\Psi}=0. 
\end{equation} where $D_{\alpha}\equiv\nabla_{\alpha}-ieA_{\alpha}$ denotes the covariant derivative operator, $A$ represents the electromagnetic four-potential, $\tilde{\Psi}$ signifies the complex scalar field with $e$ being its electric charge, and $\mathcal{M}$ represents the mass of the scalar field. To facilitate the solution of the Klein-Gordon equation, it is advantageous to introduce the incident Eddington-Finkelstein coordinate system\cite{Liu:2022cev}, which aids in understanding the BH's nature and ensures the continuity of physical quantities at the event horizon. The coordinates can be expressed as \begin{equation}
\upsilon \equiv t+y_{*} =t-\int \frac{\mathrm{d}y}{F(y)} ,
\end{equation} therefore, the non-vanishing bulk background fields are transformed into the following form \begin{equation}
\mathrm{d}s^{2} =\frac{1}{y^{2} }\Big[-F(y)\mathrm{d}\upsilon  ^{2}-2\mathrm{d}y \mathrm{d}\upsilon + \mathrm{d}\theta ^{2}+\sin ^{2}\theta \mathrm{d}\varphi ^{2} \Big],
\end{equation}
\begin{equation}
A_{\alpha}=-A(y)(\mathrm{d}\upsilon)_\alpha,
\end{equation} where $A(y)=Q(y-y_{h})$. In this case, a gauge transformation is applied to the electromagnetic four-potential. The chemical potential of the system is denoted as $\mu=y_{h}Q$. For definiteness, we choose $e=1$ and ${\mathcal{M}}^{2}=-2$. With $\tilde{\Psi}=y\Psi$, the asymptotic behavior of $\Psi$ near the AdS boundary can be expressed as follows
 \begin{eqnarray}
\Psi(\upsilon ,y,\theta, \varphi  )=J_{\mathcal{K}}(\upsilon ,\theta, \varphi )+\left \langle \mathcal{K} \right \rangle y +K(y^{2} ),
\end{eqnarray} where, according to the holographic dictionary, $J_{\mathcal{K}}$ represents the external source for the boundary field theory, and the corresponding expectation value of the dual operator, called the response function, is defined as follows\cite{Liu:2022cev} \begin{eqnarray}
\left\langle \mathcal{K}\right\rangle_{J_\mathcal{K}}= \left\langle \mathcal{K}\right\rangle-(\partial\upsilon-i\mu)J_\mathcal{K},\label{10}
\end{eqnarray} Here, $\left\langle \mathcal{K}\right\rangle$ represents the expectation value of the dual operator when the source is turned off. As defined in \cite{Klebanov:1999tb}, we choose a monochromatic and axisymmetric Gaussian wave packet as the wave source, located at the South pole $(\theta _{0} =\pi )$ of the AdS boundary, as shown in Figure 4 of reference\cite{Liu:2022cev}.

\begin{eqnarray}
J_{\mathcal{k}}  (\upsilon ,\theta )&=&e^{-i\omega \upsilon }(2\pi\eta ^{2} )^{-1} \mathrm{exp}[-\frac{(\pi -\theta )^{2} }{2\eta ^{2} }] \nonumber\\
&=&e^{-i\omega \upsilon }\sum_{l=0}^{\infty }C_{l0} X_{l0}(\theta ) \label{11} 
\end{eqnarray} where $\eta$ denotes the width of the wave  generated by the Gaussian source, and $\eta \ll \pi $. The parameter $\omega$ represents the frequency of the incident wave. $X_{l0}(\theta )$  is the spherical harmonics function,  $C_{l0}$ is the coefficients of  $X_{l0}(\theta )$, which can be expanded as \begin{equation}
C_{l0}=(-1)^{l}\sqrt{\frac{l+1/2}{2\pi} }  \mathrm{exp}[-\frac{(l+1/2)^{2}\eta ^{2}  }{2} ].
\end{equation} Based on Eq.(\ref{11}), the corresponding bulk solution is given by \begin{equation}
\Psi(\upsilon ,y,\theta)=\sum_{l=0}^{\infty } e^{-i\omega \upsilon }C_{l0}Y_{l}(y)
X_{ln}(\theta),\label{13}
\end{equation} where the radial wave function $Y_{l}$  satisfies the equation of motion as follow \begin{eqnarray}
y^{2}F(y){Y}''_{l} +(y^{2}{F}'(y)-2yF(y)+2i\omega y^{2}){Y}'_{l} \nonumber\\ +[-2i\omega y-y^{2}l(l+1)]{Y}_{l}=0.\label{yundong}
\end{eqnarray} According to the AdS/CFT correspondence, near the AdS boundary, $Y_{l}$ can be expanded as \begin{equation}
Y_{l} =1 +\left \langle \mathcal{K}\right \rangle_{l}y
+K(y^{2} ).\end{equation} Thus, the resulting response function $\left\langle \mathcal{K}\right\rangle_{J_\mathcal{K}}$ can be expanded as \begin{equation}
\left\langle \mathcal{K}\right\rangle_{J_\mathcal{K}}=\sum_{l=0}^{\infty } e^{-i\omega \upsilon }C_{l0}\left\langle \mathcal{K}\right\rangle_{J_\mathcal{K}{l}} X_{l0} (\theta ) \label{xiangying}
\end{equation} with \begin{equation}
\left\langle \mathcal{K}\right\rangle_{J_\mathcal{K}{l}}=\left\langle \mathcal{K}\right\rangle_{l}+i\tilde{\omega}.\end{equation}
Here $\tilde{\omega}=\omega+e\mu$. The function $Y_{l}$ has two boundary conditions. One at the event horizon $y=y_{h}$, and the other at the AdS boundary $y=0$. In this case, the wave source  $J_{\mathcal{K}}$ represents the asymptotic form of the scalar field at infinity, and $Y_{l} (0)=1$, as seen from  Eq.(\ref{yundong}). We obtain the corresponding numerical solution for $Y_{l}$ and extract $\left \langle \mathcal{K}\right \rangle _{l}$ using the pseudo-spectral method\cite{Hashimoto:2019jmw,Hashimoto:2018okj}. Subsequently, using the Eq.(\ref{xiangying}), we calculate the value of the total response function. Since the total response is located far from the wave source, the second term related to the wave source in Eq.(\ref{10}) can be neglected. Therefore, in Eq.(\ref{yundong}), $\left\langle \mathcal{K}\right\rangle_{J_\mathcal{K}{l}}$ can be replaced by $\left\langle \mathcal{K}\right\rangle_{l}$.

\begin{figure}
	\centering 
\includegraphics[width=0.4\textwidth, angle=0]{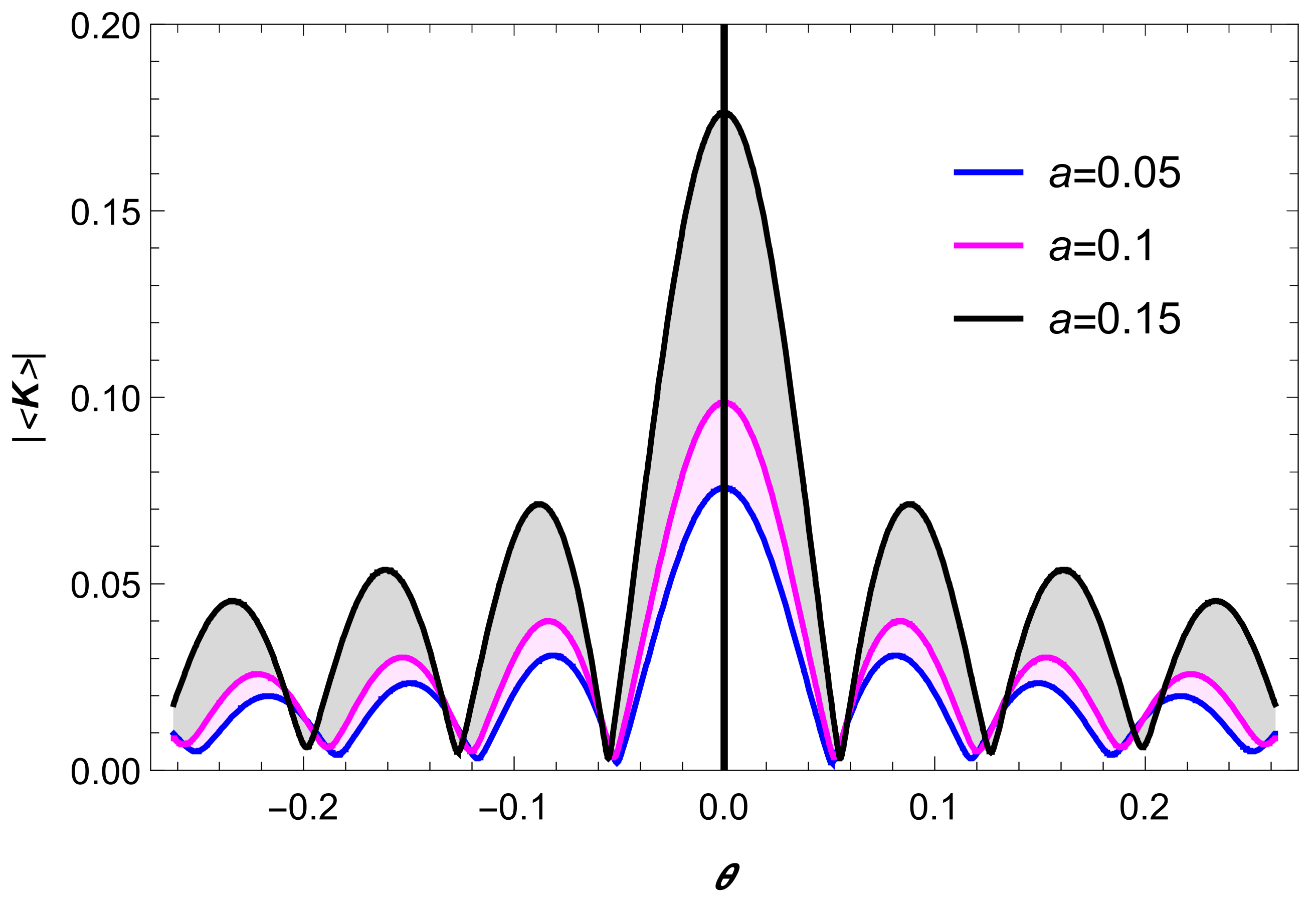}	
	\caption{Response function for different $a$ with $c=0.1$, $Q=0.1$, $\Omega=-\frac{2}{3}$, $y_{h}=5$, $e=0.5$, $\omega =90$.}
	\label{1}%
\end{figure}

\begin{figure}
	\centering 
\includegraphics[width=0.4\textwidth, angle=0]{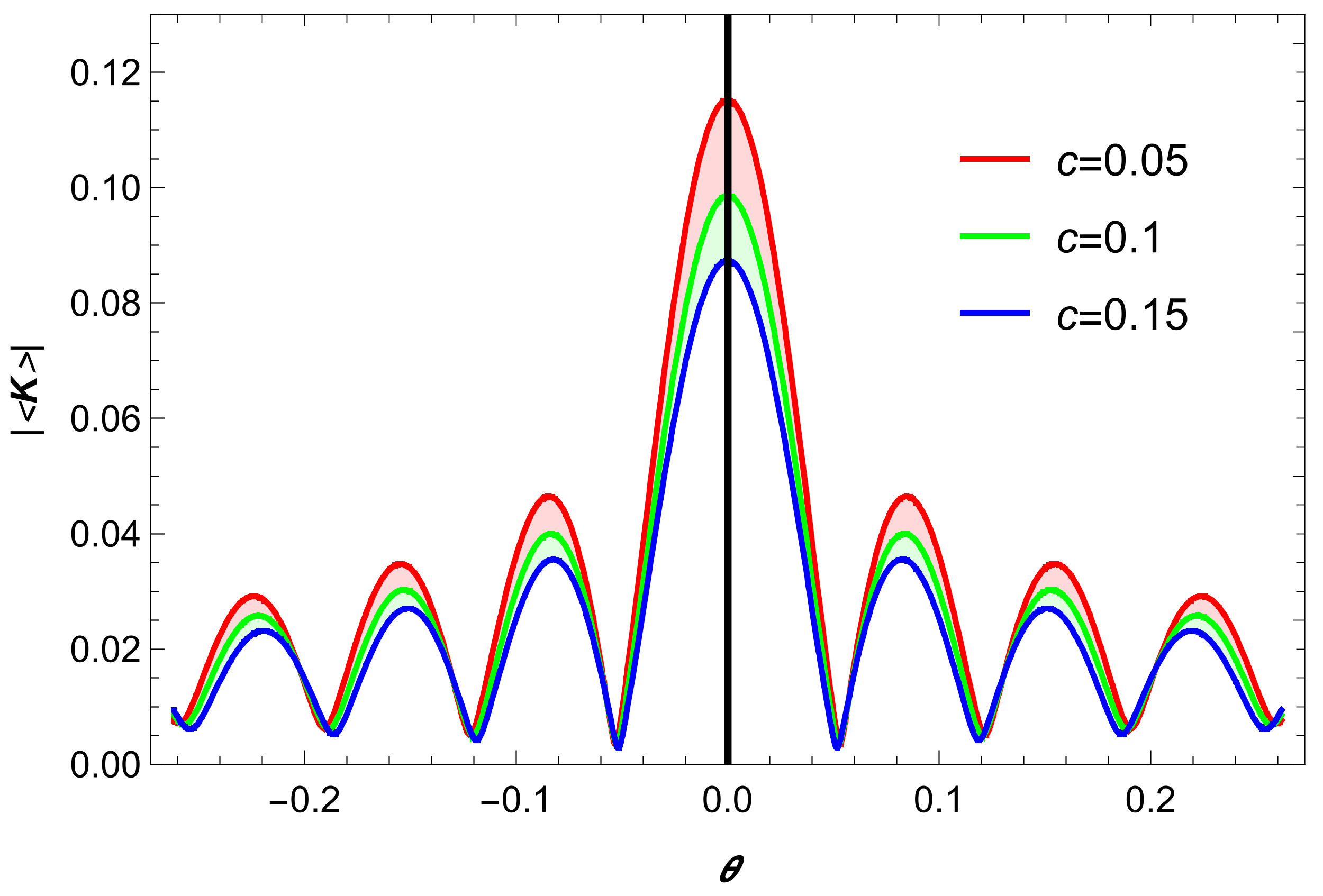}	
	\caption{Response function for different $c$ with $a=0.1$, $Q=0.1$, $\Omega=-\frac{2}{3}$, $y_{h}=5$, $e=0.5$, $\omega=90$.}
	\label{2}%
\end{figure}

\begin{figure}
	\centering 
\includegraphics[width=0.4\textwidth, angle=0]{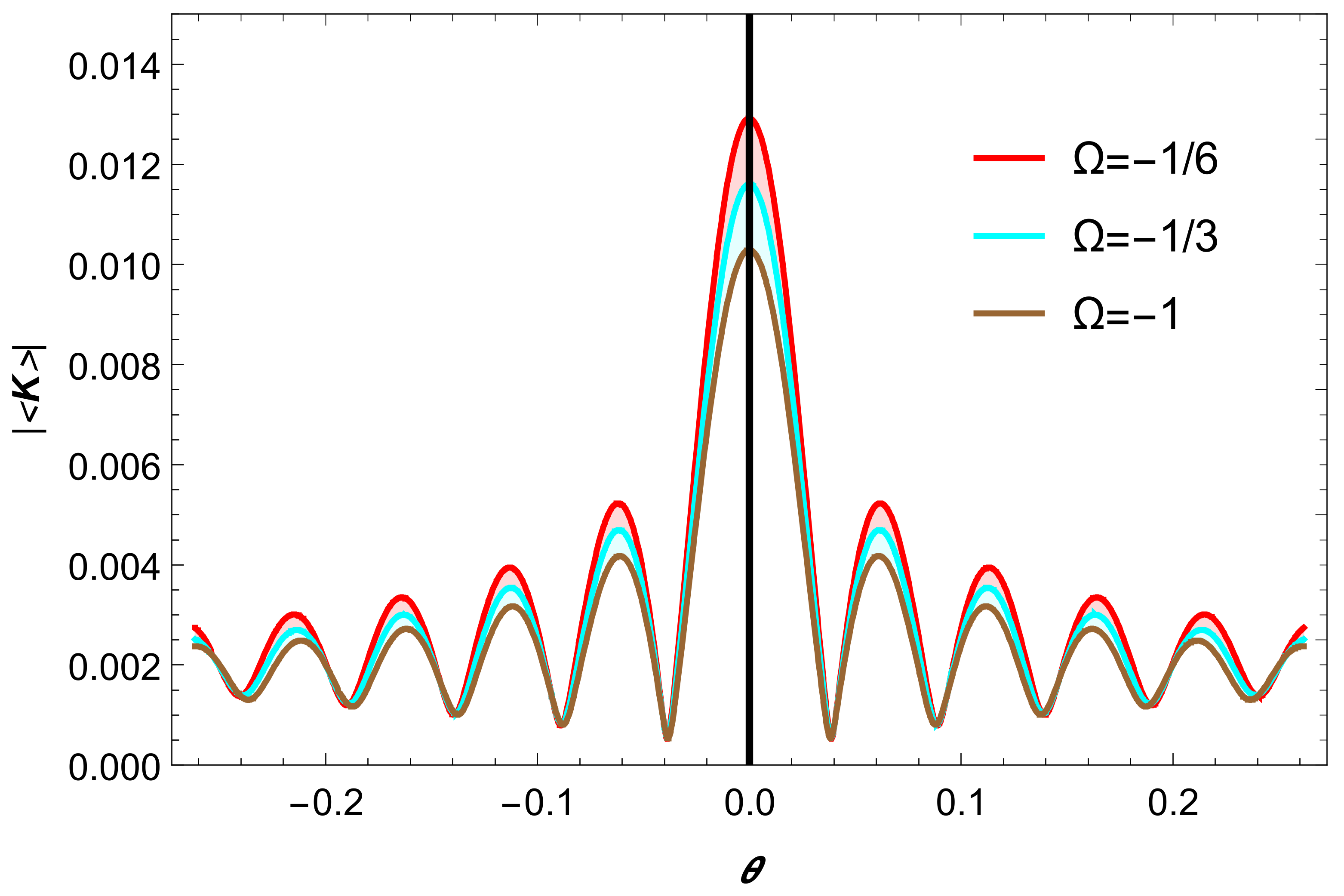}
	\caption{Response function for different $\Omega$ with $a=0.1$, $c=0.1$ 
 $Q=0.1$, $\Omega=-\frac{2}{3}$, $y_{h}=3$, $e=0.5$, $\omega=90$.}
	\label{3}%
\end{figure}

\begin{figure}
	\centering 
\includegraphics[width=0.4\textwidth, angle=0]{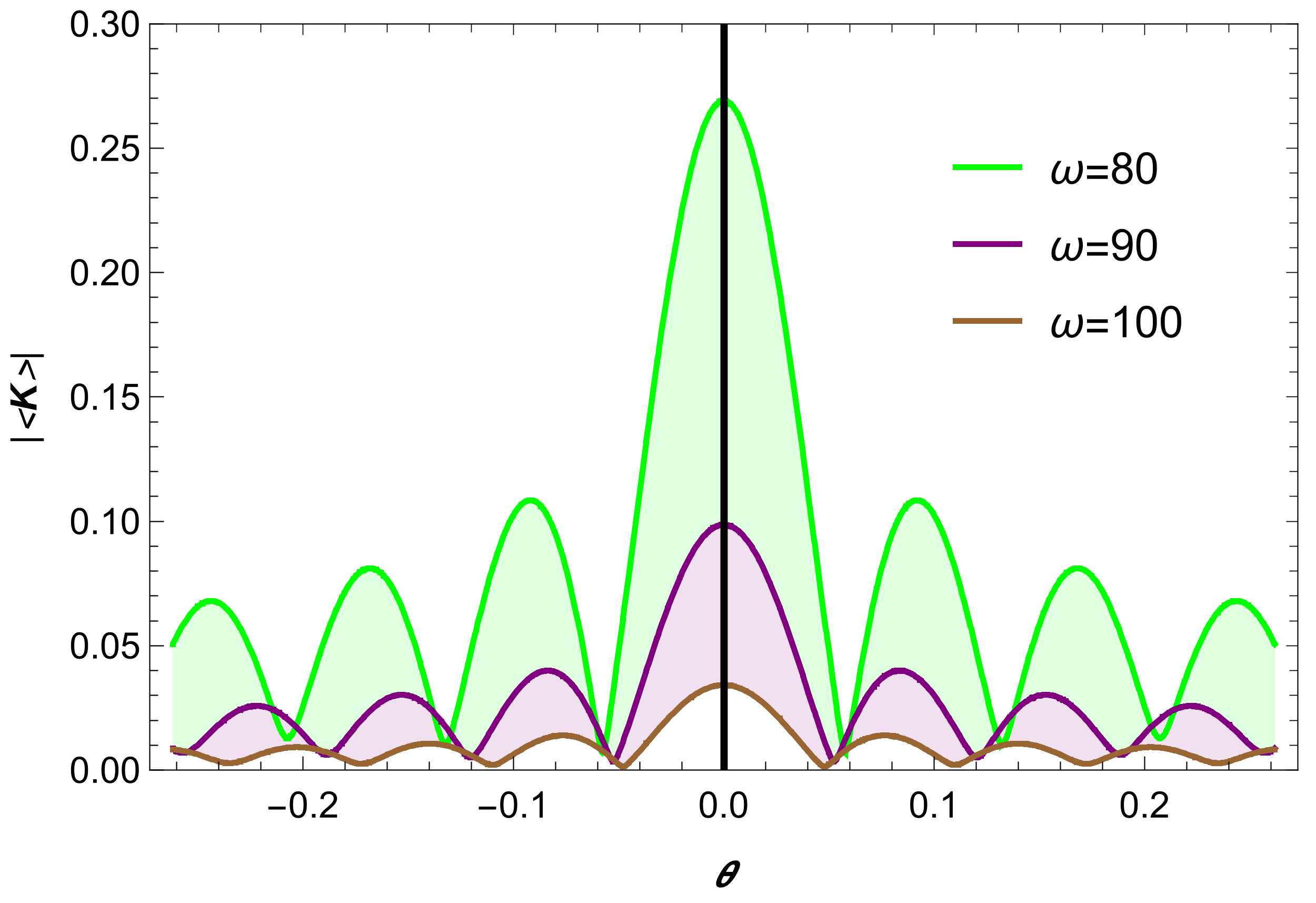}
	\caption{Response function for different $\omega$ with $a=0.1$, $c=0.1$, $\Omega=-\frac{2}{3}$, $Q=0.1$, $y_{h}=5$, $e=0.5$.}
	\label{4}%
\end{figure}
\begin{figure}
	\centering 
\includegraphics[width=0.4\textwidth, angle=0]{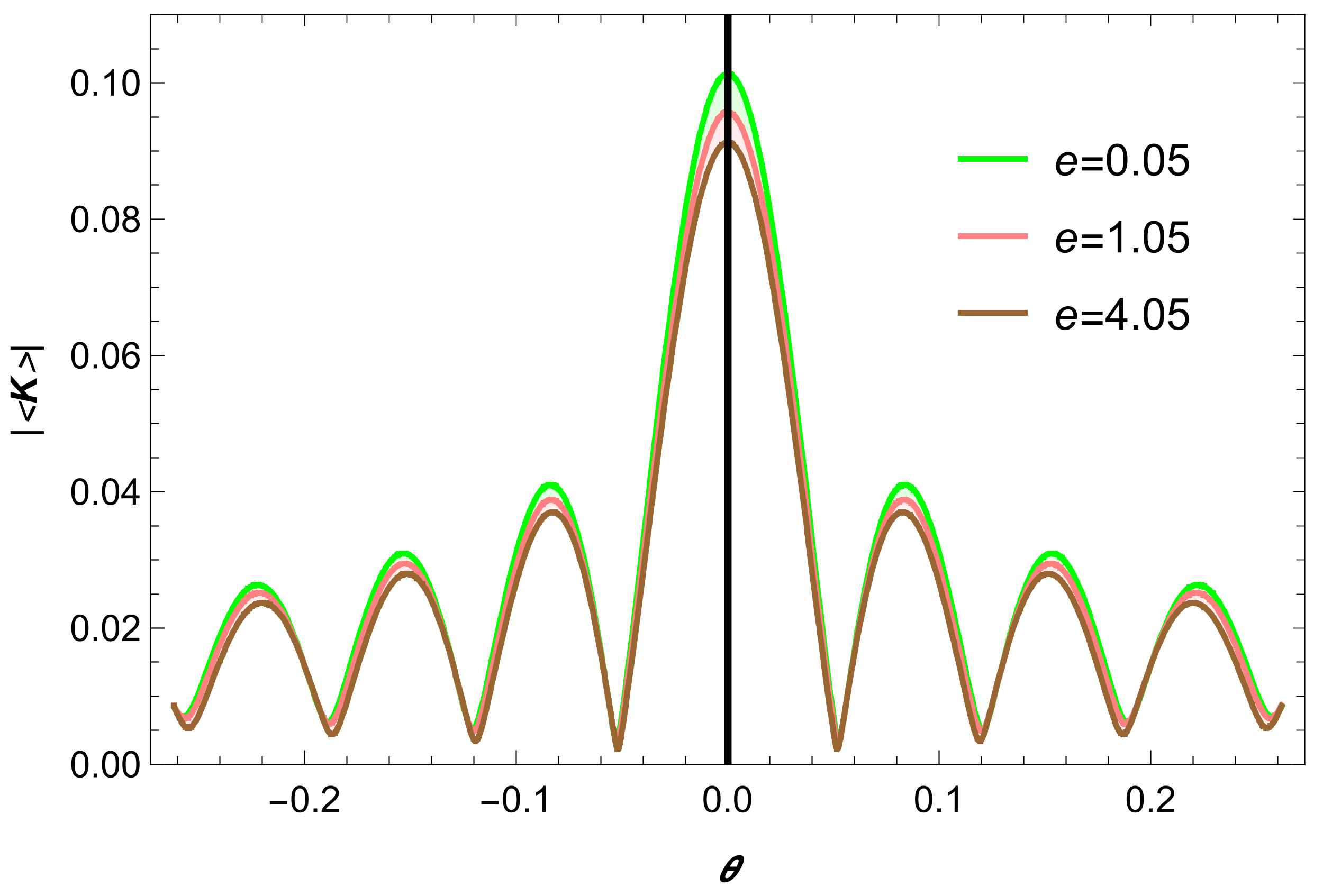}	
	\caption{Response function for different $e$  with $a=0.1$, $c=0.1$, $\Omega=-\frac{2}{3}$, $Q=0.1$, $y_{h}=5$, $\omega=90$.}
	\label{5}%
\end{figure}
\begin{figure}
	\centering 
\includegraphics[width=0.4\textwidth, angle=0]{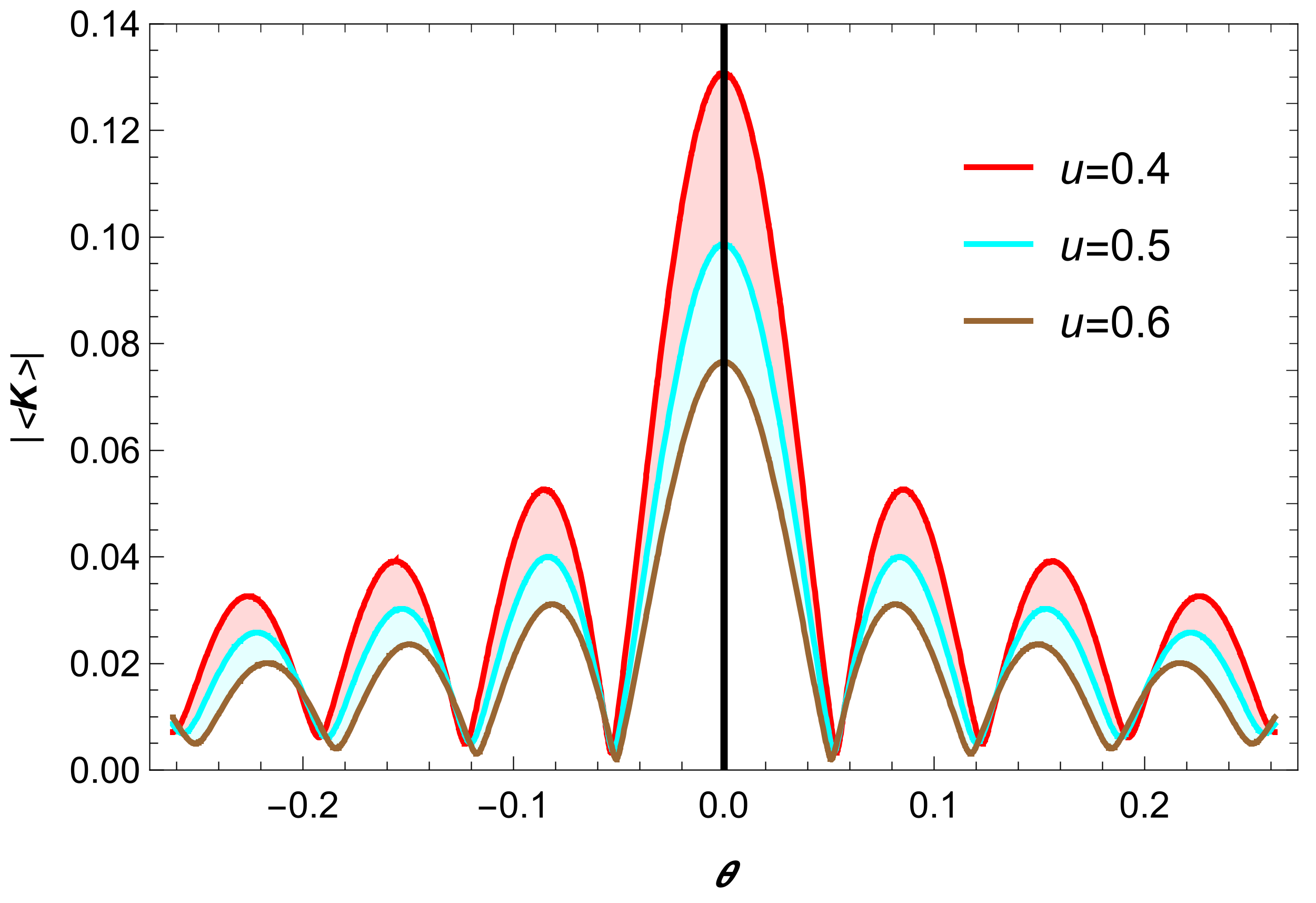}	
	\caption{Response function for different $\mu$  with $a=0.1$, $c=0.1$, $\Omega=-\frac{2}{3}$, $y_{h}=5$, $e=0.5$, $\omega=90$, from top to bottom,
the values of $\mu$ correspond to $Q=0.08, 
0.12, 0.16$, respectively.
}
	\label{6}%
\end{figure}

\begin{figure}
	\centering 
\includegraphics[width=0.4\textwidth, angle=0]{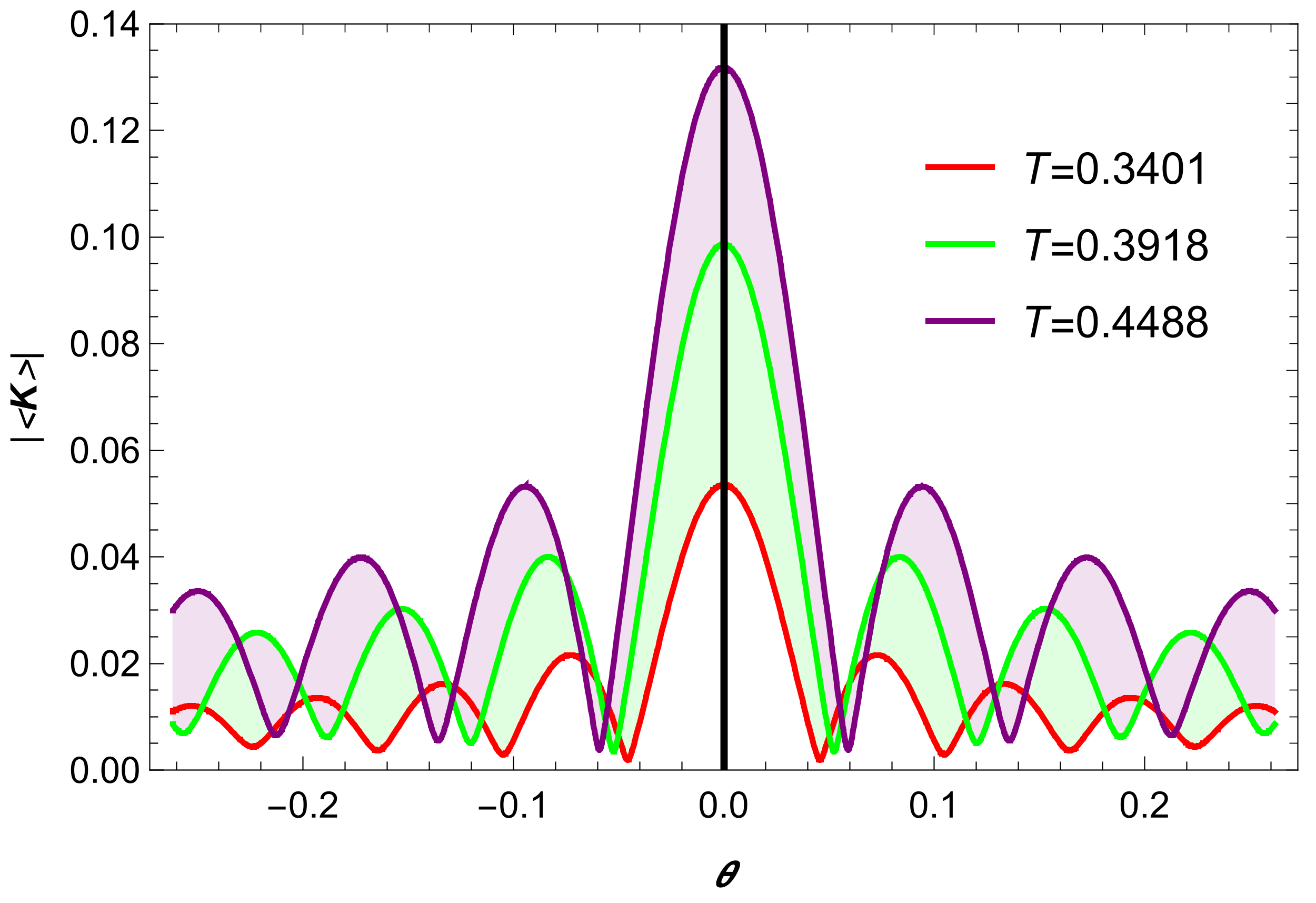}	
	\caption{Response function for different $T$  with $a=0.1$, $c=0.1$, $\Omega=-\frac{2}{3}$, $e=0.5$, $\omega=90$, from top to bottom,
the values of $T$ correspond to $y_{h}=4, 
5, 6$, respectively.
}
	\label{7}%
\end{figure}

Figure \ref{1} through Figure \ref{7} illustrate the response functions for different parameters. Specifically, Figure \ref{1} clearly shows that the amplitude of the response function increases as the quantum correction $a$ increases. Figure \ref{2} indicates that the amplitude decreases as the cosmological fluid parameter $c$ increases. Figure \ref{3} demonstrates that the amplitude increases as the equation of state parameter $\Omega$ increases. Figure \ref{4} shows a decrease in amplitude as the wave source frequency $\omega$ increases. Figure \ref{5} suggests that the amplitude decreases as the magnitude of the charge $e$ increases. Figure \ref{6} shows a reduction in amplitude as the chemical potential $\mu$ increases. Finally, Figure \ref{7} demonstrates that the amplitude increases with higher temperatures $T$. After analyzing the response function under various parameter conditions, we conclude that the response function is essentially a diffraction pattern resulting from the scattering of a wave source by a BH. To gain deeper insights into the information related to BH shadows, we should specifically introduce an optical observation setup with a convex lens, designed to observe the Einstein ring.

\section{The formation of Einstein ring,}
To observe Einstein rings, it is necessary to introduce an optical setup incorporating a convex lens \cite{Liu:2022cev}. The convex lens serves to convert plane waves into spherical waves, enabling observations at the boundary with a specific angle $\theta _{obs}$. By rotating the coordinates $(\theta, \varphi)$, a new coordinate system is obtained, which satisfies the following relationship\cite{Liu:2022cev}. \begin{equation} \cos\varphi ^{'}  +i\cos\theta ^{'}  =e^{i\theta _{obs} }  (\sin\theta  \cos\varphi +i\cos\theta  ), \end{equation} where $\theta^{'}=0, \varphi^{'}=0$ correspond to the observational center. A Cartesian coordinate system $(x,y,z)$ is established such that  $(x,y)=(\theta^{'} \cos\varphi ^{'}  ,\theta^{'} \sin\varphi ^{'})$ at the boundary where the observer is located, for a virtual optical system. The convex lens is adjusted on a two-dimensional plane $(x,y)$, with the focal length of the lens and the corresponding radius denoted by $f$ and $d$, respectively. Further, the coordinates on the spherical screen are defined as $(x,y,z)=(x_{SC},y_{SC},z_{SC} )$, satisfying $x^{2} _{SC}+y^{2} _{SC}+z^{2} _{SC} =f^{2} $. The relationship between the incident wave $\Psi (\vec{x})$ before passing through the convex lens and the outgoing wave  $\Psi _{T} (\vec{x})$ after passing through the convex lens is given by \begin{equation}
\Psi _{T} (\vec{x} )=e^{-i\tilde{\omega }\frac{\left |\vec{x} \right |^2 }{2f} } \Psi (\vec{x} ).
\end{equation} The wave function on the screen can be expressed as\cite{Liu:2022cev} \begin{eqnarray}
\Psi _{SC} (\vec{x}_{SC}  )&=&\int_{\left |\vec{x}  \right |\le d}d^{2} x\Psi _{T} (\vec{x} )
e^{i\tilde{\omega } R}\nonumber\\&\propto &\int_{\left |\vec{x}  \right |\le d}d^{2} x\Psi(\vec{x} )e^{-i\frac{\tilde{\omega }}{f}
  \vec{x}\cdot \vec{x}_{SC} }\nonumber\\&=&\int d^{2} x\Psi (\vec{x} )\sigma (\vec{x})e^{-i\frac{\tilde{\omega }}{f}
  \vec{x}\cdot \vec{x}_{SC} },\label{bo}
\end{eqnarray} where $R$ is the distance from the point $(x,y,0)$ on the lens to the point $(x^{2} _{SC},y^{2} _{SC},z^{2} _{SC} )$ on the screen. In this equation, $\sigma (\vec{x})$ is the window function, defined as \begin{equation}
\sigma (\vec{x}): = 
\begin{cases}
  1,~~~~0\le\left |\vec{x}  \right | \le d;  \\
  0,~~~~~~~~~~~\left |\vec{x}  \right | >  d.
\end{cases} 
\end{equation}
From Eq.(\ref{bo}), it is evident  that the observed wave on the screen is related to the incident wave through a Fourier transform. In this study, the response function is considered as the incident wave $\Psi (\vec{x}$), holographic images can be observed on this screen using Eq.(\ref{bo}). The effects of relevant physical parameters on holographic images are investigated, with the source width  $\eta=0.02$ and radius of convex lens $d=0.6$.

\subsection{Impact of parameters dependent on spacetime structure on Einstein rings}

Figure {\ref8} explores the impact of the observer's position $\theta_{obs}$ on the Einstein rings. At $\theta_{obs}=0$, the observer is located at the North Pole of the AdS boundary. As $\theta_{obs}$ increases from $0$ to $\frac{\pi}{3}$ and then to $\frac{\pi}{2}$, the ring gradually disappears, with only bright spots remaining when $\theta_{obs}$ reaches $\frac{\pi}{2}$. Therefore, the shape of the Einstein ring is highly dependent on the observer's position. Figure {\ref9} examines the impact of parameter $a$ on Einstein rings, showing that as $a$ increases, the ring radius of the Einstein ring decreases. This effect is more clearly observed in Figure {\ref10}. The x-coordinate values corresponding to the peaks in Figure {\ref10} represent the ring radii of the Einstein rings, while the y-axis represents the luminosity. It is evident that when $a=0.03, 0.08, 0.13, 0.18$, the x-coordinates corresponding to the peaks are $0.52, 0.51, 0.50$ and $0.45$, respectively, indicating that as parameter $a$ increases, the ring radius of the Einstein ring decreases.

\begin{figure}[htbp]
  \centering
  \begin{subfigure}[b]{0.48\columnwidth}
    \centering
    \includegraphics[width=\textwidth,height=0.8\textwidth]{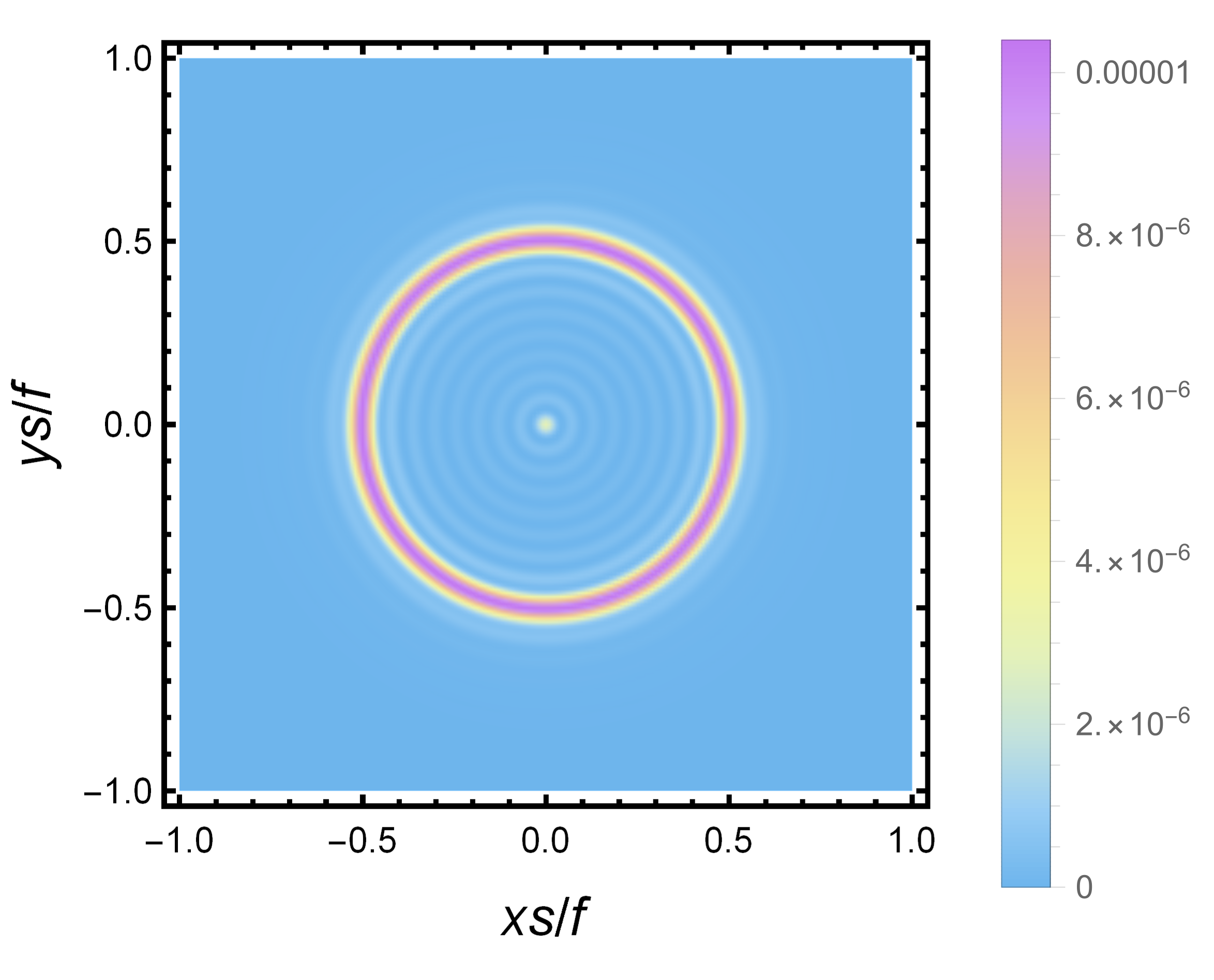}
    \caption{$\theta_{obs}=0$}
  \end{subfigure}
  \hfill
  \begin{subfigure}[b]{0.48\columnwidth}
    \centering
    \includegraphics[width=\textwidth,height=0.8\textwidth]{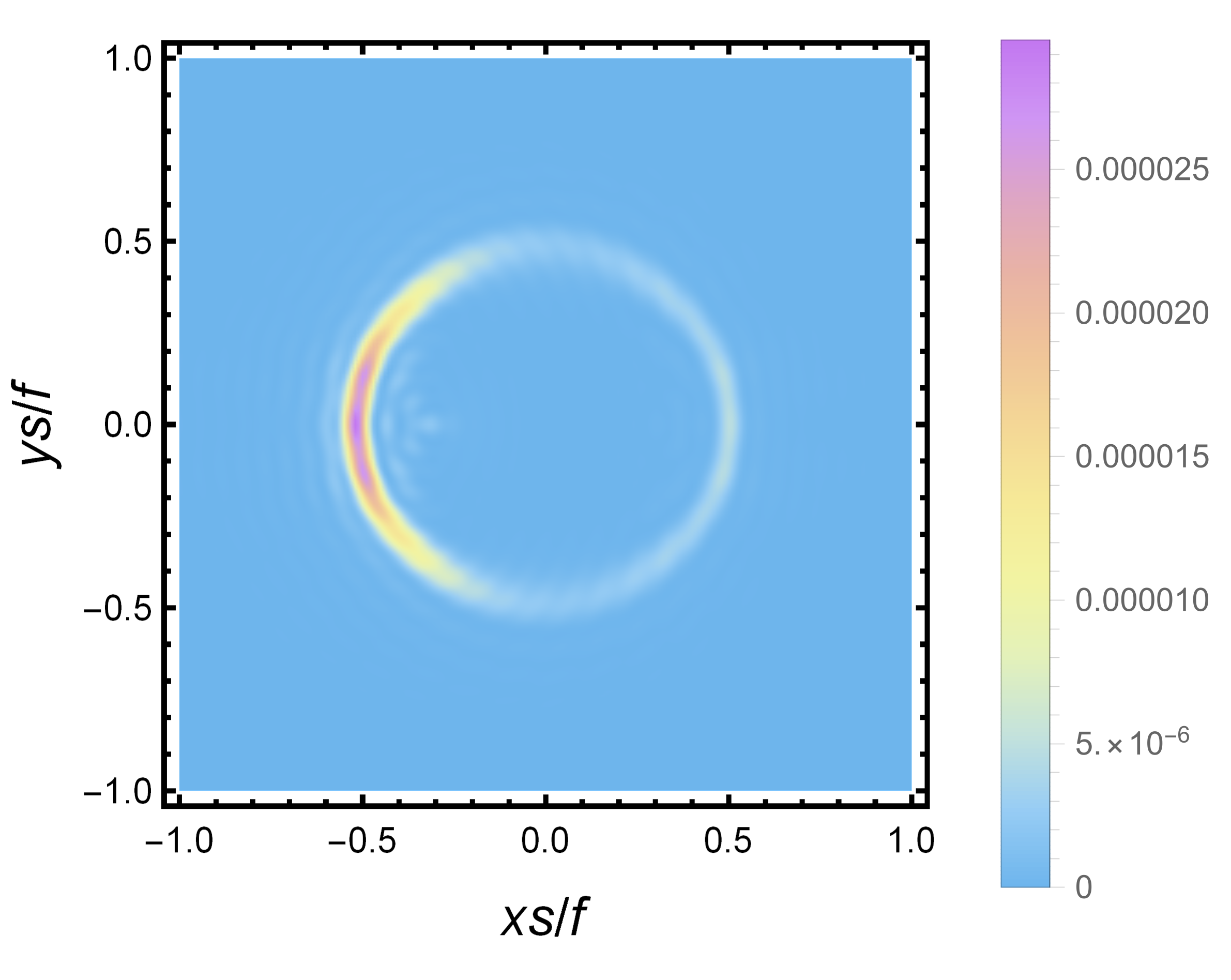}
    \caption{$\theta_{obs}=30$}
  \end{subfigure}
\begin{subfigure}[b]{0.48\columnwidth}
    \centering
    \includegraphics[width=\textwidth,height=0.8\textwidth]{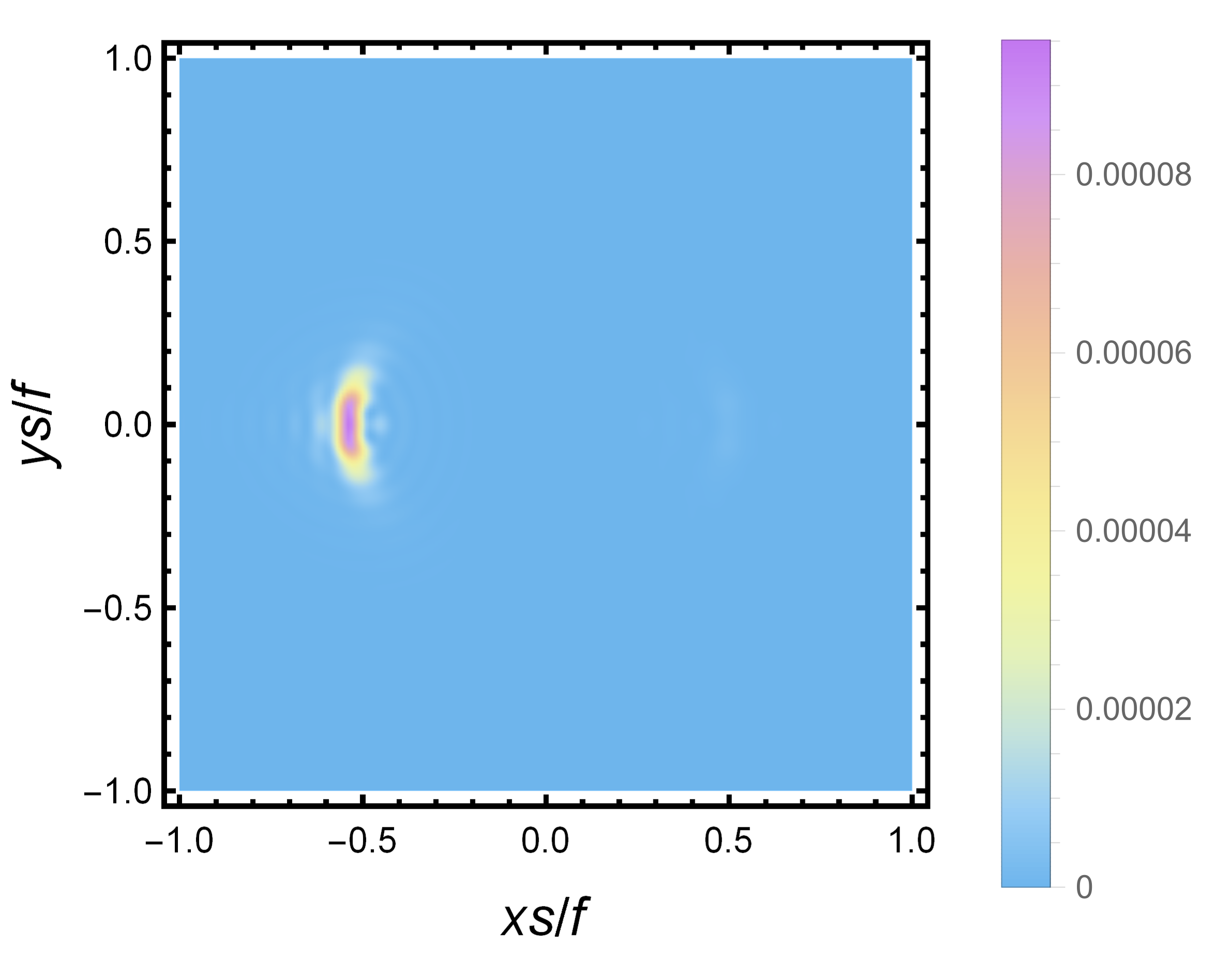}
    \caption{$\theta_{obs}=60$}
  \end{subfigure}
  \hfill
  \begin{subfigure}[b]{0.48\columnwidth}
    \centering
    \includegraphics[height=0.8\textwidth,width=\textwidth]{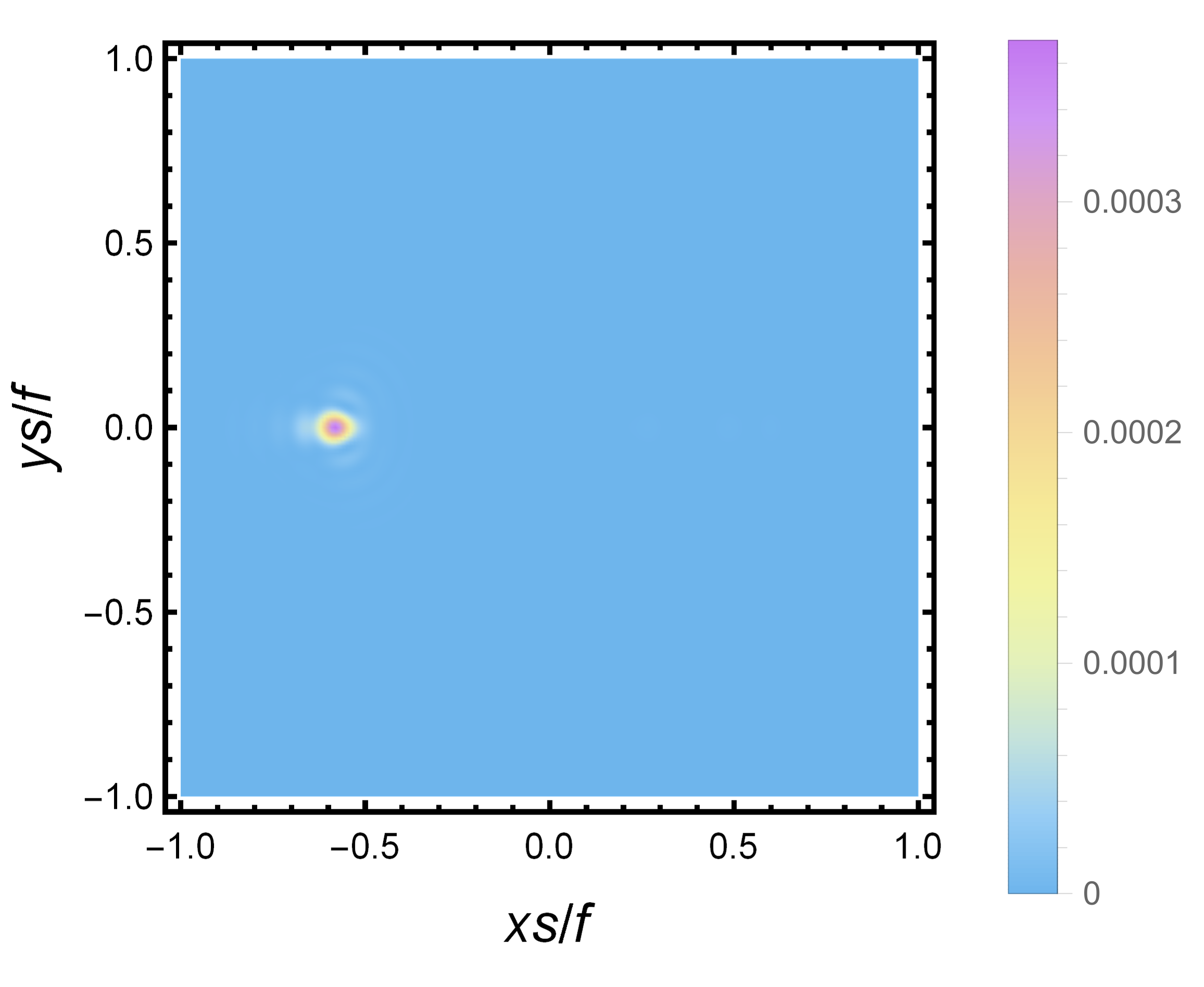}
    \caption{$\theta_{obs}=90$}
  \end{subfigure}
  \caption{Effect of the observer's position $\theta_{obs}$ on the Einstein ring, where $a=0.1$, $c=0.05$, $\Omega=-\frac{2}{3}$, $Q=0.1$, $e=0.5$, $y_{h}=5$, $\omega=90$.}
  \label{8}%
\end{figure}

\begin{figure}[htbp]
  \centering
  \begin{subfigure}[b]{0.48\columnwidth}
    \centering
    \includegraphics[width=\textwidth,height=0.8\textwidth]{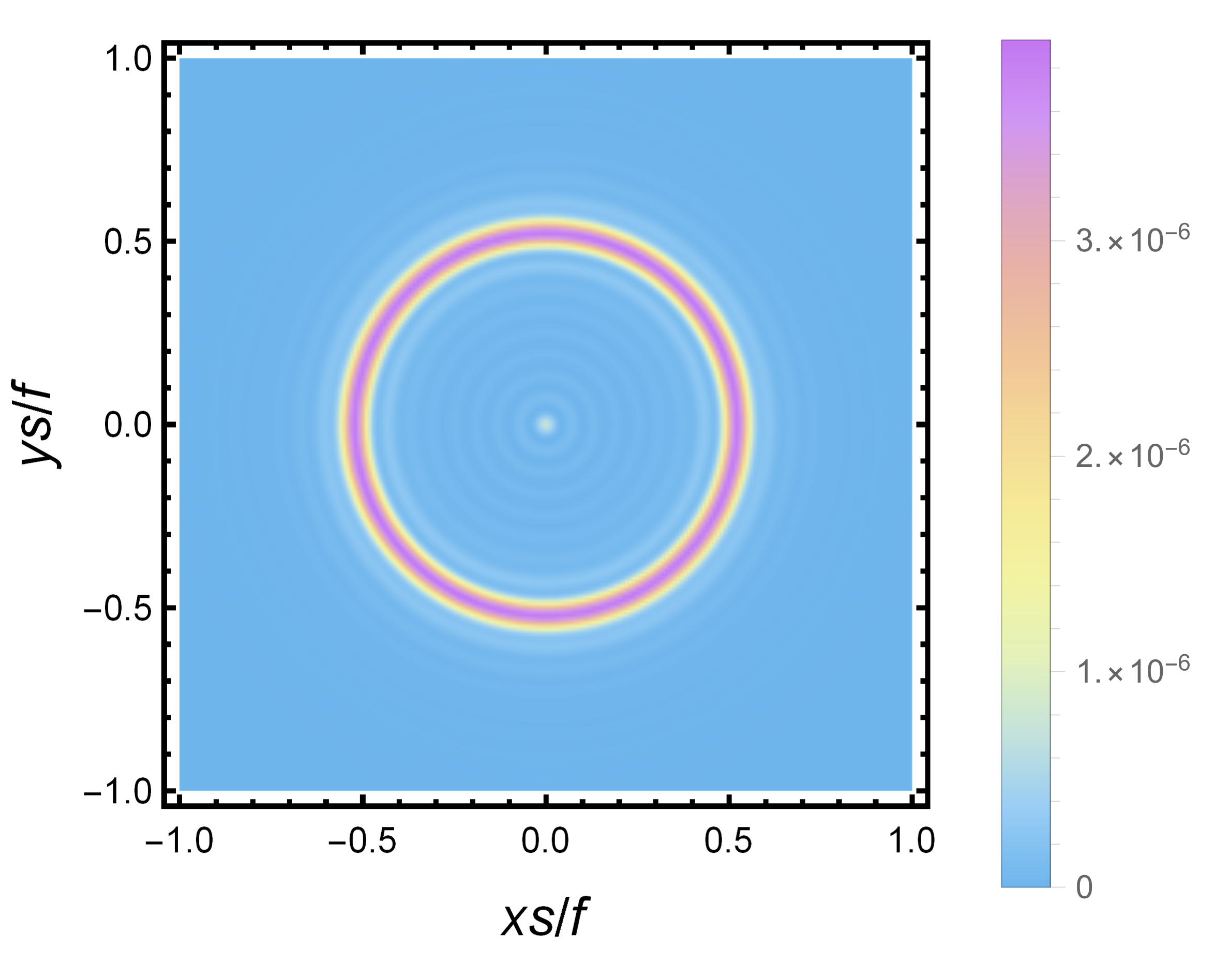}
    \caption{$a=0.03$}
  \end{subfigure}
  \hfill
  \begin{subfigure}[b]{0.48\columnwidth}
    \centering
    \includegraphics[width=\textwidth,height=0.8\textwidth]{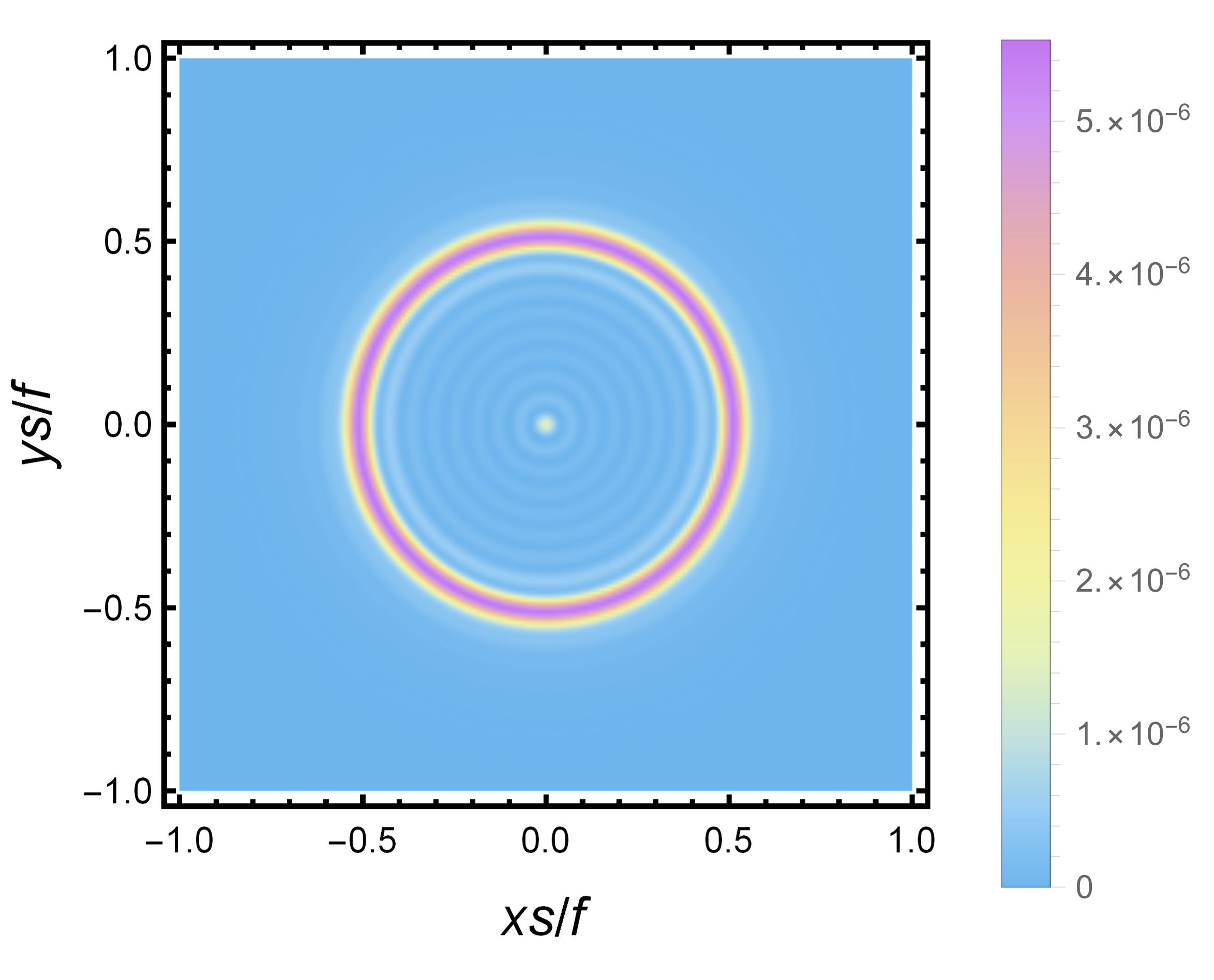}
    \caption{$a=0.08$}
  \end{subfigure}
\begin{subfigure}[b]{0.48\columnwidth}
    \centering
    \includegraphics[width=\textwidth,height=0.8\textwidth]{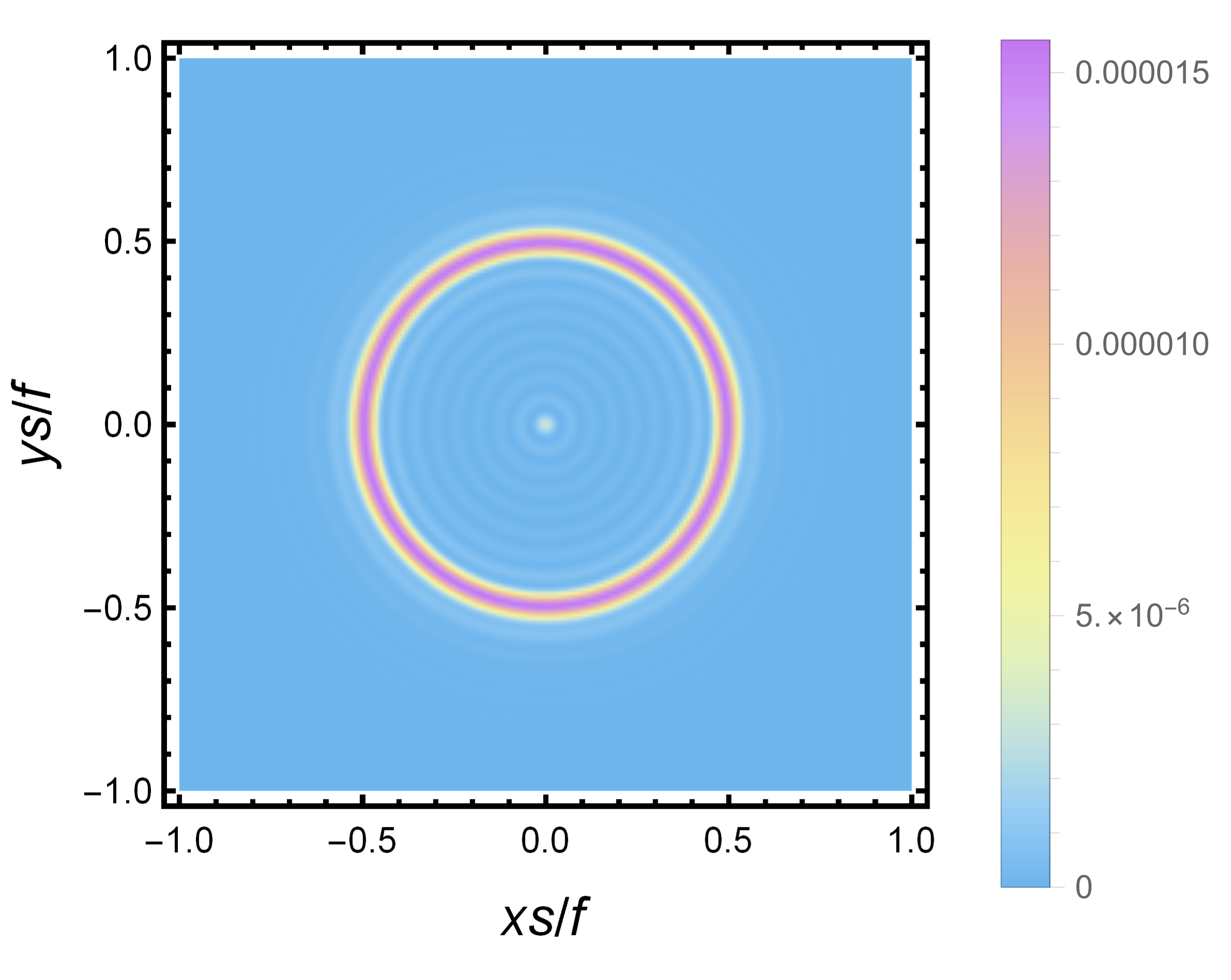}
    \caption{$a=0.13$}
  \end{subfigure}
  \hfill
  \begin{subfigure}[b]{0.48\columnwidth}
    \centering
    \includegraphics[height=0.8\textwidth,width=\textwidth]{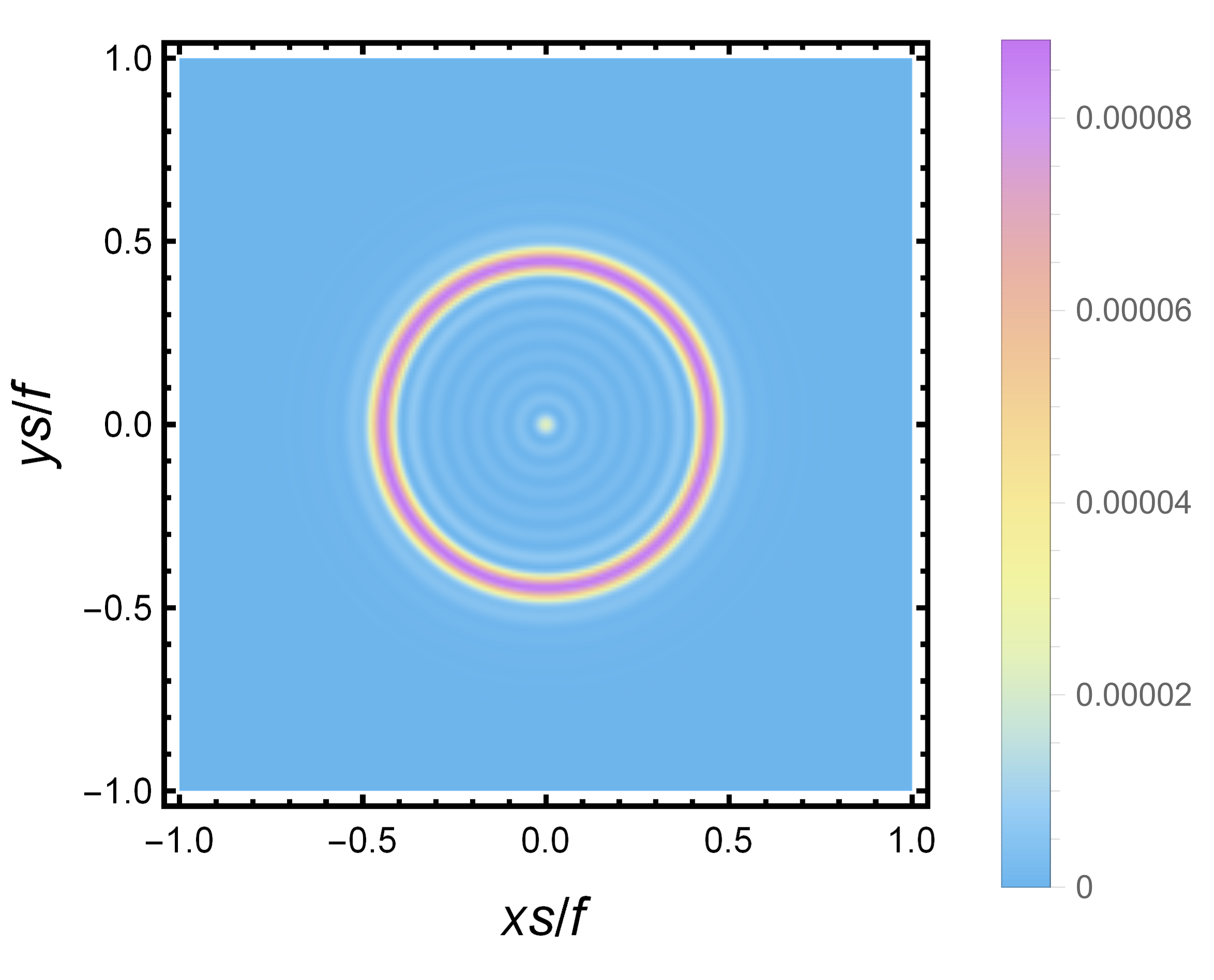}
    \caption{$a=0.18$}
  \end{subfigure}
  \caption{Effect of $a$ on the Einstein ring, where $\theta_{obs}=0$, $c=0.1$, $\Omega=-\frac{2}{3}$, $Q=0.1$, $e=0.5$, $y_{h}=5$, $\  \omega=90$.}
  \label{9}%
\end{figure}

\begin{figure}[htbp]
  \centering
  \begin{subfigure}[b]{0.48\columnwidth}
    \centering
    \includegraphics[width=\textwidth,height=0.8\textwidth]{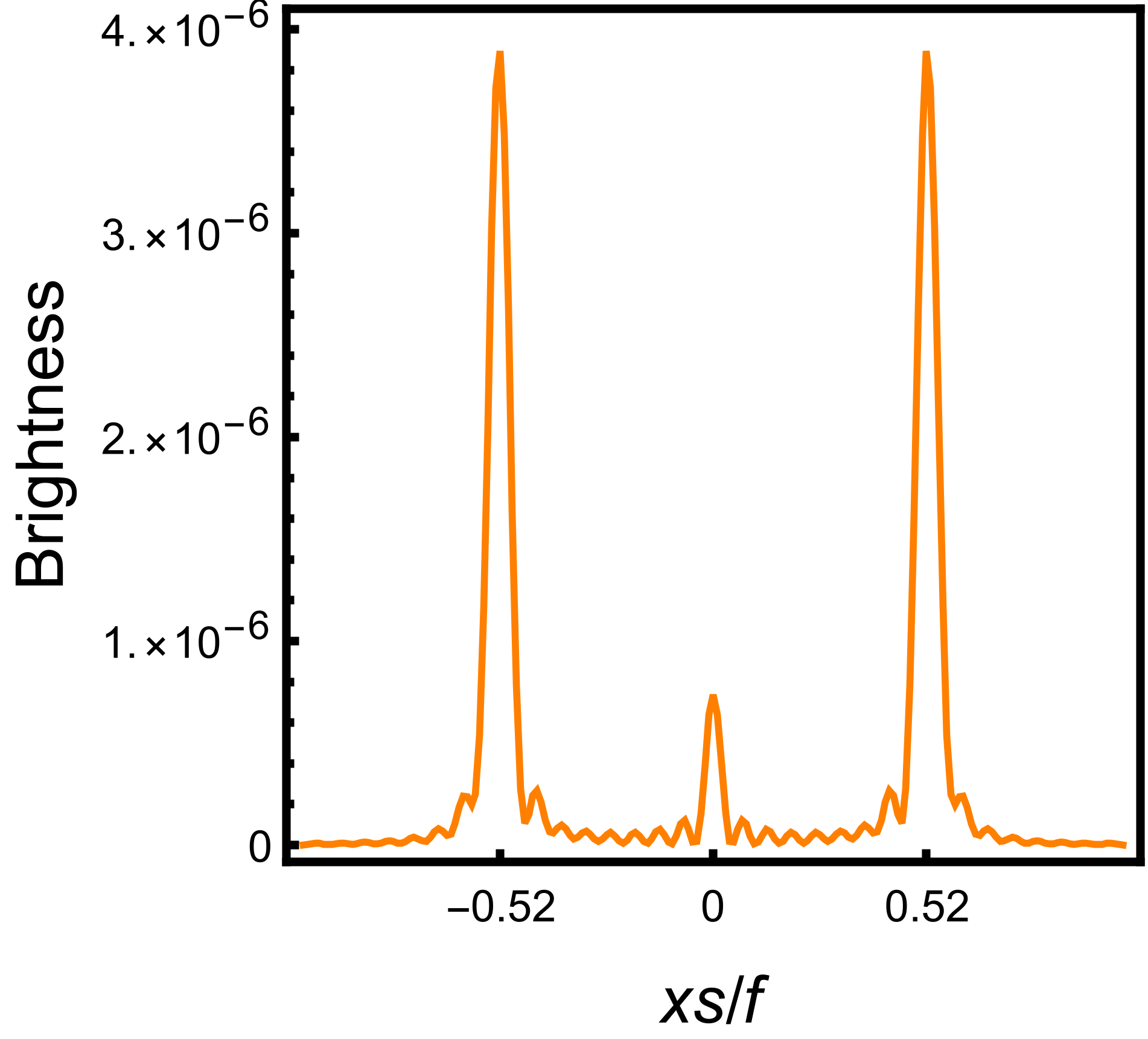}
    \caption{$a=0.03$}
  \end{subfigure}
  \hfill
  \begin{subfigure}[b]{0.48\columnwidth}
    \centering
    \includegraphics[width=\textwidth,height=0.8\textwidth]{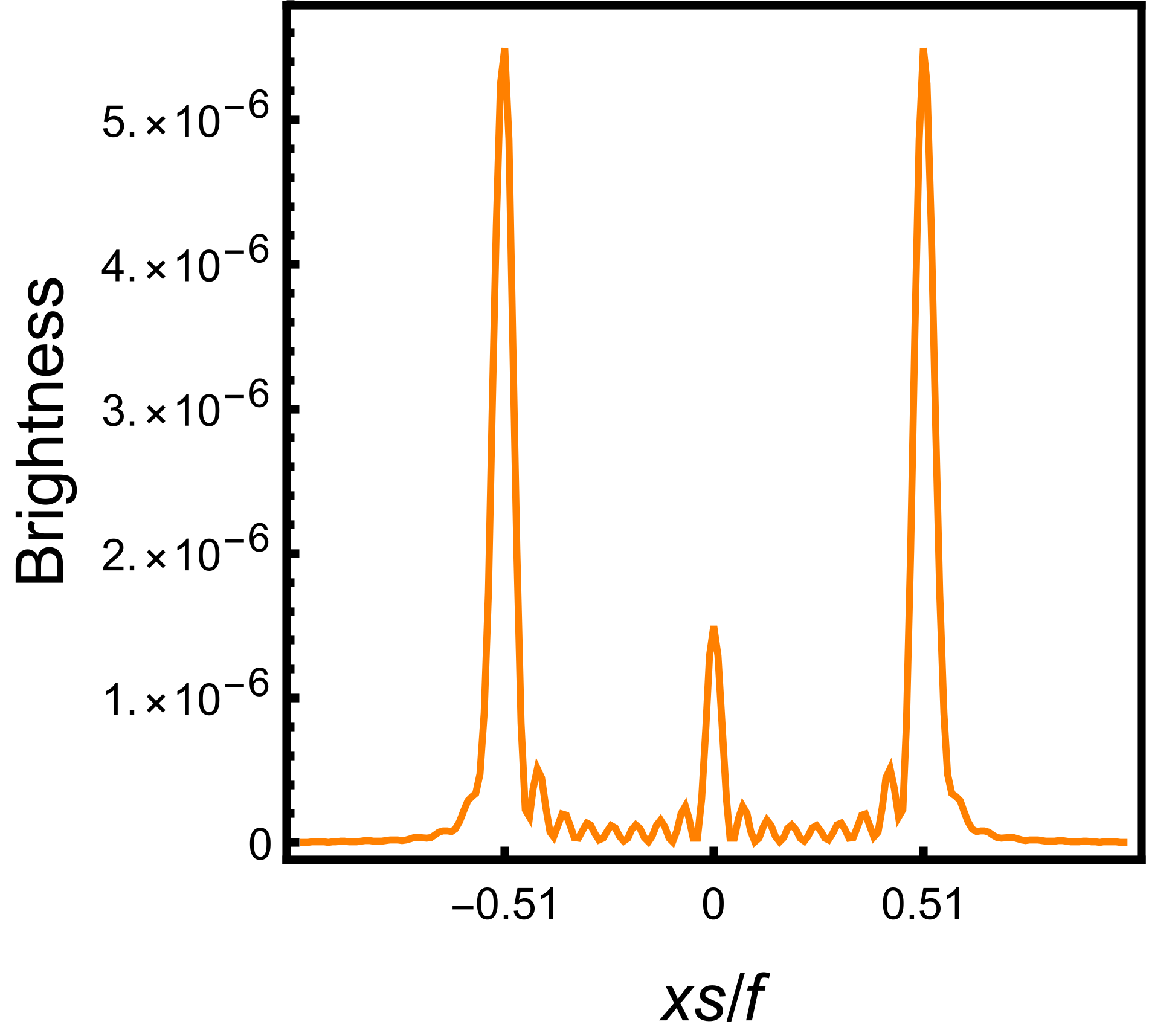}
    \caption{$a=0.08$}
  \end{subfigure}
\begin{subfigure}[b]{0.48\columnwidth}
    \centering
    \includegraphics[width=\textwidth,height=0.8\textwidth]{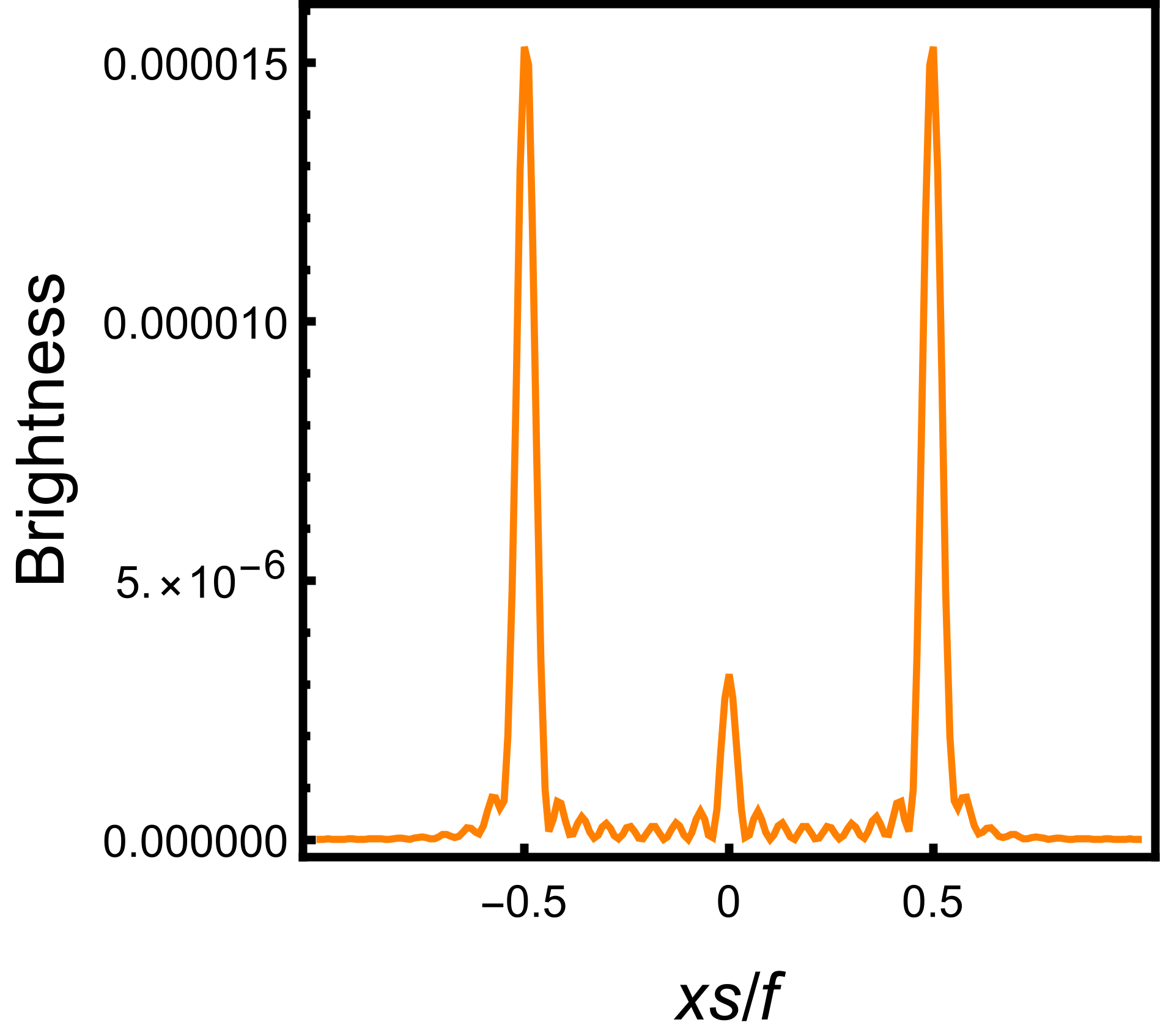}
    \caption{$a=0.13$}
  \end{subfigure}
  \hfill
  \begin{subfigure}[b]{0.48\columnwidth}
    \centering
    \includegraphics[height=0.8\textwidth,width=\textwidth]{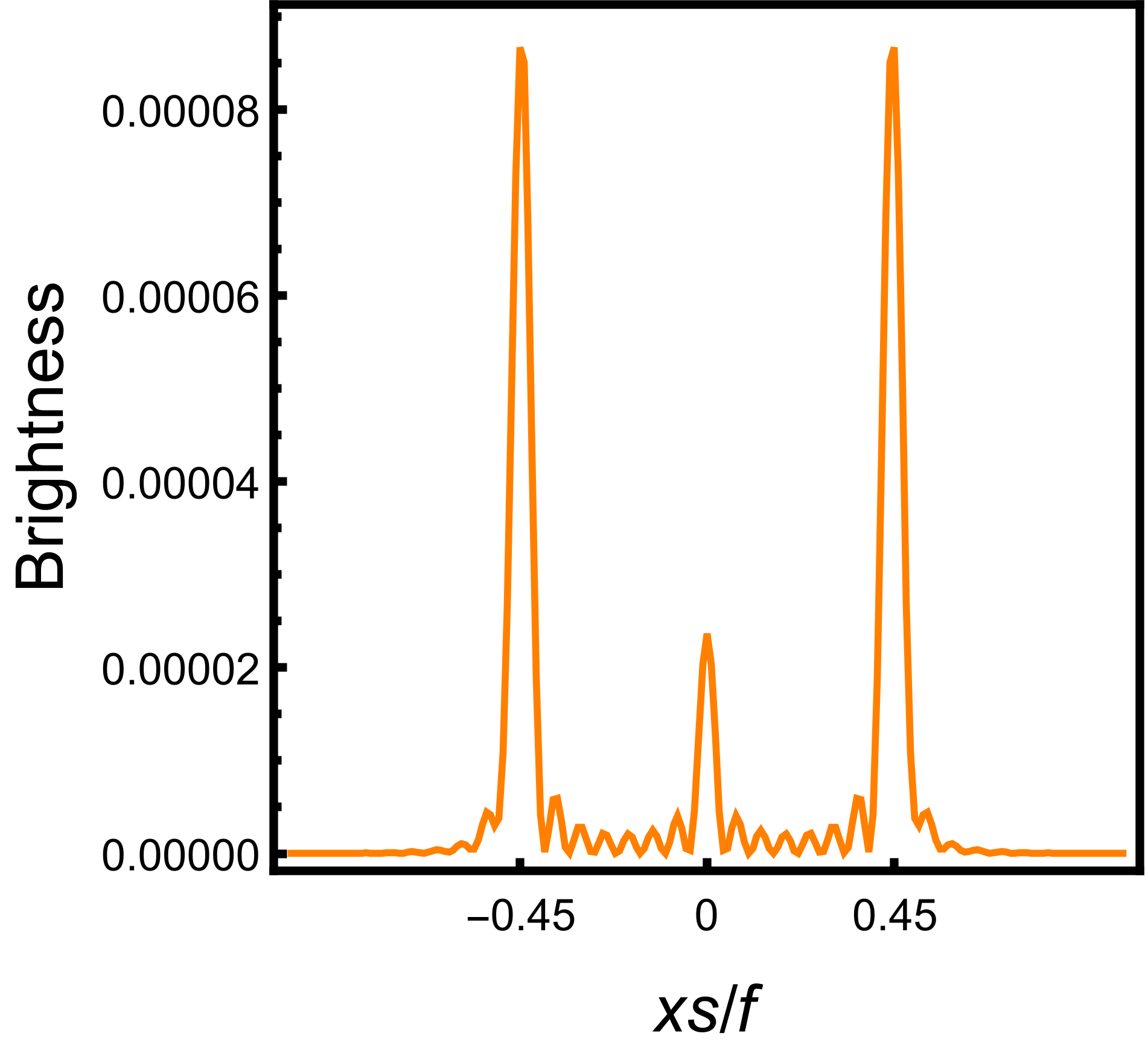}
    \caption{$a=0.18$}
  \end{subfigure}
  \caption{Effect of $a$ on the brightness, where $\theta_{obs}=0$, $c=0.1$, $\Omega=-\frac{2}{3}$, $Q=0.1$, $e=0.5$, $y_{h}=5$, $\omega=90$.}
  \label{10}%
\end{figure}

Similarly, Figure \ref{11} examines the impact of parameter $c$ on Einstein rings, showing that as $c$ increases, the ring radius of the Einstein ring increases. Figure \ref{12} shows the corresponding luminosity diagram. The luminosity reaches its maximum when $c=0.05$, $xs/f=0.5$. As $c$ gradually increases, the x-coordinate peaks of the luminosity shift, indicating that the ring radius increases with $c$. Figure \ref{13} explores the impact of parameter $\Omega$ on Einstein rings. While this impact may not be immediately obvious in the direct graph presentation, a more comprehensive analysis, in conjunction with the luminosity diagram in Figure\ref{14}, reveals that as the $\Omega$ increases, the radius of the Einstein ring decreases. Specifically, when $\Omega=-1$ and the luminosity reaches its maximum, $xs/f=0.7$. Furthermore, when $\Omega=-1/6$ and $\Omega=1/3$, and the luminosity peaks, $xs/f=0.68$ and $xs/f=0.61$, respectively. This observation further reinforces the negative correlation between $\Omega$ and the radius of the Einstein ring.

The impact of wave sources, such as the wave source frequency $\omega$, on the Einstein ring can also be analyzed. As shown in Figure \ref{15}, a clear trend emerges where the resolution of the Einstein ring improves as the wave source frequency increases. At $\omega=30$, multiple diffraction rings are observable, indicating significant interference effects. However, as the frequency rises to $\omega=120$, these additional diffraction rings gradually fade, leaving only the primary Einstein ring distinct. This transition can be attributed to the increasing dominance of geometric optics at higher frequencies. To further support this understanding, Figure \ref{16} illustrates the correlation between luminosity and  wave source frequency. It is evident that high frequencies are crucial for effectively observing and resolving the radius of the ring.

\begin{figure}[htbp]
  \centering
  \begin{subfigure}[b]{0.48\columnwidth}
    \centering
    \includegraphics[width=\textwidth,height=0.8\textwidth]{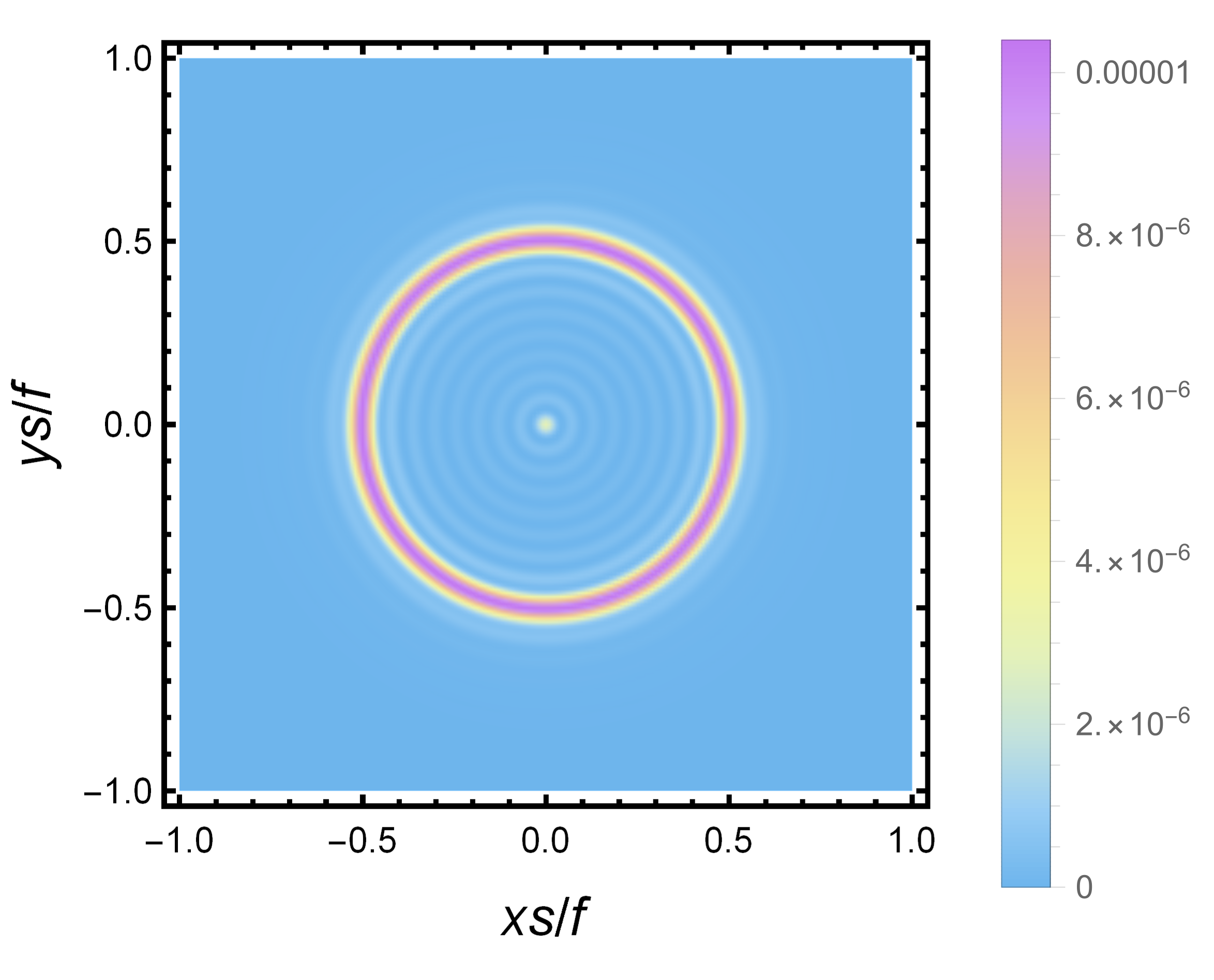}
    \caption{$c=0.05$}
  \end{subfigure}
  \hfill
  \begin{subfigure}[b]{0.48\columnwidth}
    \centering
    \includegraphics[width=\textwidth,height=0.8\textwidth]{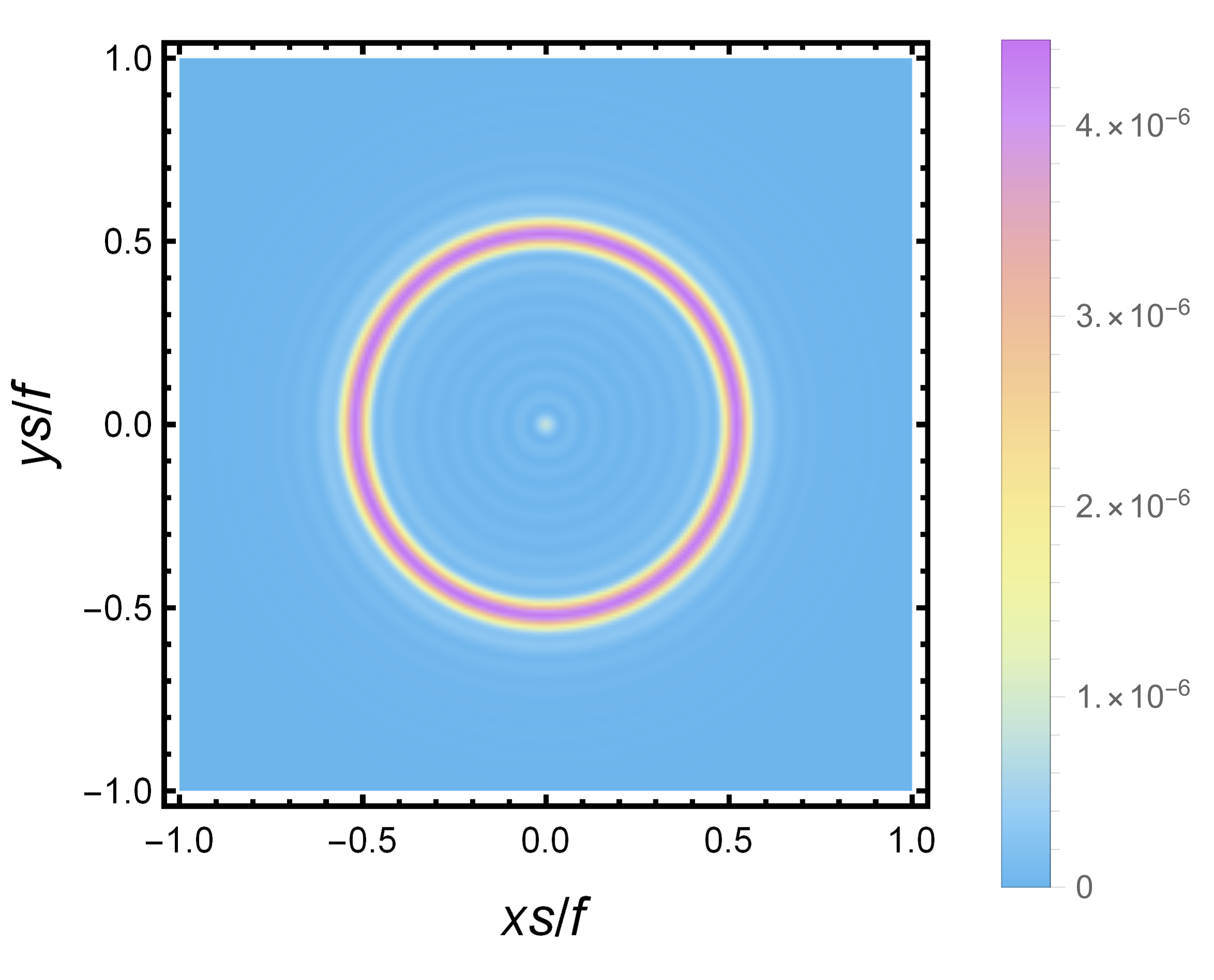}
    \caption{$c=0.25$}
  \end{subfigure}
\begin{subfigure}[b]{0.48\columnwidth}
    \centering
    \includegraphics[width=\textwidth,height=0.8\textwidth]{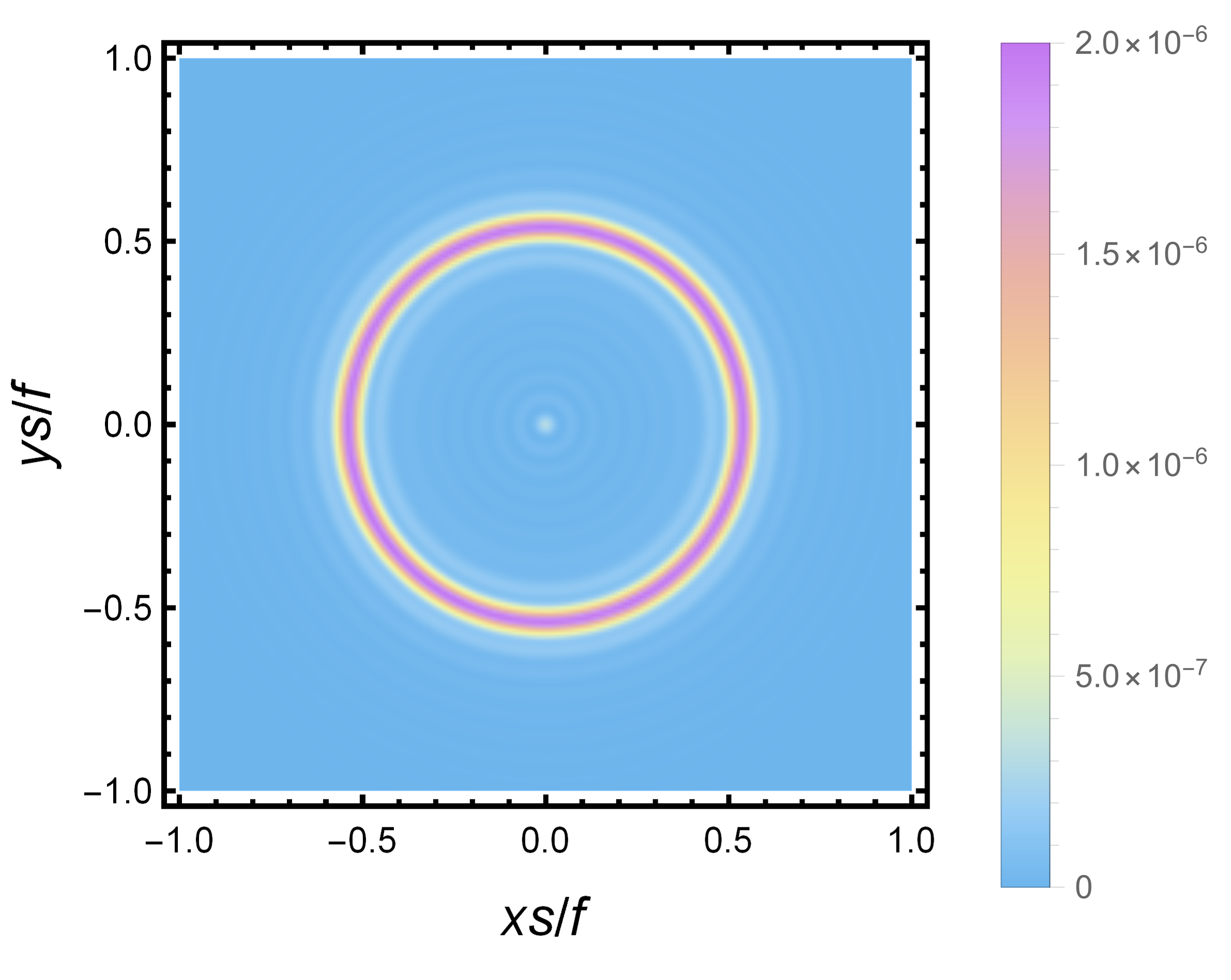}
    \caption{$c=0.45$}
  \end{subfigure}
  \hfill
  \begin{subfigure}[b]{0.48\columnwidth}
    \centering
    \includegraphics[height=0.8\textwidth,width=\textwidth]{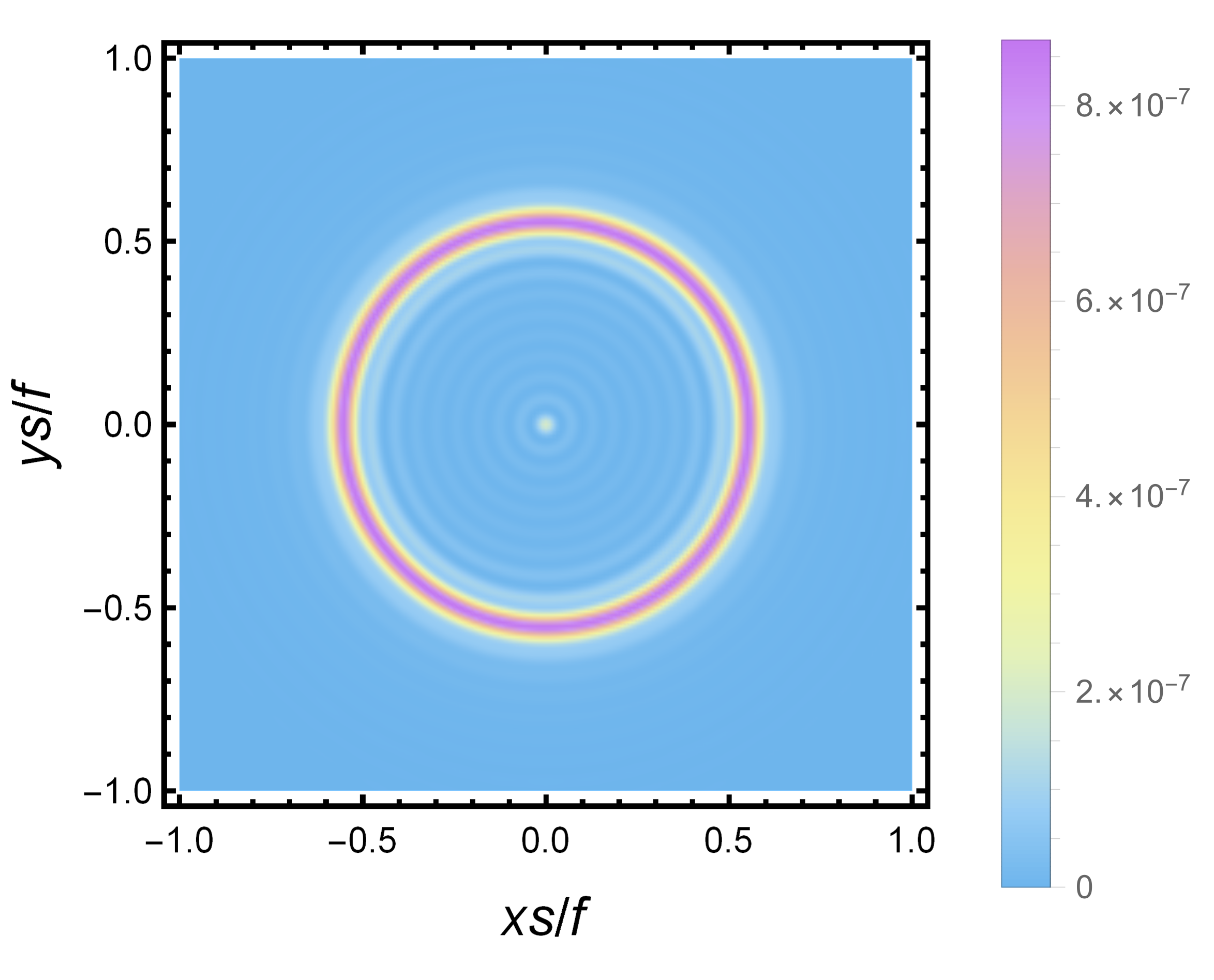}
    \caption{$c=0.65$}
  \end{subfigure}
  \caption{Effect of $c$ on the Einstein ring, where $a=0.1$, $\theta_{obs}=0$, $\Omega=-\frac{2}{3}$, $Q=0.1$, $e=0.5$, $y_{h}=5$, $\omega=90$.}
  \label{11}%
\end{figure}

\begin{figure}[htbp]
  \centering
  \begin{subfigure}[b]{0.48\columnwidth}
    \centering
    \includegraphics[width=\textwidth,height=0.8\textwidth]{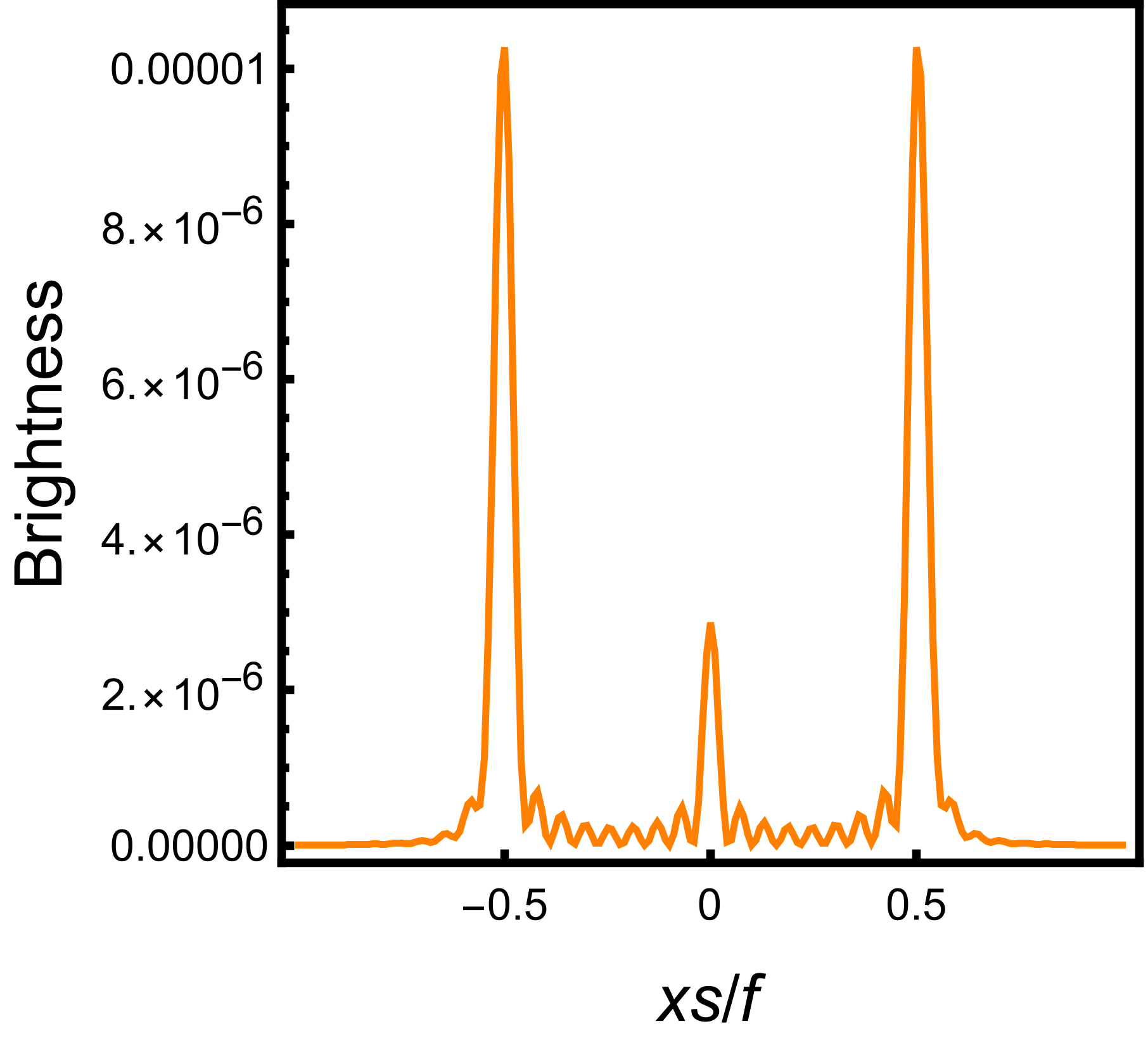}
    \caption{$c=0.05$}
  \end{subfigure}
  \hfill
  \begin{subfigure}[b]{0.48\columnwidth}
    \centering
    \includegraphics[width=\textwidth,height=0.8\textwidth]{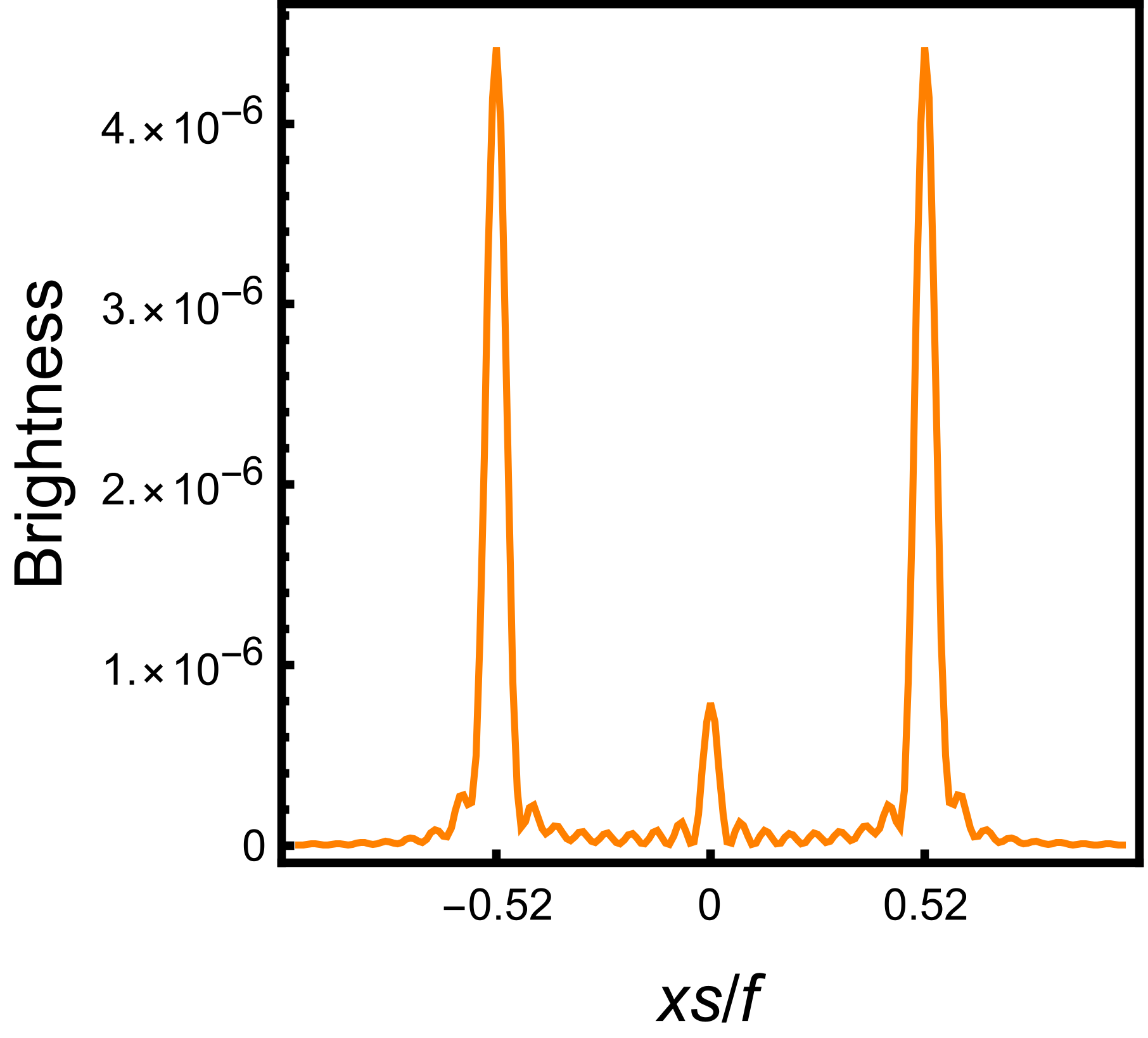}
    \caption{$c=0.25$}
  \end{subfigure}
\begin{subfigure}[b]{0.48\columnwidth}
    \centering
    \includegraphics[width=\textwidth,height=0.8\textwidth]{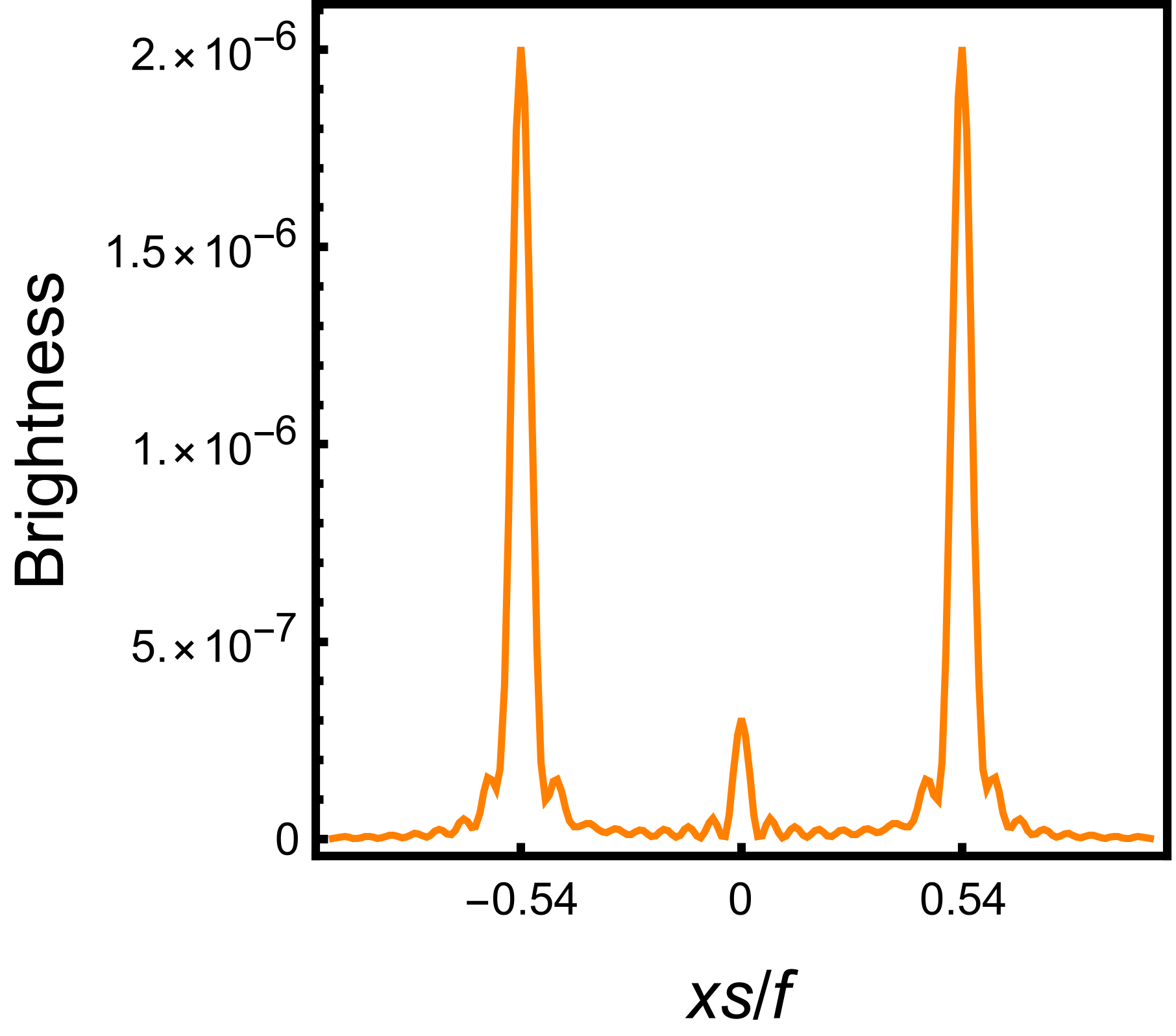}
    \caption{$c=0.45$}
  \end{subfigure}
  \hfill
  \begin{subfigure}[b]{0.48\columnwidth}
    \centering
    \includegraphics[height=0.8\textwidth,width=\textwidth]{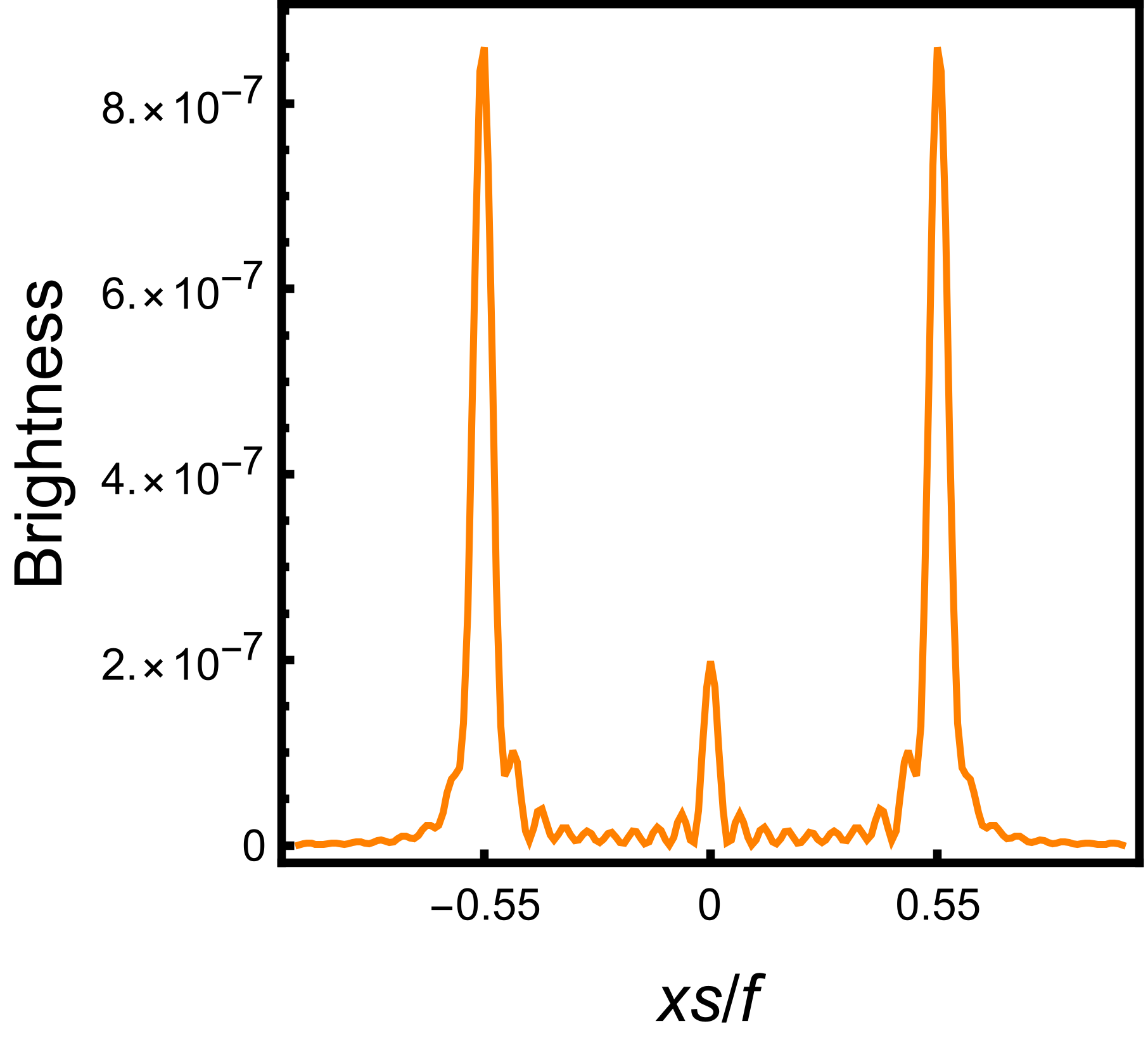}
    \caption{$c=0.65$}
  \end{subfigure}
  \caption{Effect of $c$ on the brightness, where $\theta_{obs}=0$, $a=0.1$, $\Omega=-\frac{2}{3}$, $Q=0.1$, $e=0.5$, $y_{h}=5$, $\omega=90$.}
  \label{12}%
\end{figure}

\begin{figure}[htbp]
  \centering
  \begin{subfigure}[b]{0.48\columnwidth}
    \centering
    \includegraphics[width=\textwidth,height=0.8\textwidth]{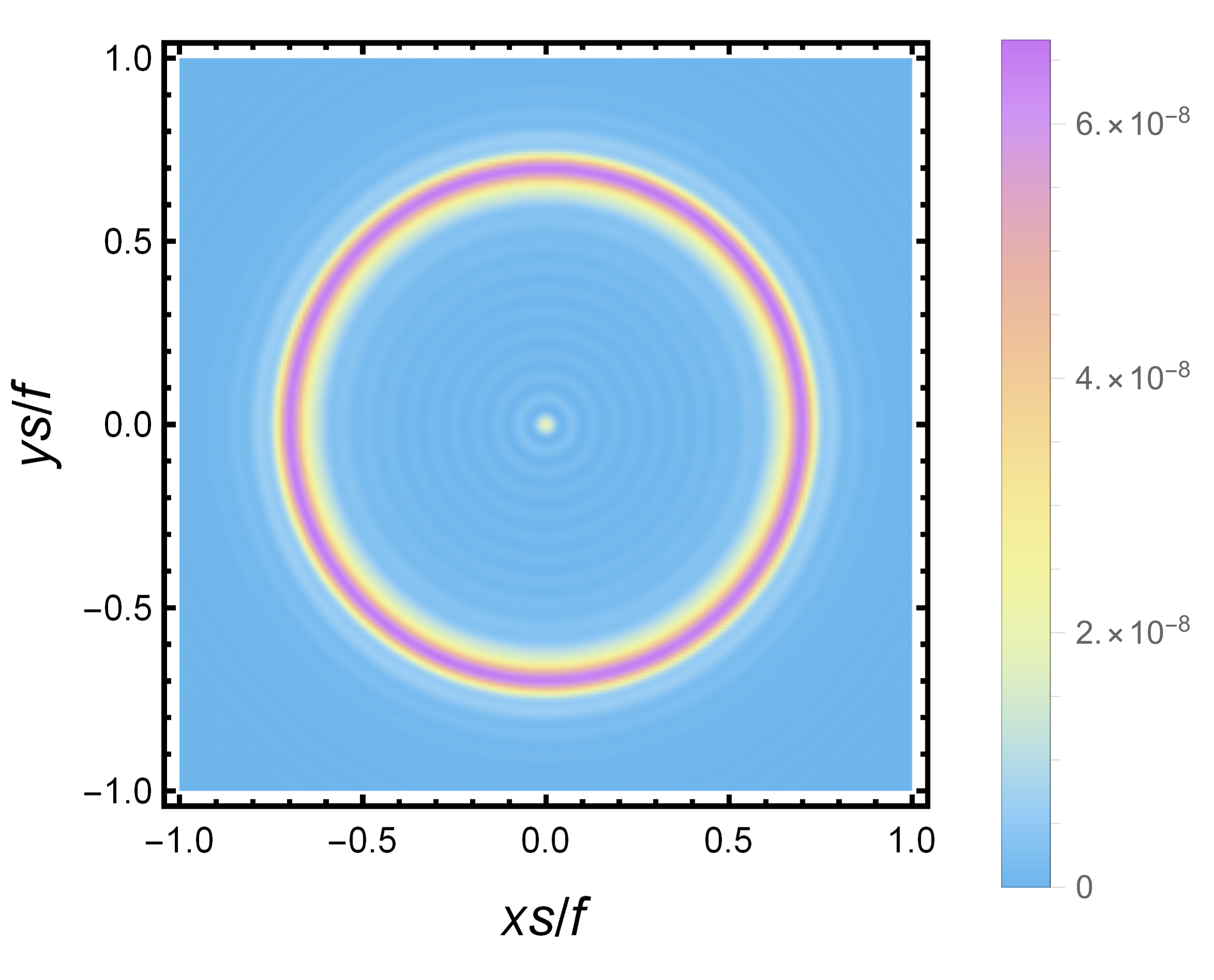}
    \caption{$\Omega=-1$}
  \end{subfigure}
  \hfill
  \begin{subfigure}[b]{0.48\columnwidth}
    \centering
    \includegraphics[width=\textwidth,height=0.8\textwidth]{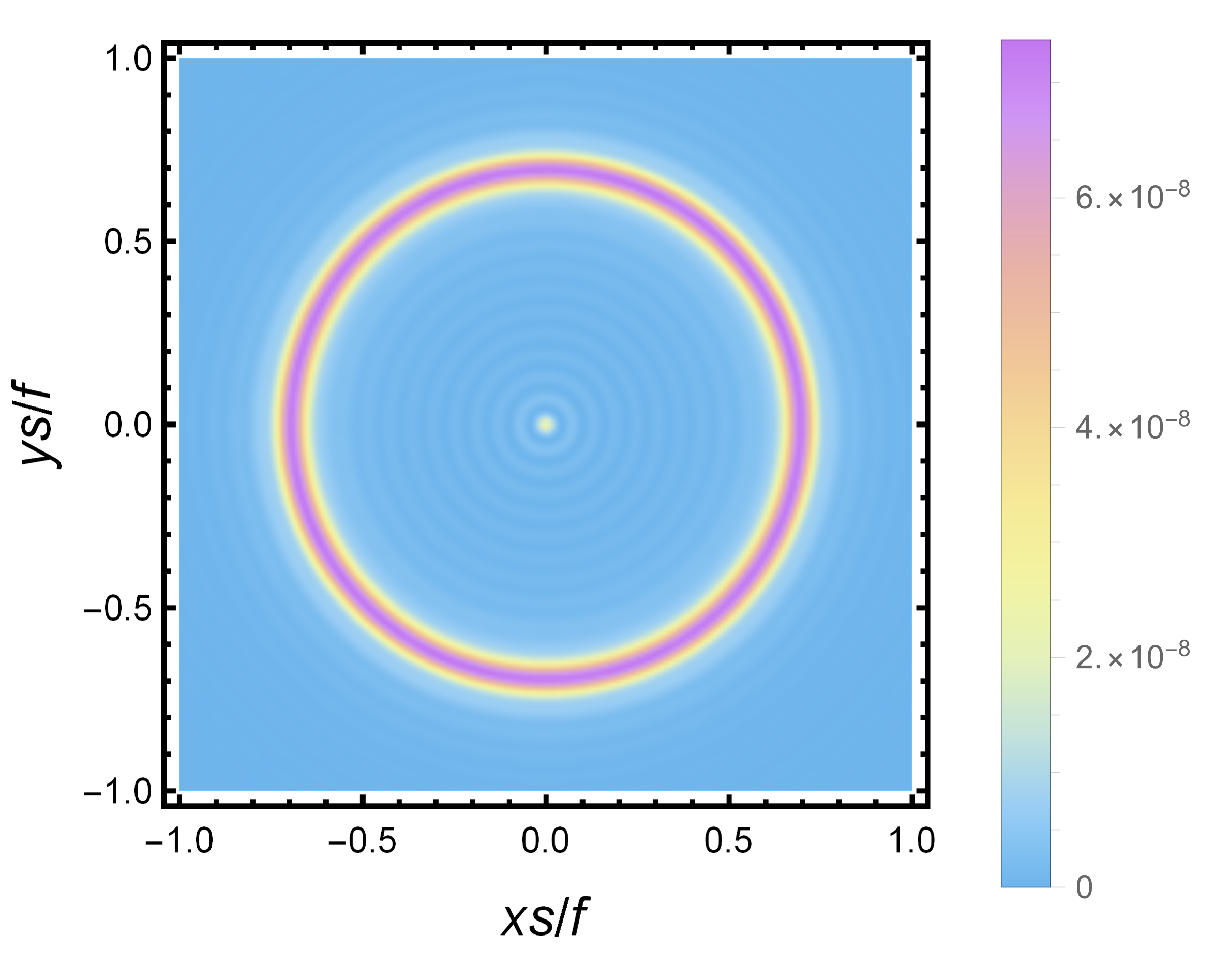}
    \caption{$\Omega=-1/3$}
  \end{subfigure}
\begin{subfigure}[b]{0.48\columnwidth}
    \centering
    \includegraphics[width=\textwidth,height=0.8\textwidth]{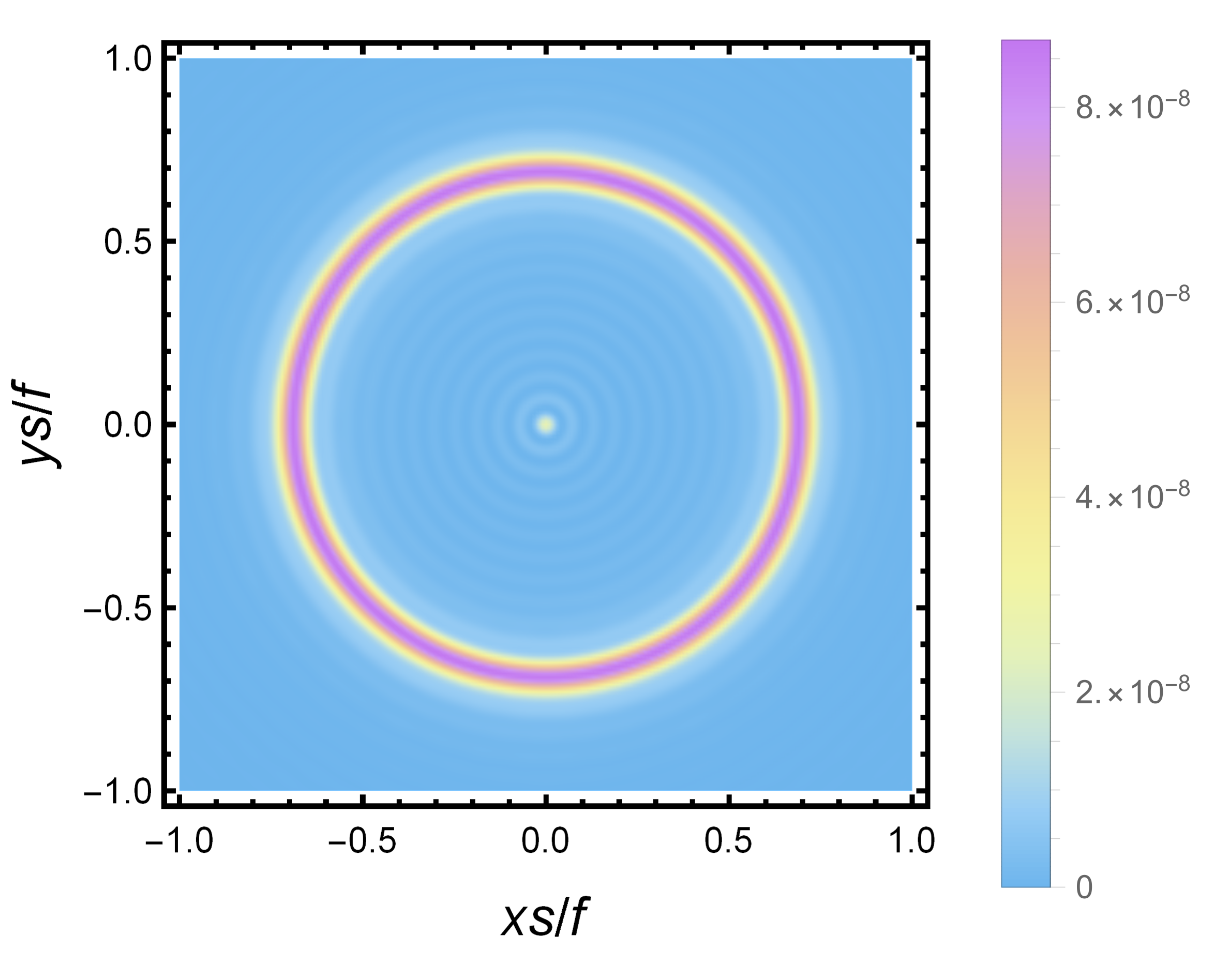}
    \caption{$\Omega=-1/6$}
  \end{subfigure}
  \hfill
  \begin{subfigure}[b]{0.48\columnwidth}
    \centering
    \includegraphics[height=0.8\textwidth,width=\textwidth]{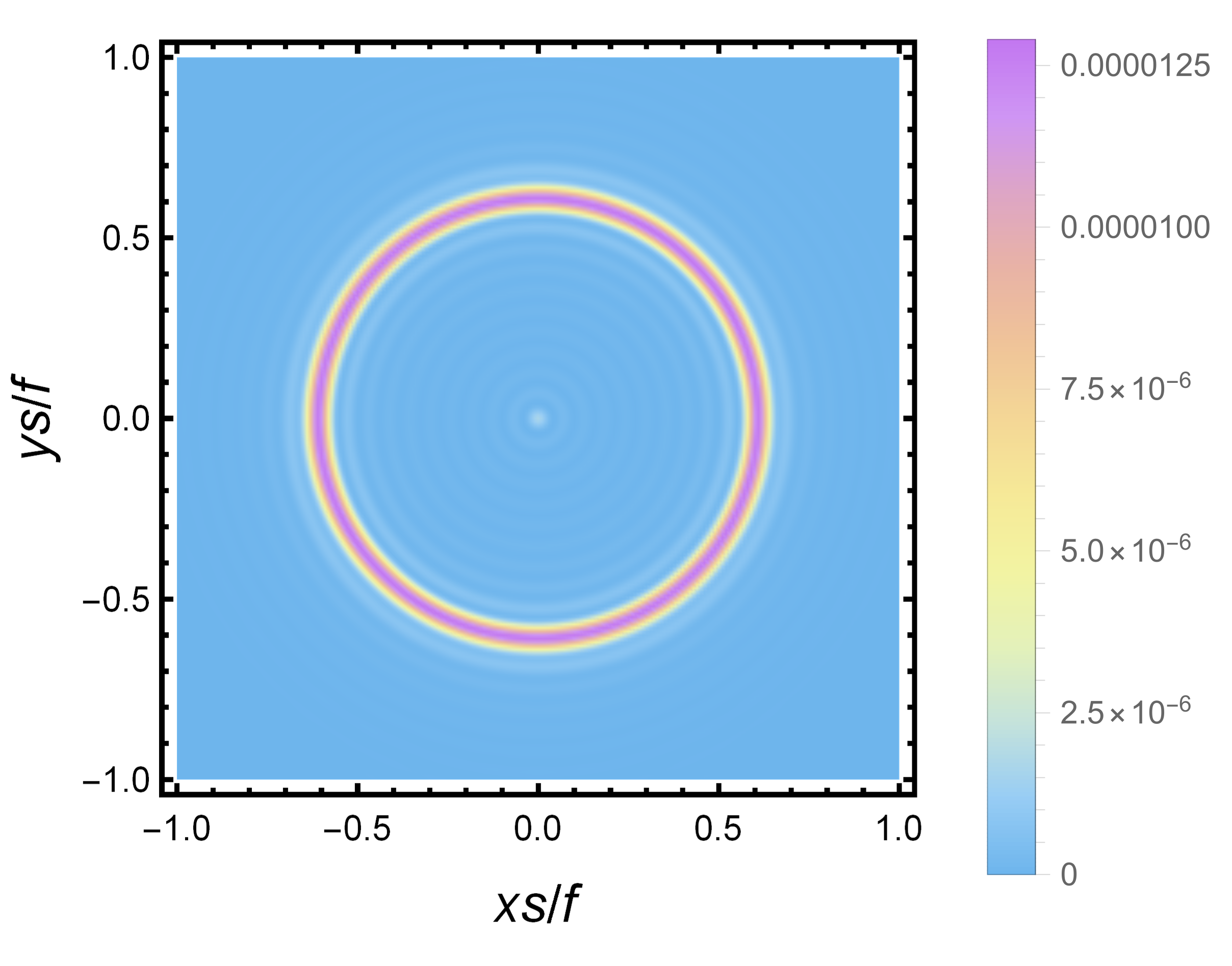}
    \caption{$\Omega=1/3$}
  \end{subfigure}
  \caption{Effect of $\Omega$ on the Einstein ring, where $\theta_{obs}=0$, $a=0.1$, $c=0.1$, $Q=0.1$, $e=0.5$, $y_{h}=3$, $\omega=90$.}
  \label{13}%
\end{figure}

\begin{figure}[htbp]
  \centering
  \begin{subfigure}[b]{0.48\columnwidth}
    \centering
    \includegraphics[width=\textwidth,height=0.8\textwidth]{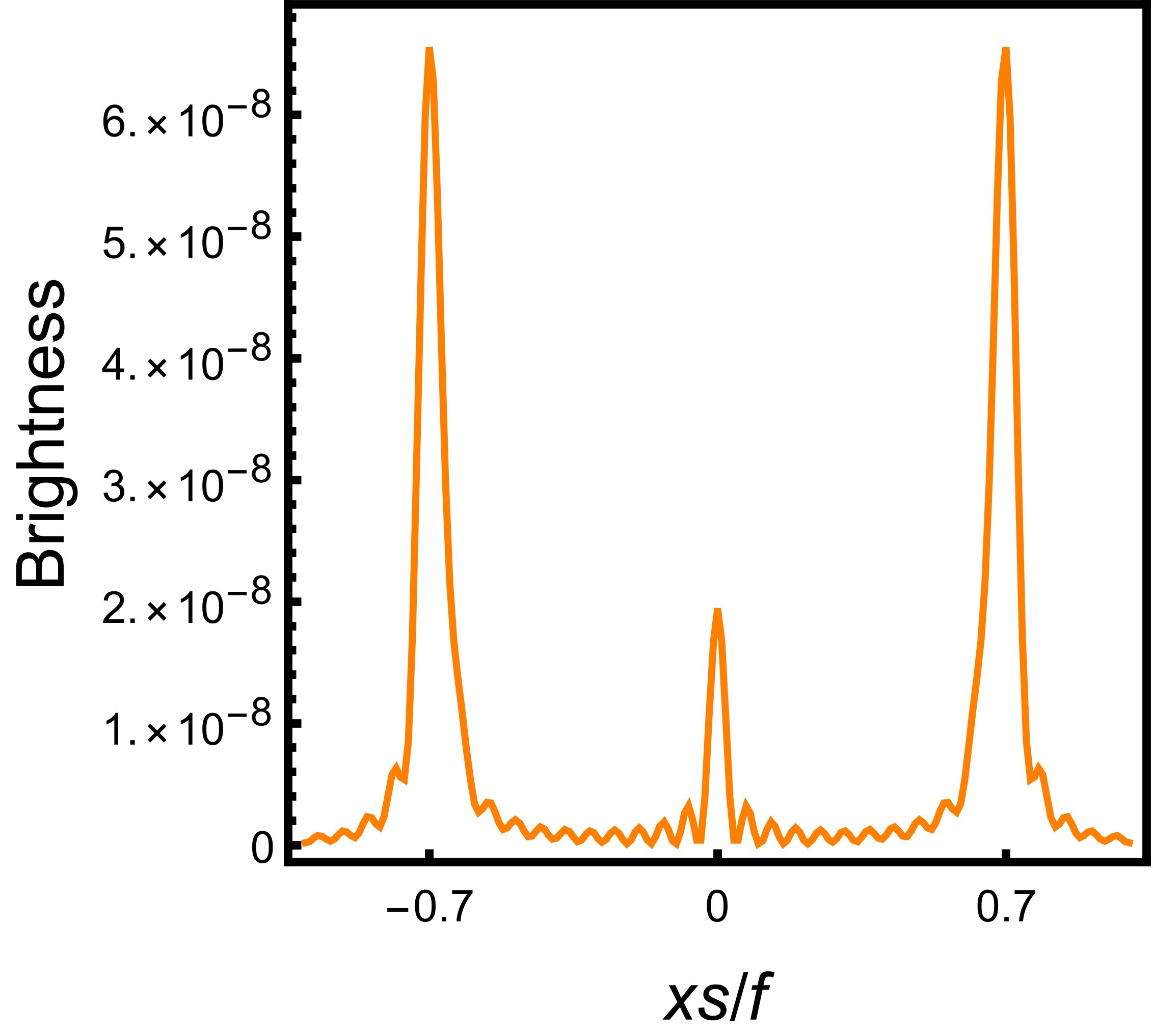}
    \caption{$\Omega=-1$}
  \end{subfigure}
  \hfill
  \begin{subfigure}[b]{0.48\columnwidth}
    \centering
    \includegraphics[width=\textwidth,height=0.8\textwidth]{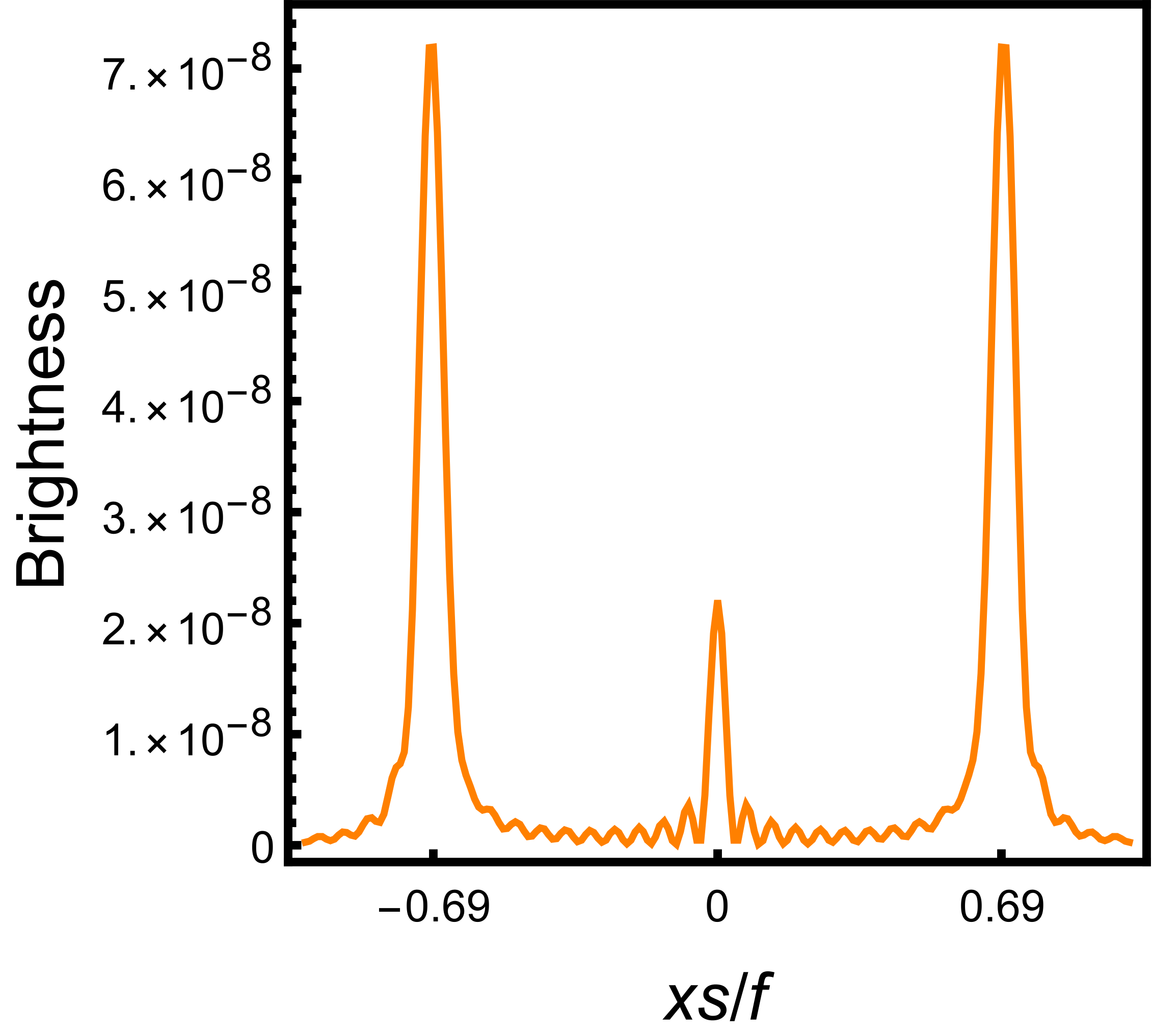}
    \caption{$\Omega=-1/3$}
  \end{subfigure}
\begin{subfigure}[b]{0.48\columnwidth}
    \centering
    \includegraphics[width=\textwidth,height=0.8\textwidth]{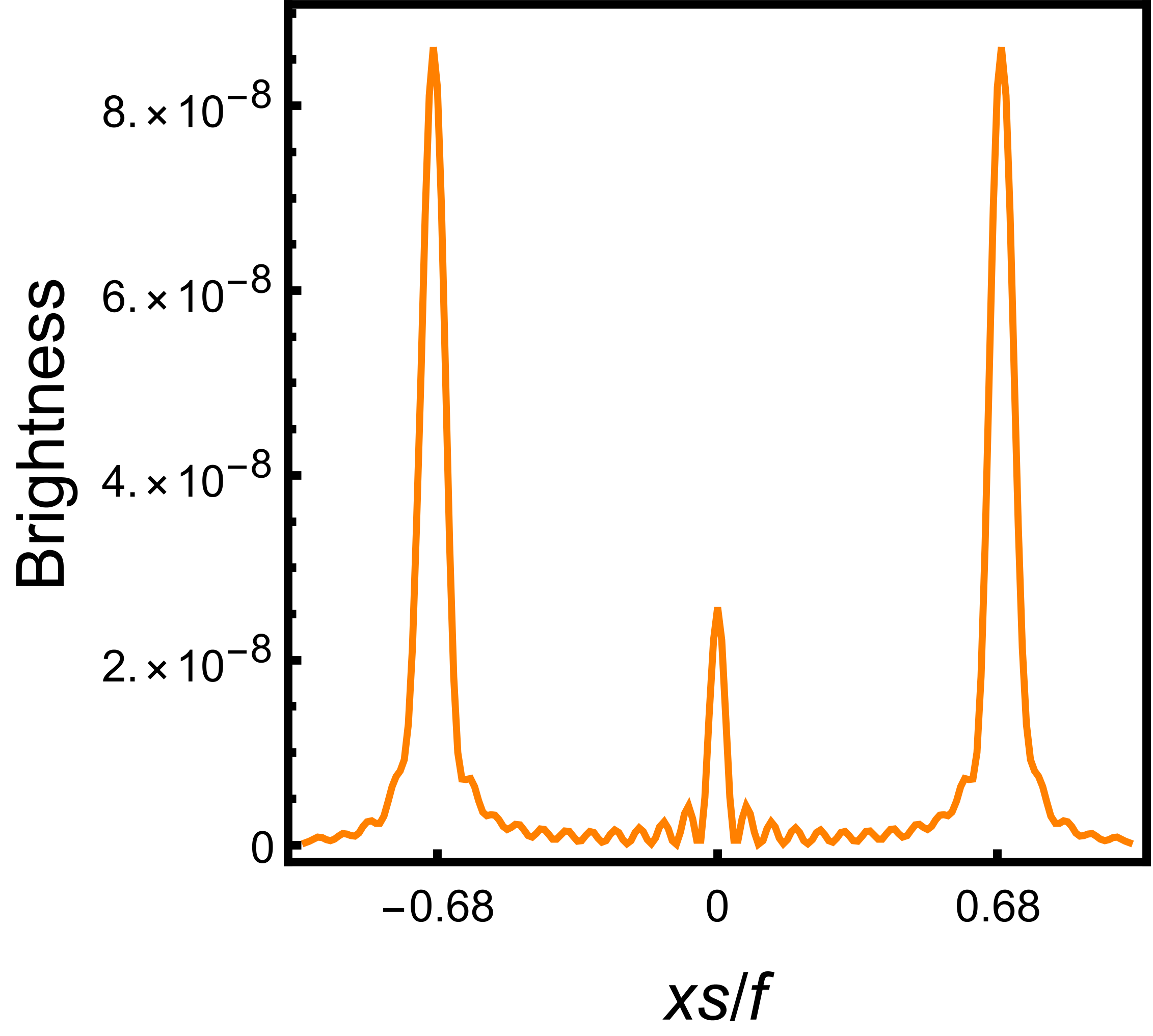}
    \caption{$\Omega=-1/6$}
  \end{subfigure}
  \hfill
  \begin{subfigure}[b]{0.48\columnwidth}
    \centering
    \includegraphics[height=0.8\textwidth,width=\textwidth]{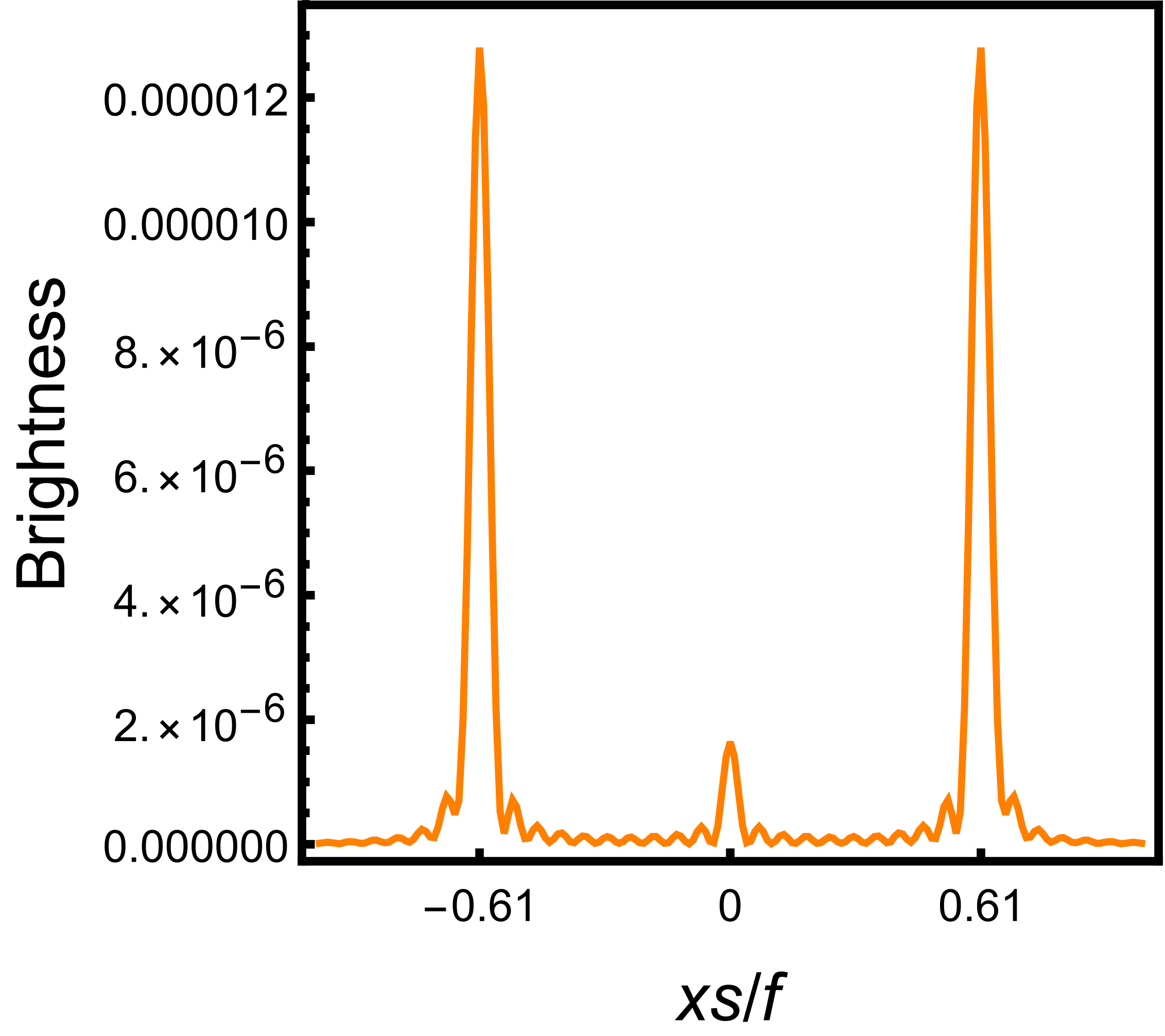}
    \caption{$\Omega=1/3$}
  \end{subfigure}
  \caption{Effect of $\Omega$ on the brightness, where $\theta_{obs}=0$, $a=0.1$, $c=0.1$, $Q=0.1$, $e=0.5$, $y_{h}=3$, $\omega=90$.}
  \label{14}%
\end{figure}

\begin{figure}[htbp]
  \centering
  \begin{subfigure}[b]{0.48\columnwidth}
    \centering
    \includegraphics[width=\textwidth,height=0.8\textwidth]{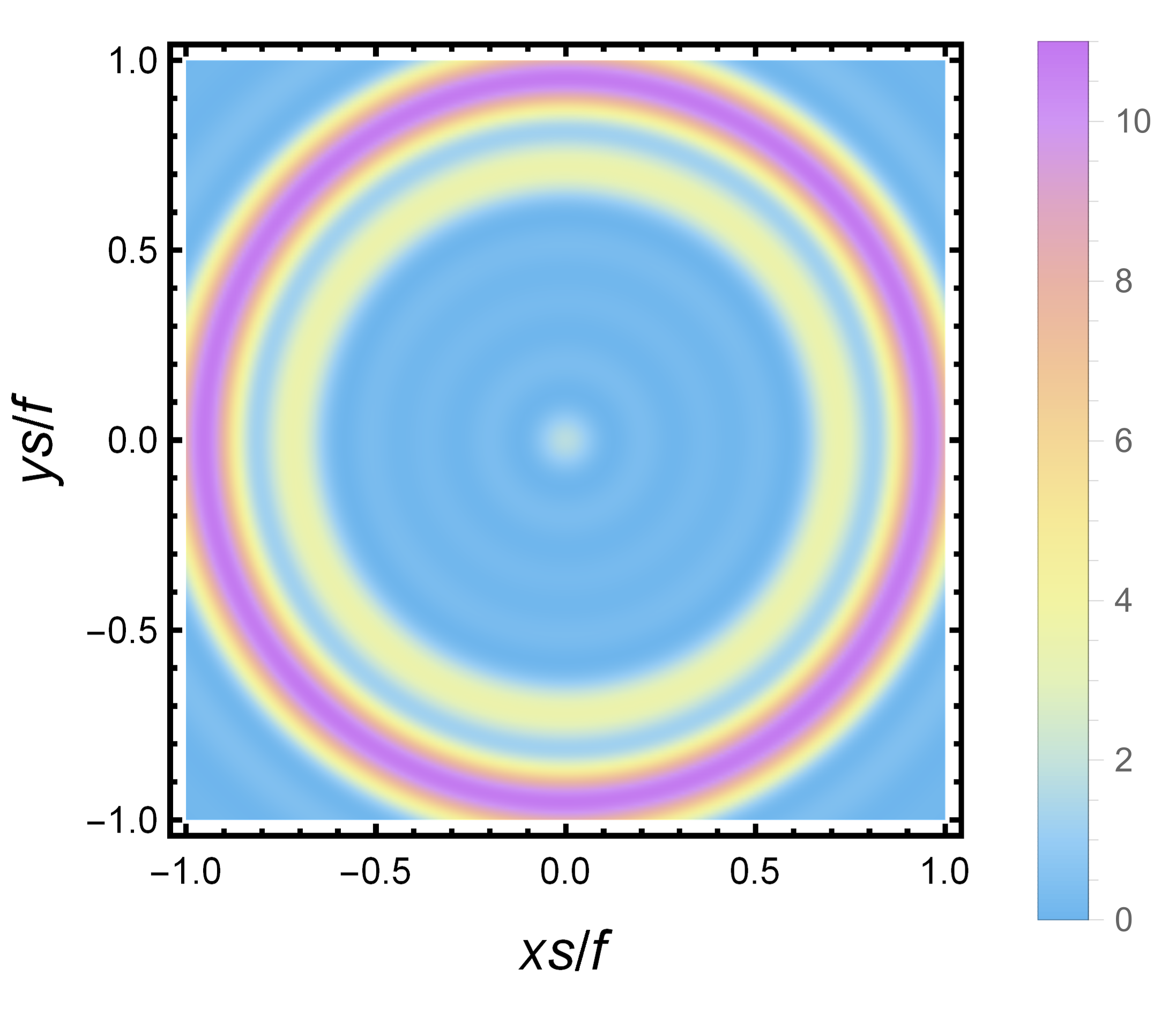}
    \caption{$\omega=30$}
  \end{subfigure}
  \hfill
  \begin{subfigure}[b]{0.48\columnwidth}
    \centering
    \includegraphics[width=\textwidth,height=0.8\textwidth]{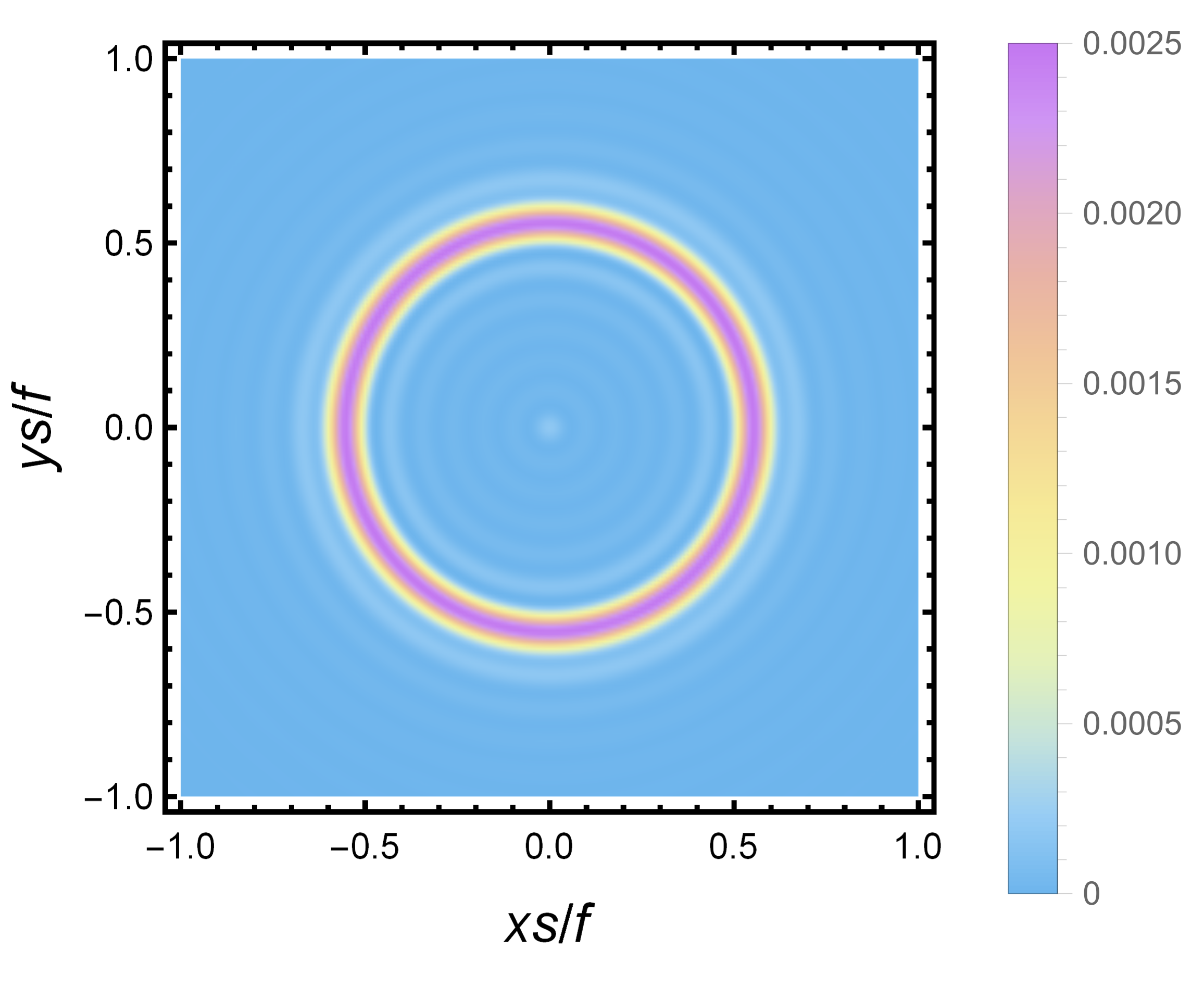}
    \caption{$\omega=60$}
  \end{subfigure}
\begin{subfigure}[b]{0.48\columnwidth}
    \centering
    \includegraphics[width=\textwidth,height=0.8\textwidth]{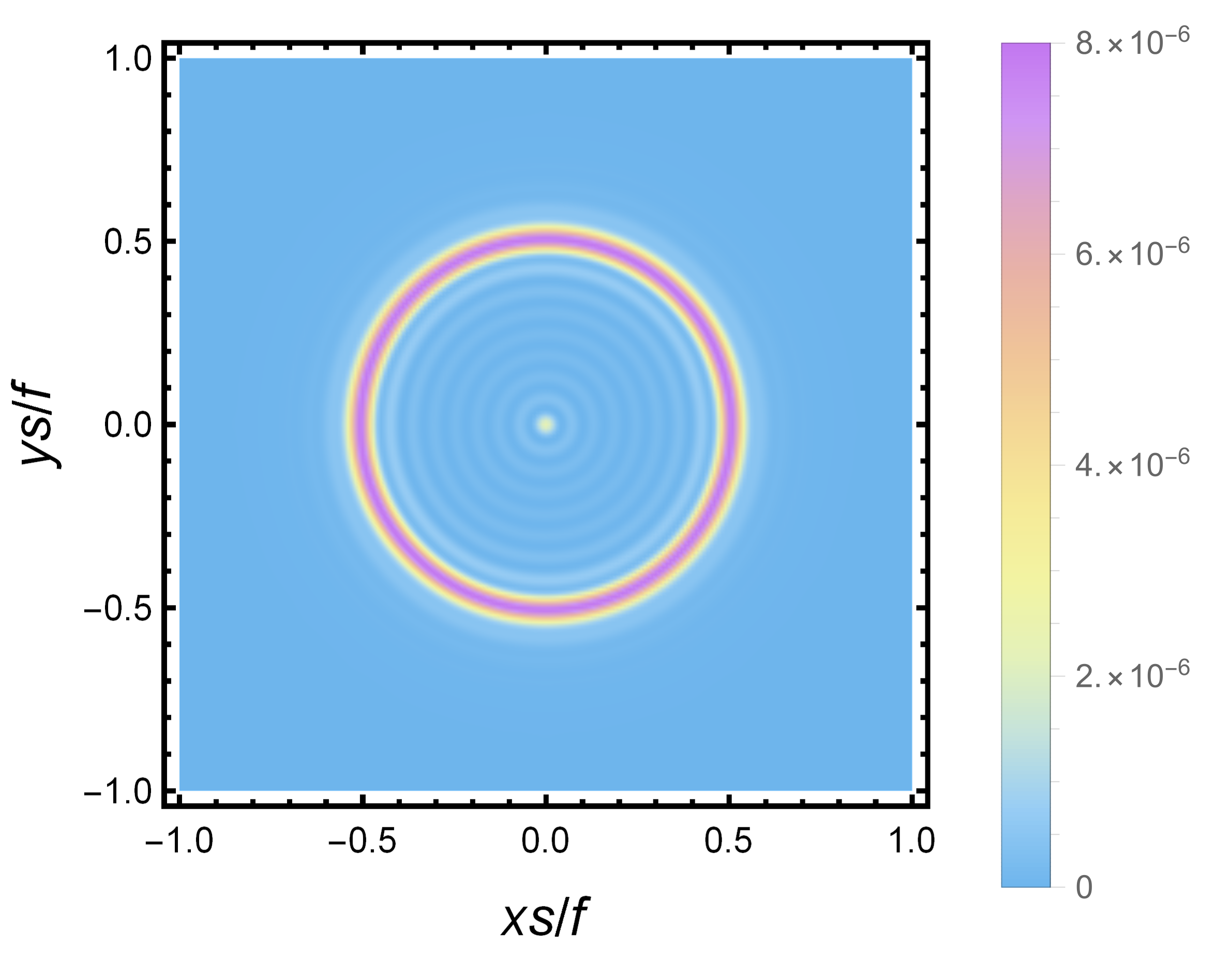}
    \caption{$\omega=90$}
  \end{subfigure}
  \hfill
  \begin{subfigure}[b]{0.48\columnwidth}
    \centering
    \includegraphics[height=0.8\textwidth,width=\textwidth]{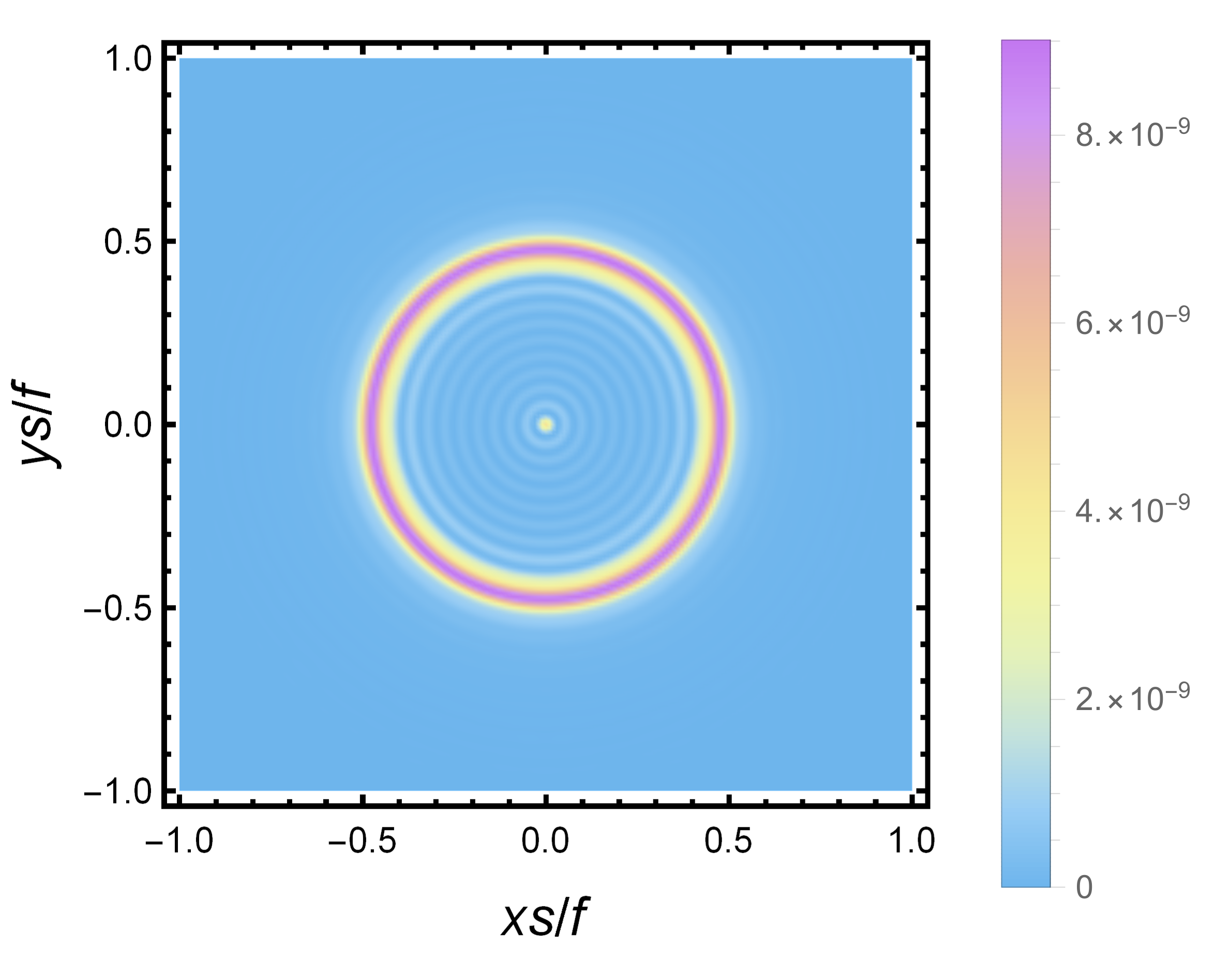}
    \caption{$\omega=120$}
  \end{subfigure}
  \caption{Effect of $\omega$ on the Einstein ring, where $\theta_{obs}=0$, $a=0.1$, $c=0.1$, $\Omega=-\frac{2}{3}$, $Q=0.1$, $e=0.5$, $y_{h}=5$.}
  \label{15}%
\end{figure}

\begin{figure}[htbp]
  \centering
  \begin{subfigure}[b]{0.48\columnwidth}
    \centering
    \includegraphics[width=\textwidth,height=0.8\textwidth]{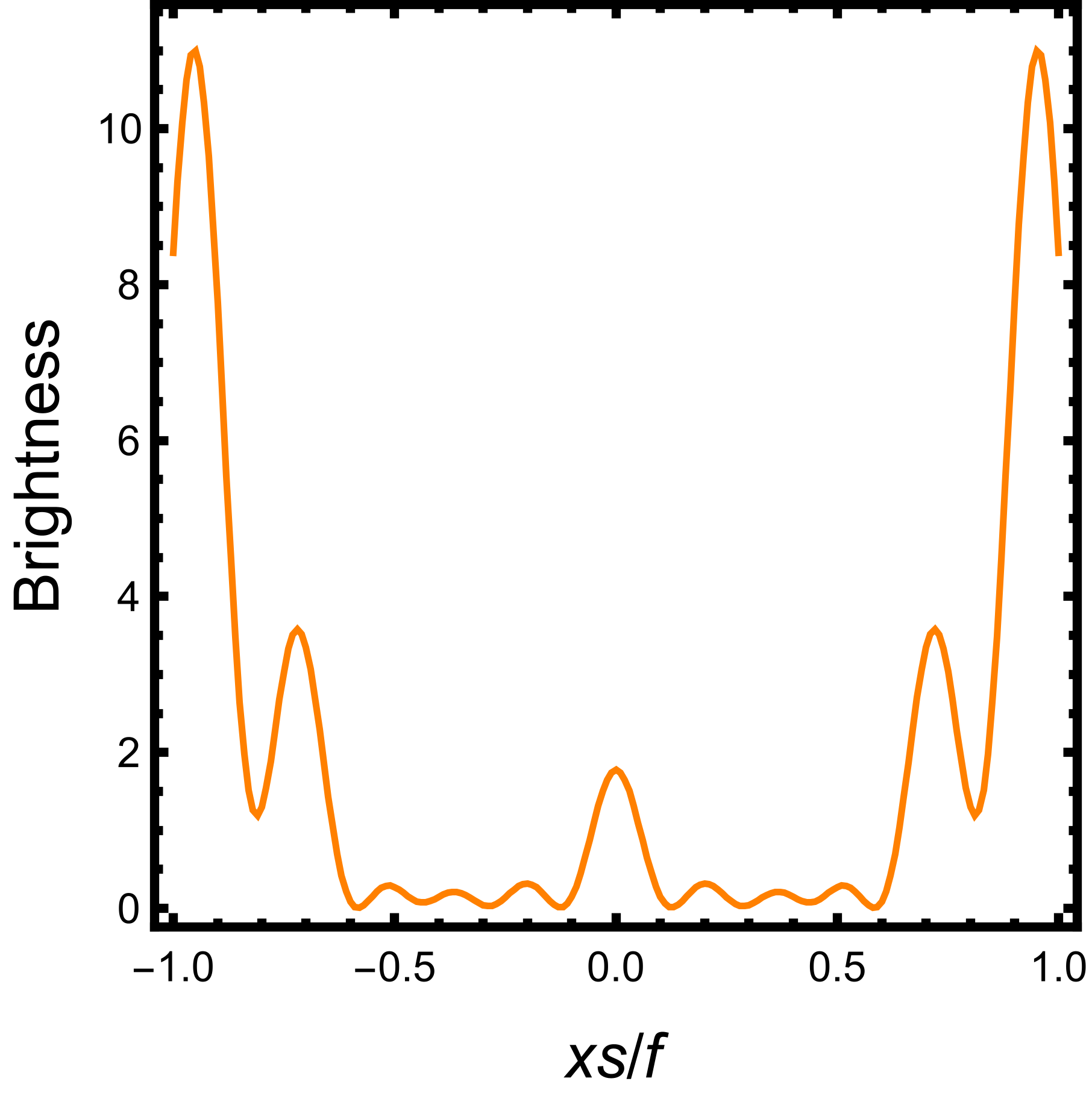}
    \caption{$\omega=30$}
  \end{subfigure}
  \hfill
  \begin{subfigure}[b]{0.48\columnwidth}
    \centering
    \includegraphics[width=\textwidth,height=0.8\textwidth]{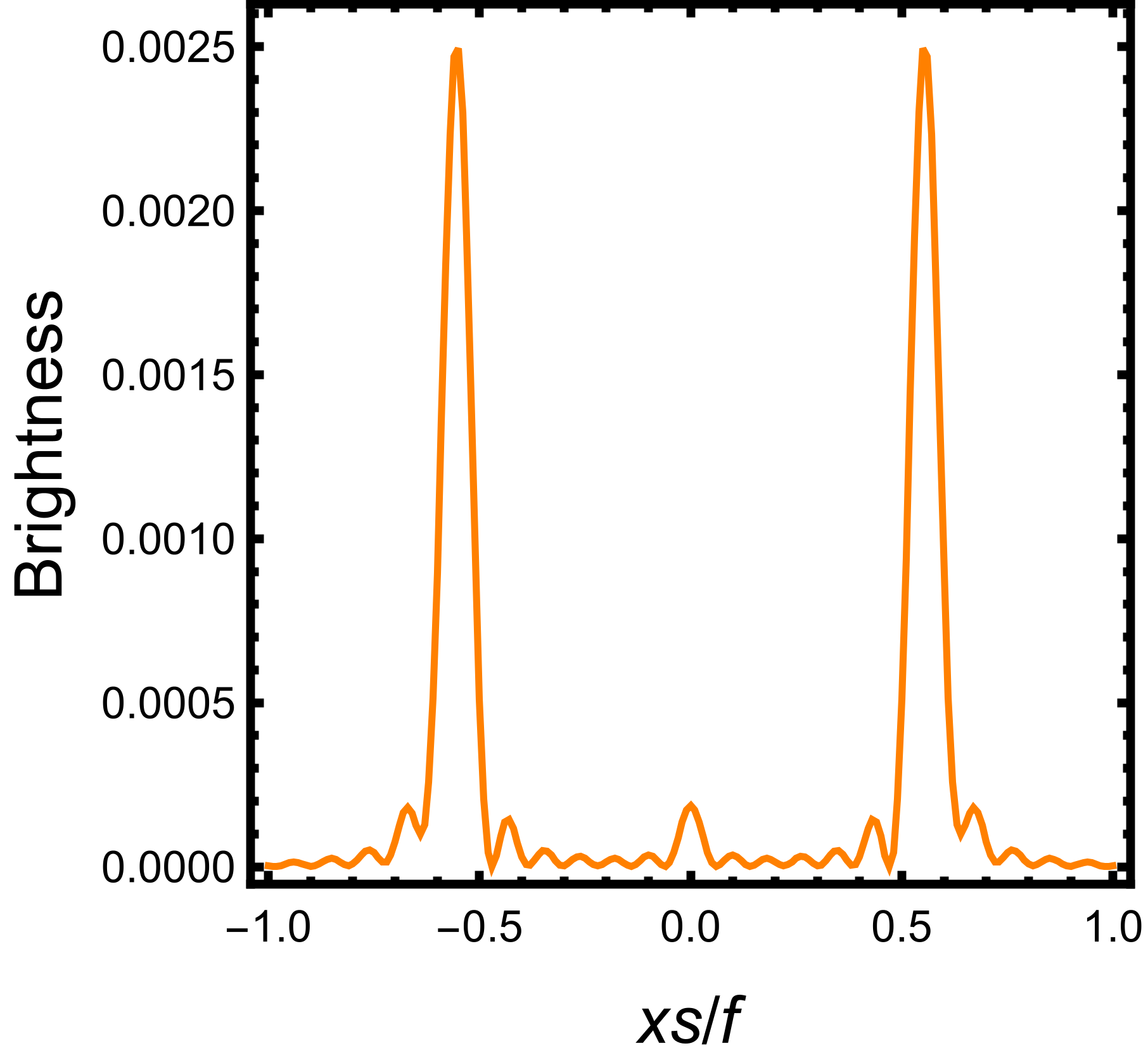}
    \caption{$\omega=60$}
  \end{subfigure}
\begin{subfigure}[b]{0.48\columnwidth}
    \centering
    \includegraphics[width=\textwidth,height=0.8\textwidth]{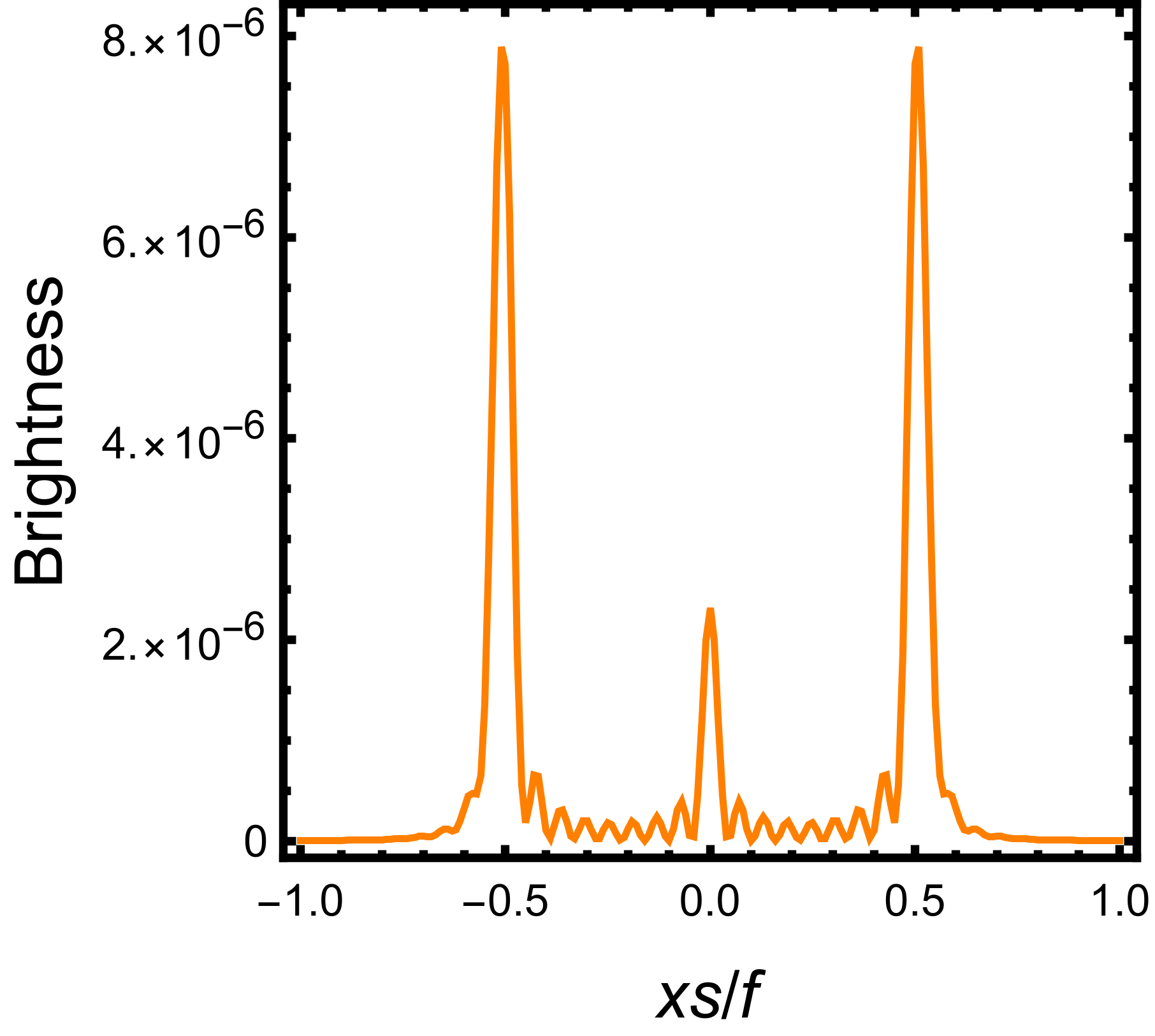}
    \caption{$\omega=90$}
  \end{subfigure}
  \hfill
  \begin{subfigure}[b]{0.48\columnwidth}
    \centering
    \includegraphics[height=0.8\textwidth,width=\textwidth]{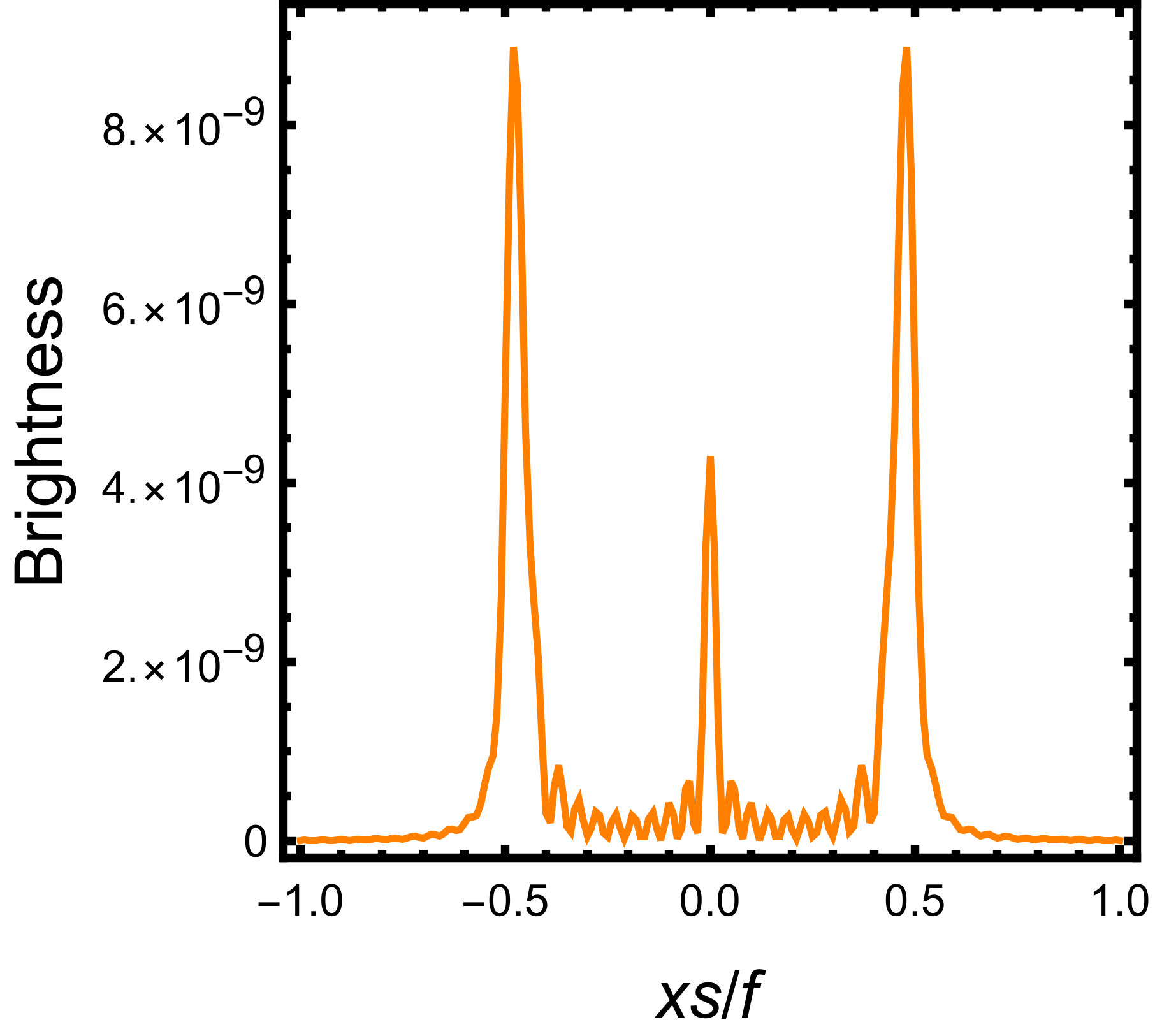}
    \caption{$\omega=120$}
  \end{subfigure}
  \caption{Effect of $\omega$ on the brightness, where $\theta_{obs}=0$, $a=0.1$, $c=0.1$, $\Omega=-\frac{2}{3}$, $Q=0.1$, $e=0.5$, $y_{h}=5$.}
  \label{16}%
\end{figure}

\subsection{Impact of temperature and chemical potential on Einstein rings}

\begin{figure}[htbp]
  \centering
  \begin{subfigure}[b]{0.48\columnwidth}
    \centering
    \includegraphics[width=\textwidth,height=0.8\textwidth]{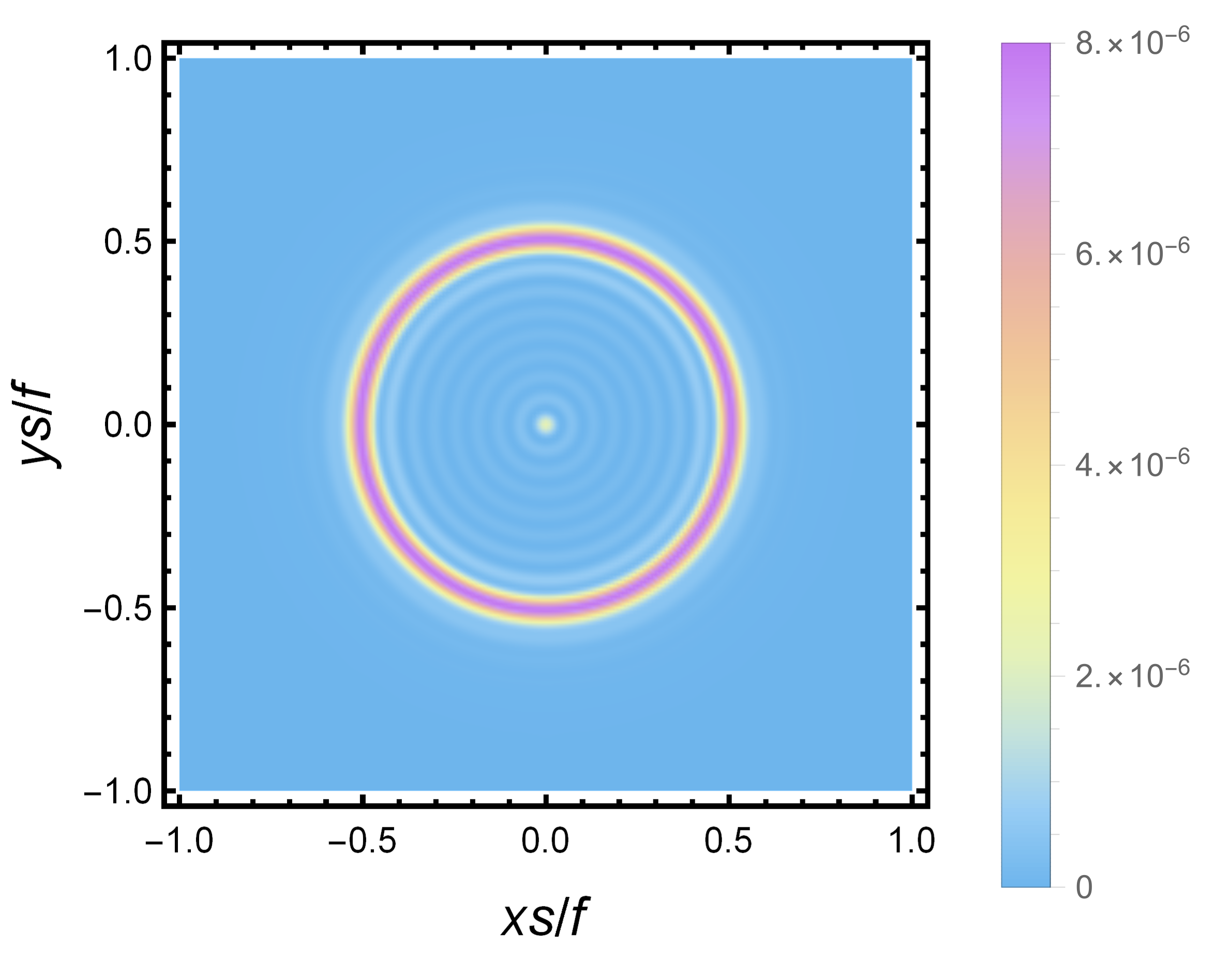}
    \caption{$T=0.3918$}
  \end{subfigure}
  \hfill
  \begin{subfigure}[b]{0.48\columnwidth}
    \centering
    \includegraphics[width=\textwidth,height=0.8\textwidth]{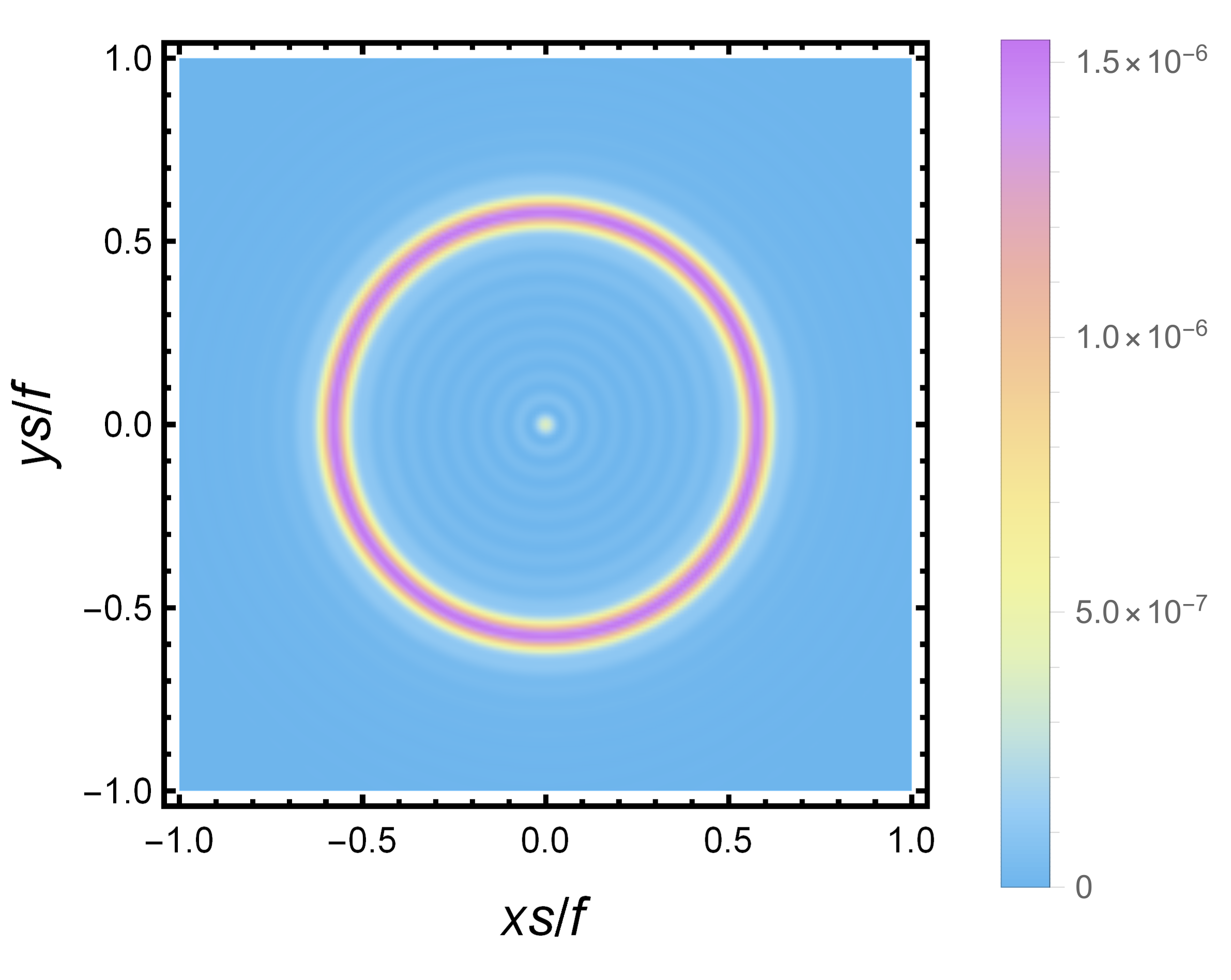}
    \caption{$T=0.3232$}
  \end{subfigure}
\begin{subfigure}[b]{0.48\columnwidth}
    \centering
    \includegraphics[width=\textwidth,height=0.8\textwidth]{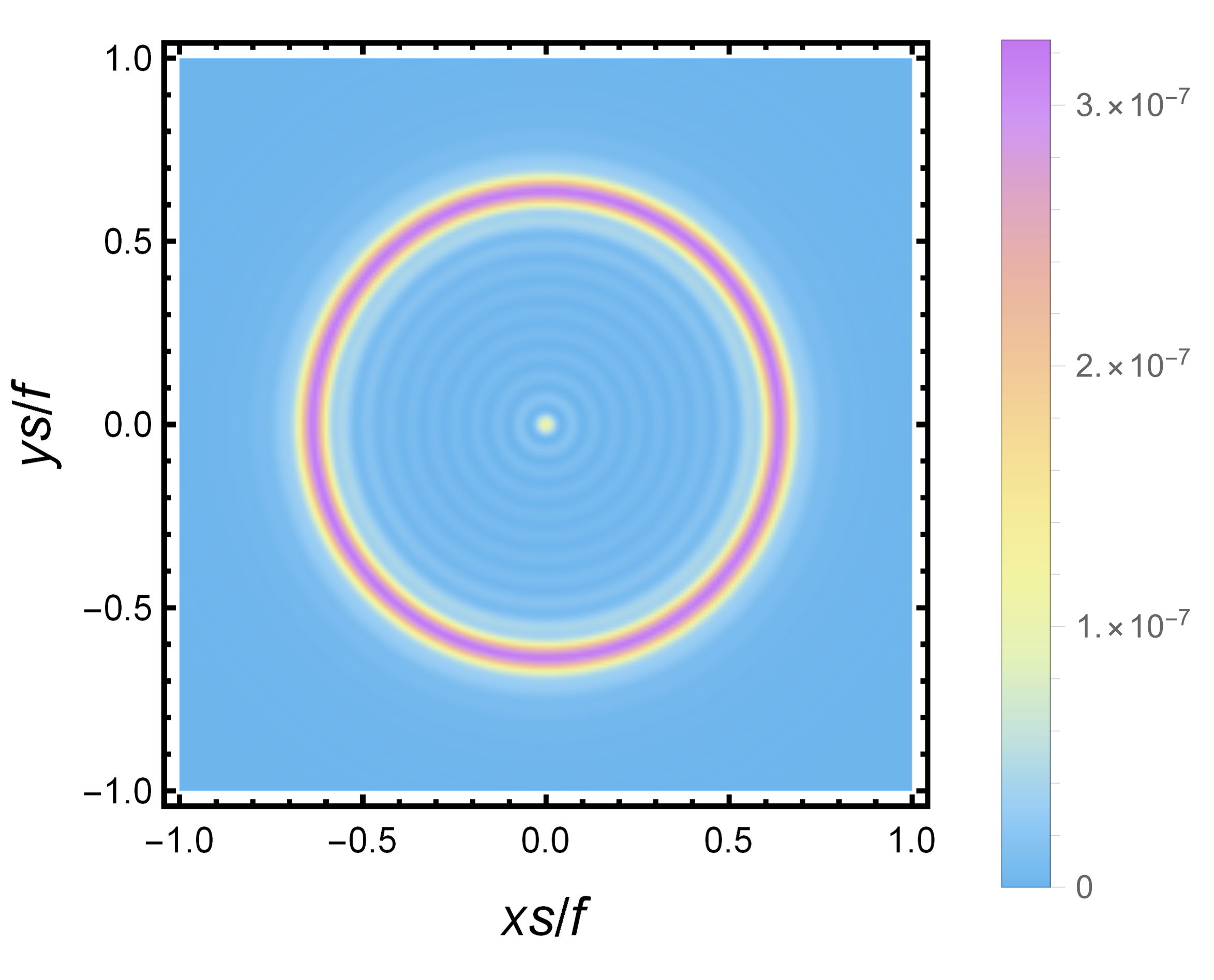}
    \caption{$T=0.2842$}
  \end{subfigure}
  \hfill
  \begin{subfigure}[b]{0.48\columnwidth}
    \centering
    \includegraphics[height=0.8\textwidth,width=\textwidth]{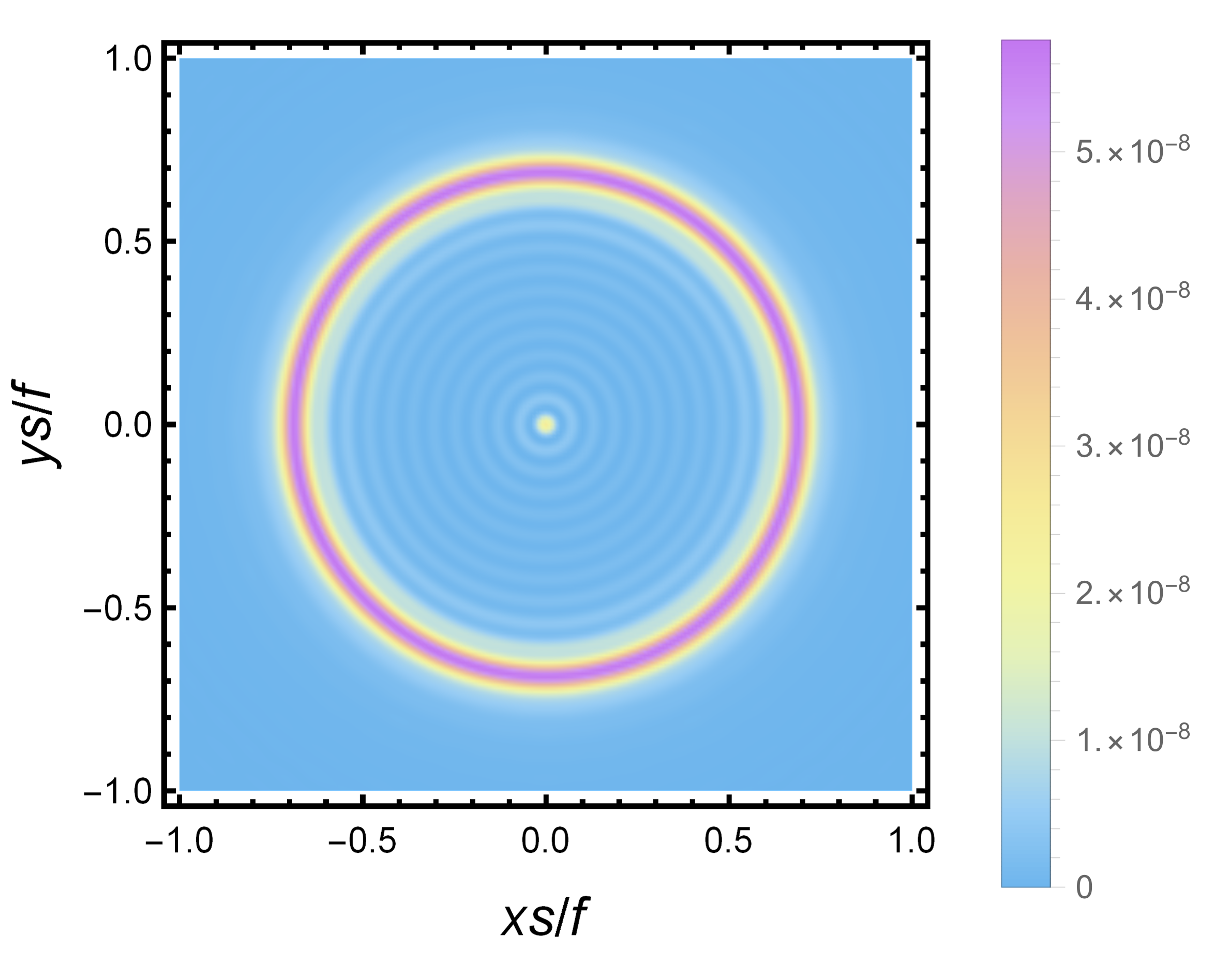}
    \caption{$T=0.2601$}
  \end{subfigure}
  \caption{Effect of $T$ on the Einstein ring, where $\theta_{obs}=0$, $a=0.1$, $c=0.1$, $\Omega=-\frac{2}{3}$, $e=0.5$, $y_{h}=5$, $\omega=90$, The temperature is from high to low, corresponding to $Q=0.1,0.12,0.14,0.16$, respectively.}
  \label{17}%
\end{figure}

\begin{figure}[htbp]
  \centering
  \begin{subfigure}[b]{0.48\columnwidth}
    \centering
    \includegraphics[width=\textwidth,height=0.8\textwidth]{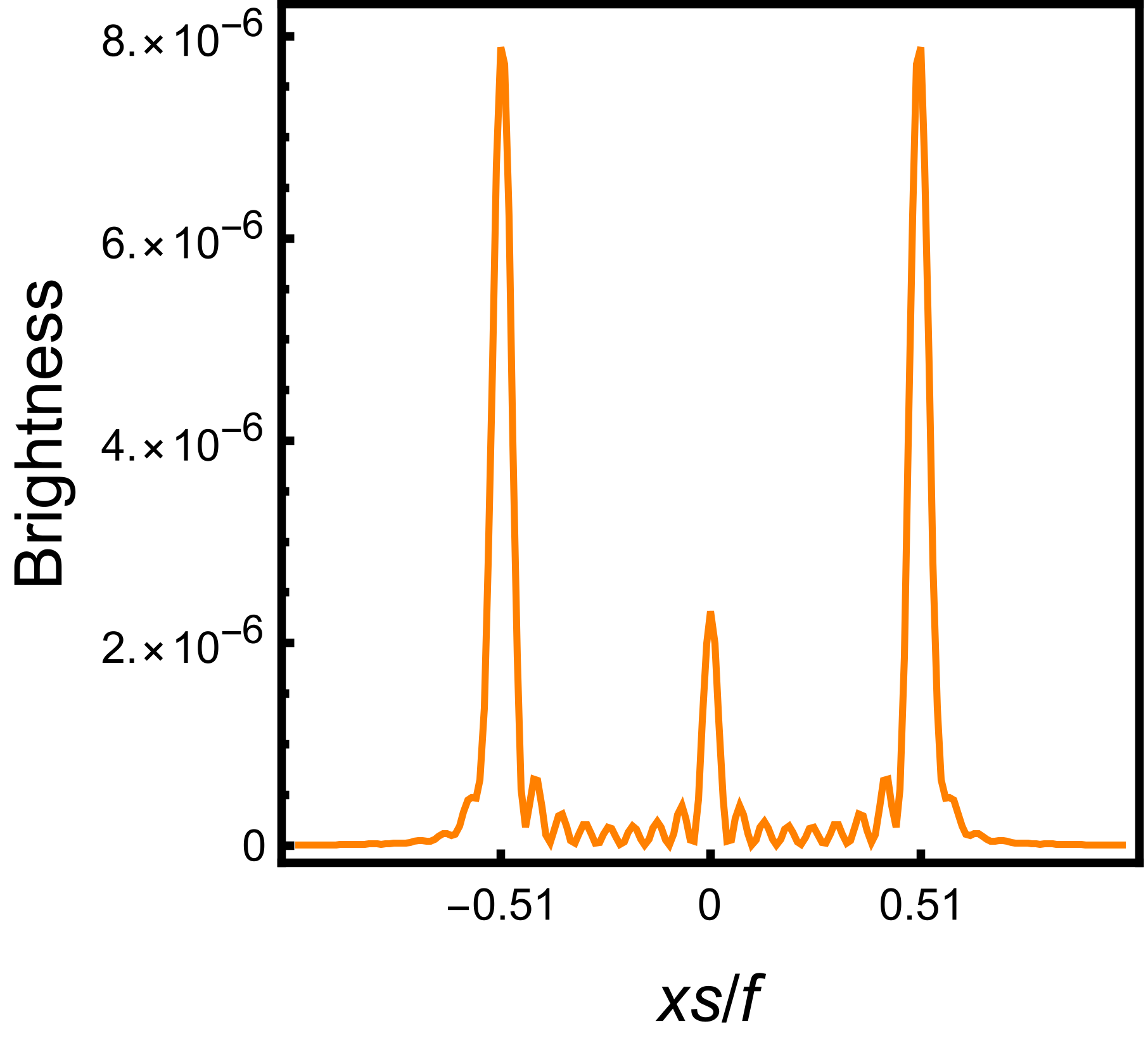}
    \caption{$T=0.3918$}
  \end{subfigure}
  \hfill
  \begin{subfigure}[b]{0.48\columnwidth}
    \centering
    \includegraphics[width=\textwidth,height=0.8\textwidth]{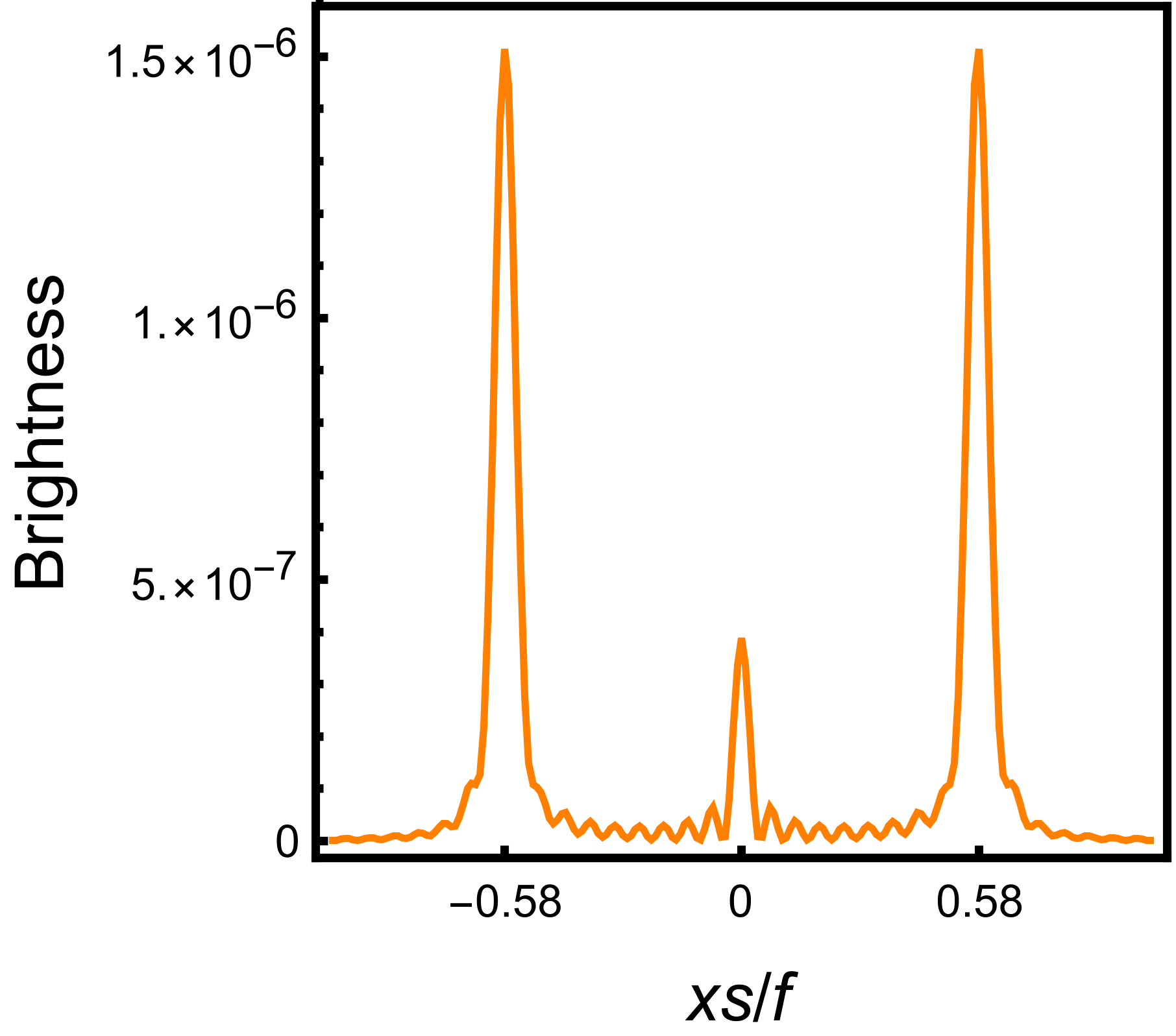}
    \caption{$T=0.3232$}
  \end{subfigure}
\begin{subfigure}[b]{0.48\columnwidth}
    \centering
    \includegraphics[width=\textwidth,height=0.8\textwidth]{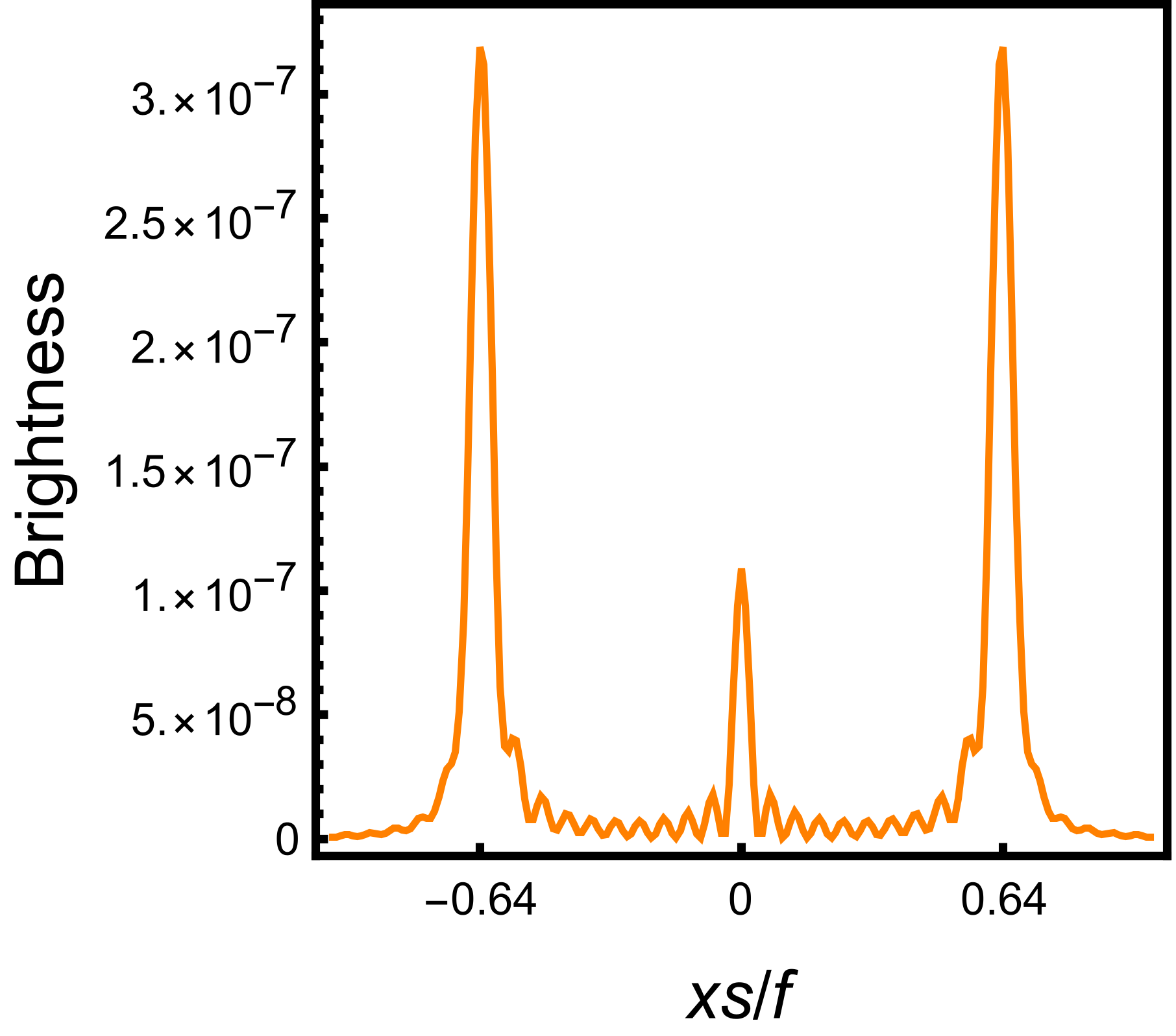}
    \caption{$T=0.2842$}
  \end{subfigure}
  \hfill
  \begin{subfigure}[b]{0.48\columnwidth}
    \centering
    \includegraphics[height=0.8\textwidth,width=\textwidth]{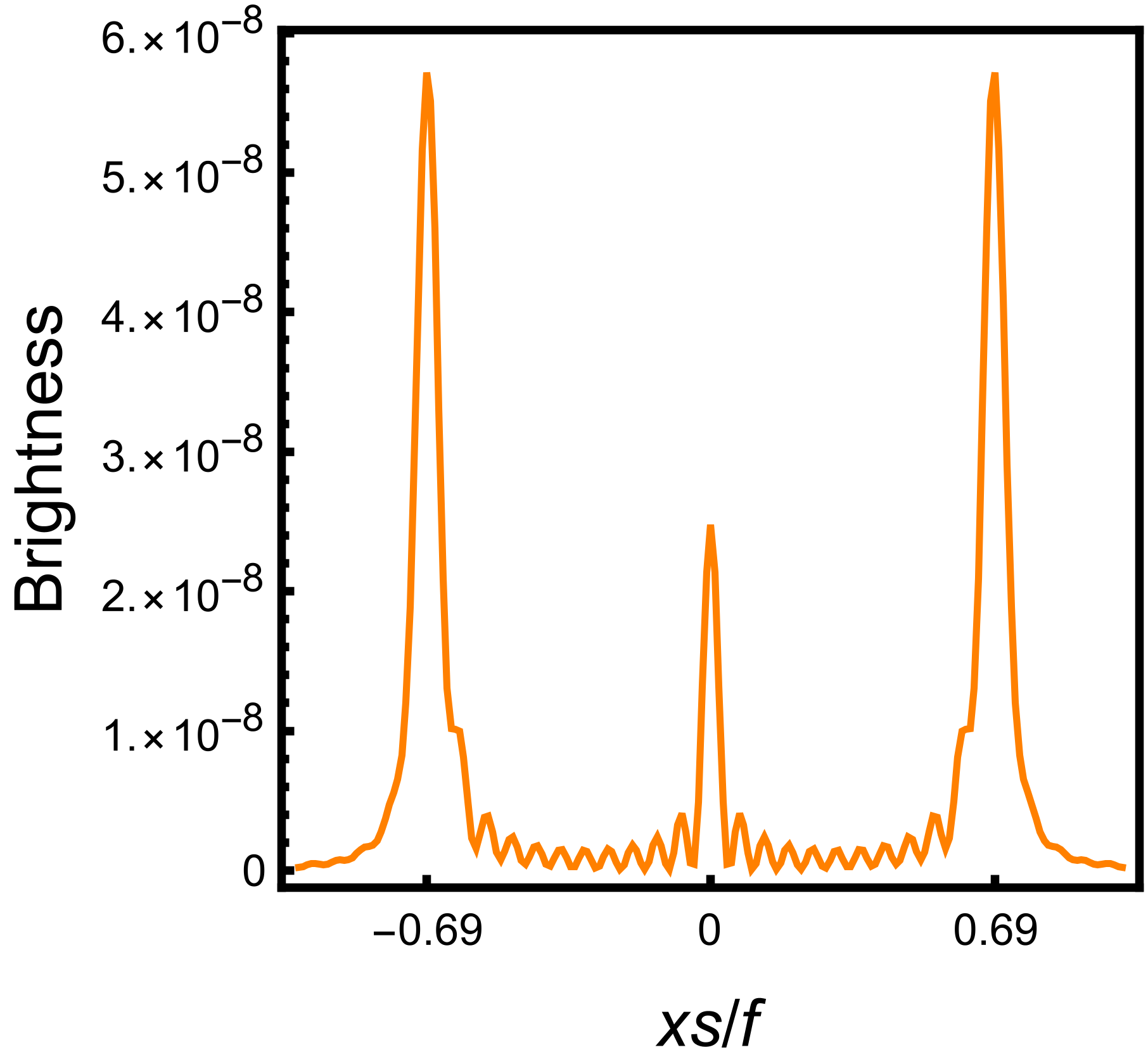}
    \caption{$T=0.2601$}
  \end{subfigure}
  \caption{Effect of $T$ on the brightness, where $\theta_{obs}=0$, $a=0.1$, $c=0.1$, $\Omega=-\frac{2}{3}$, $e=0.5$, $y_{h}=5$, $\omega=90$.}
  \label{18}%
\end{figure}

This section examines the impact of temperature $T$ on Einstein rings under the condition of a chemical potential $\mu=0.5$. As shown in Figure \ref{17}, a decrease in temperature leads to an increase in the radius of the Einstein ring. Additionally, Figure \ref{18} presents the luminosity diagram, where the luminosity reaches its peak values at $T=0.3918$, $xs/f=0.51$. Notably, as the temperature $T$ gradually decreases, the x-coordinate values corresponding to the luminosity peaks increase, indicating that the radius of the Einstein ring enlarges as the temperature decreases.

We also consider the impact of the chemical potential $\mu$ on Einstein rings, while keeping the temperature fixed at $0.5$. As illustrated in Figure \ref{19}, as the chemical potential increases, the radius of the ring decreases. This trend is more clearly observed in Figure \ref{20}. When $\mu=0.1$, the luminosity is maximized at  $xs/f=0.48$, whereas when $\mu=0.7$, the luminosity is maximized at $xs/f=0.41$. In other words, as $\mu$ gradually increases, the x-coordinate corresponding to the maximum luminosity decreases, indicating that an increase in $\mu$ leads to a reduction in the radius of the Einstein ring.

\begin{figure}[htbp]
  \centering
  \begin{subfigure}[b]{0.48\columnwidth}
    \centering
    \includegraphics[width=\textwidth,height=0.8\textwidth]{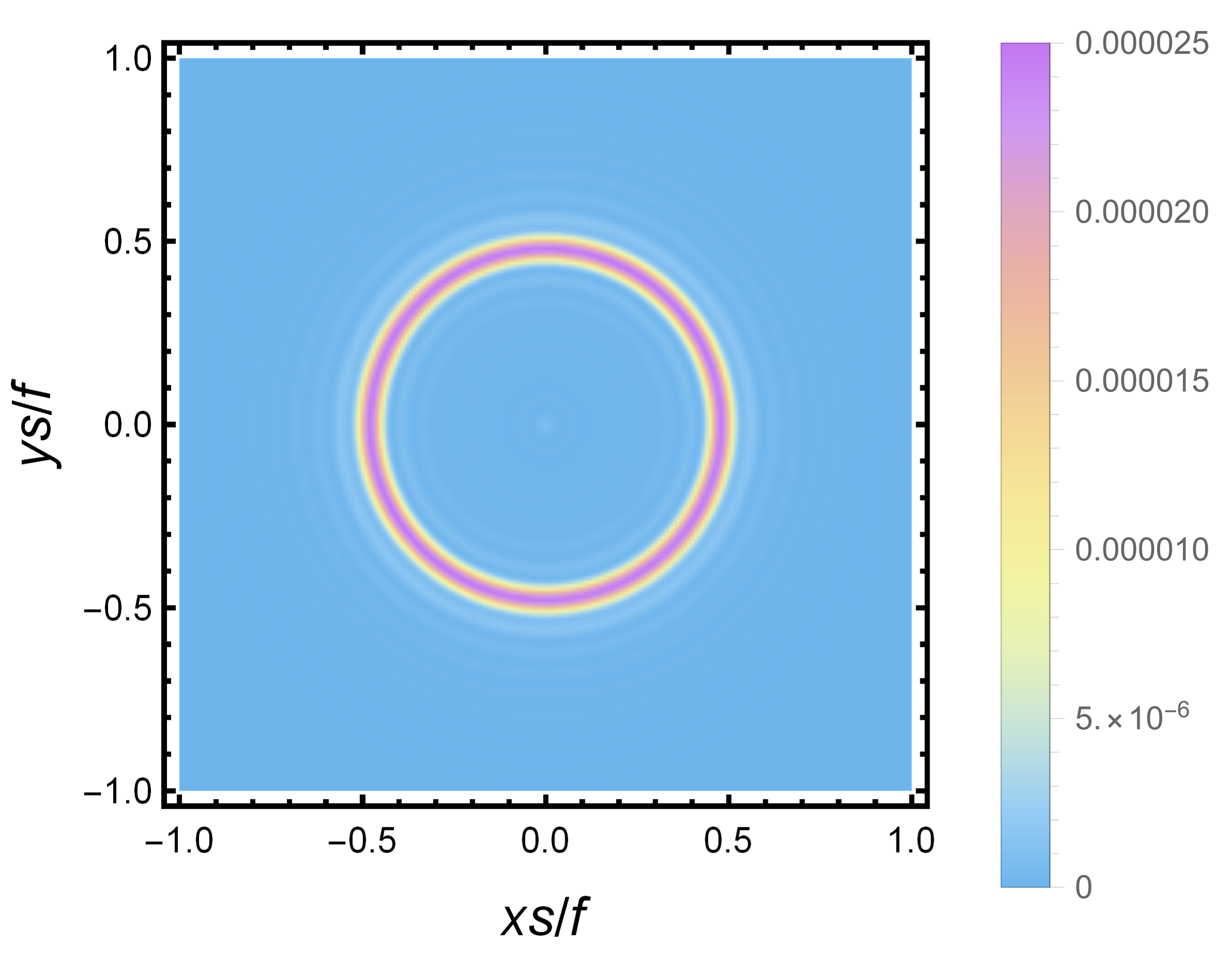}
    \caption{$\mu=0.1$}
  \end{subfigure}
  \hfill
  \begin{subfigure}[b]{0.48\columnwidth}
    \centering
    \includegraphics[width=\textwidth,height=0.8\textwidth]{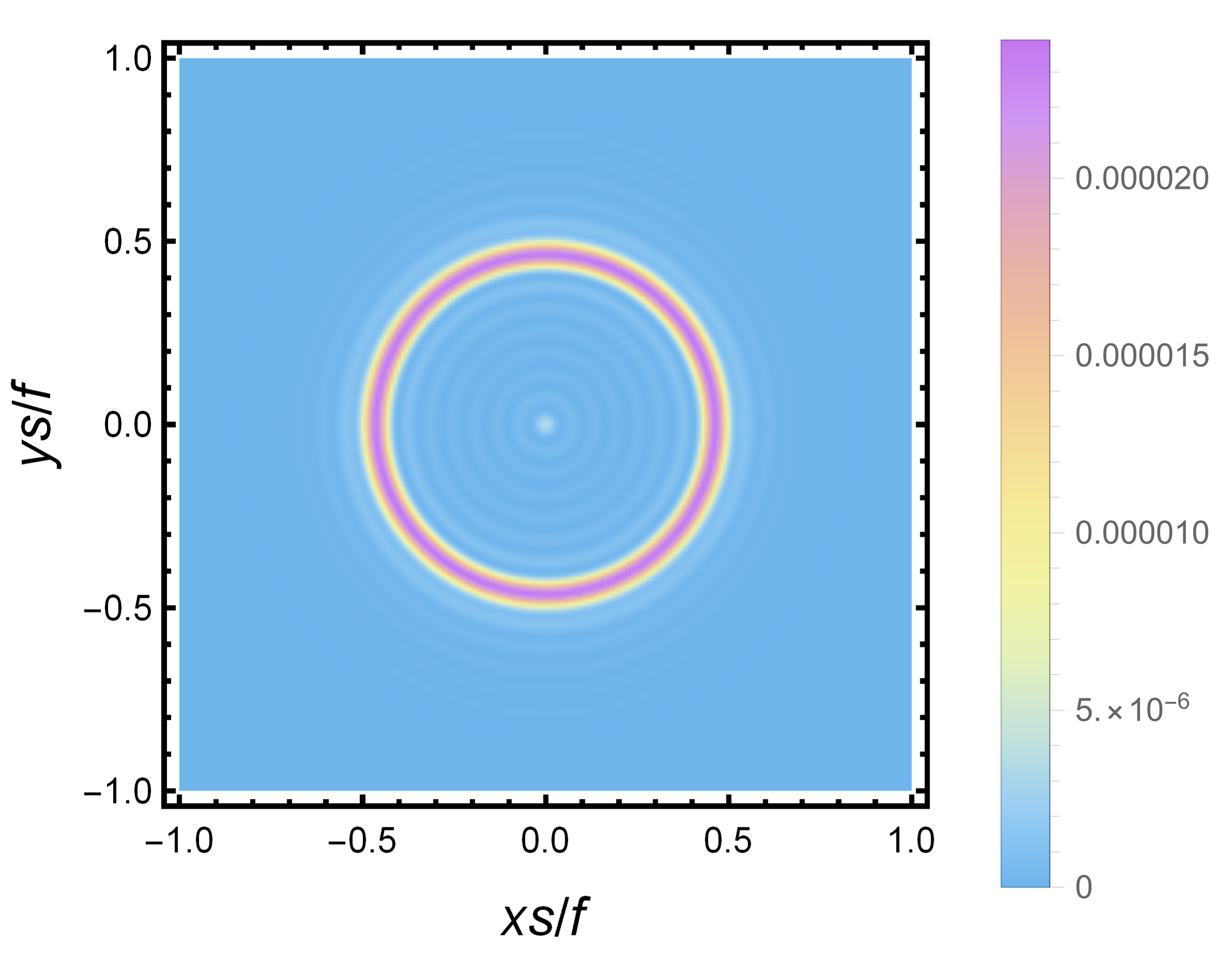}
    \caption{$\mu=0.3$}
  \end{subfigure}
\begin{subfigure}[b]{0.48\columnwidth}
    \centering
    \includegraphics[width=\textwidth,height=0.8\textwidth]{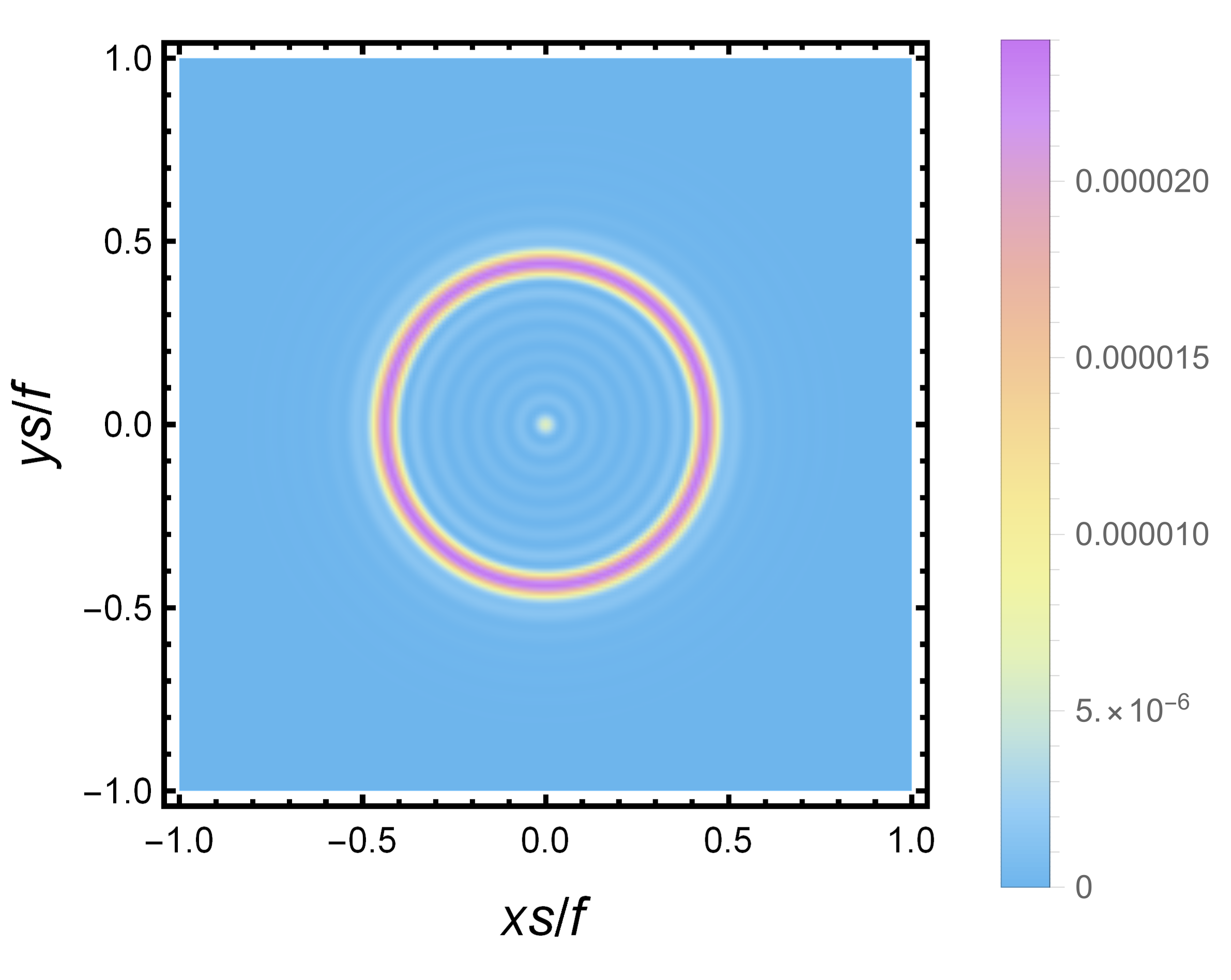}
    \caption{$\mu=0.5$}
  \end{subfigure}
  \hfill
  \begin{subfigure}[b]{0.48\columnwidth}
    \centering
    \includegraphics[height=0.8\textwidth,width=\textwidth]{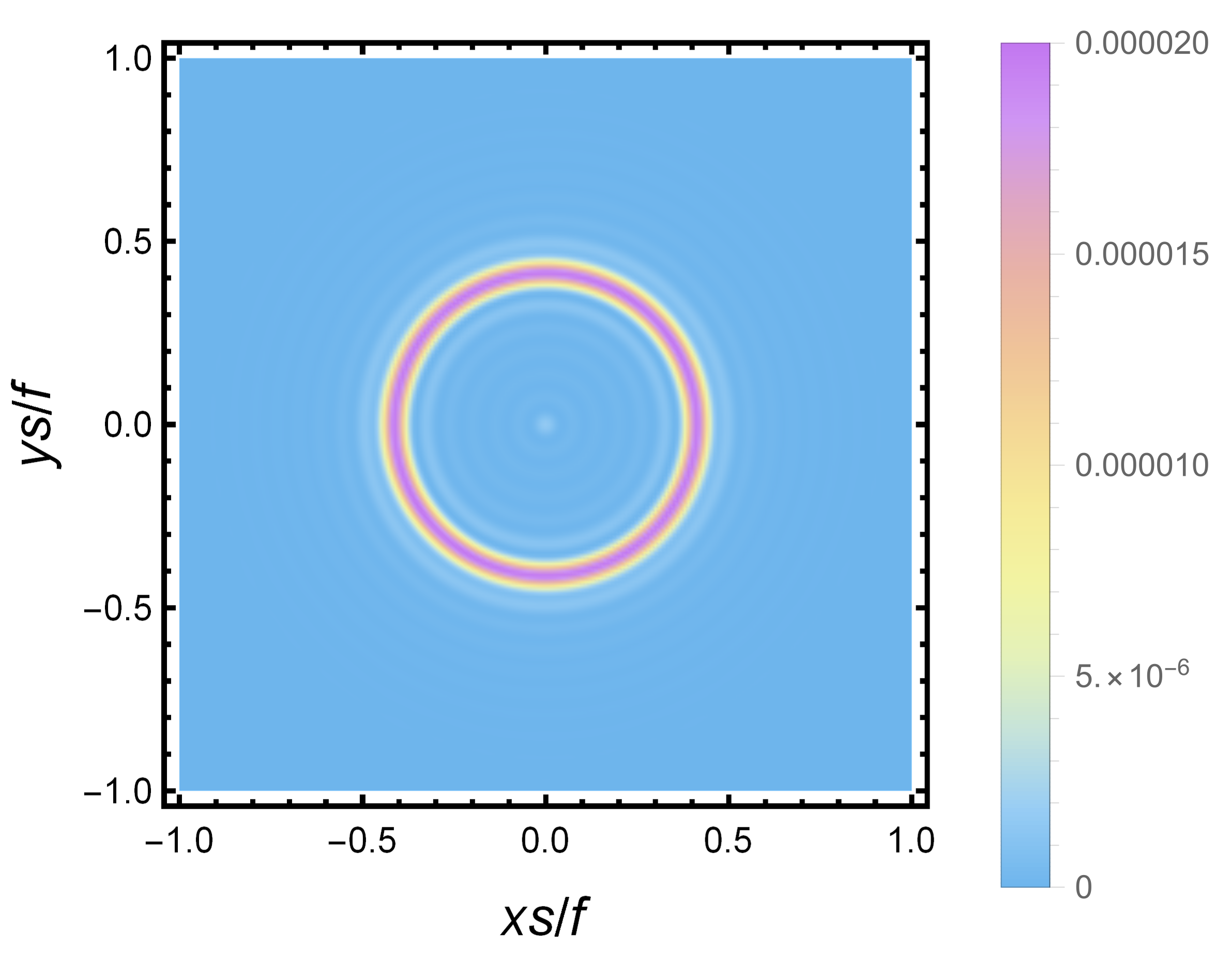}
    \caption{$\mu=0.7$}
  \end{subfigure}
  \caption{Effect of $\mu$ on the Einstein ring, where $\theta_{obs}=0$, $a=0.1$, $c=0.1$, $\Omega=-\frac{2}{3}$, $e=0.5$, $y_{h}=5$, $\omega=90$, The chemical potential vaules change from low to high, corresponding to $Q=0.0196,0.0556,0.0835,0.1022$, respectively.}
  \label{19}%
\end{figure}

\begin{figure}[htbp]
  \centering
  \begin{subfigure}[b]{0.48\columnwidth}
    \centering
    \includegraphics[width=\textwidth,height=0.8\textwidth]{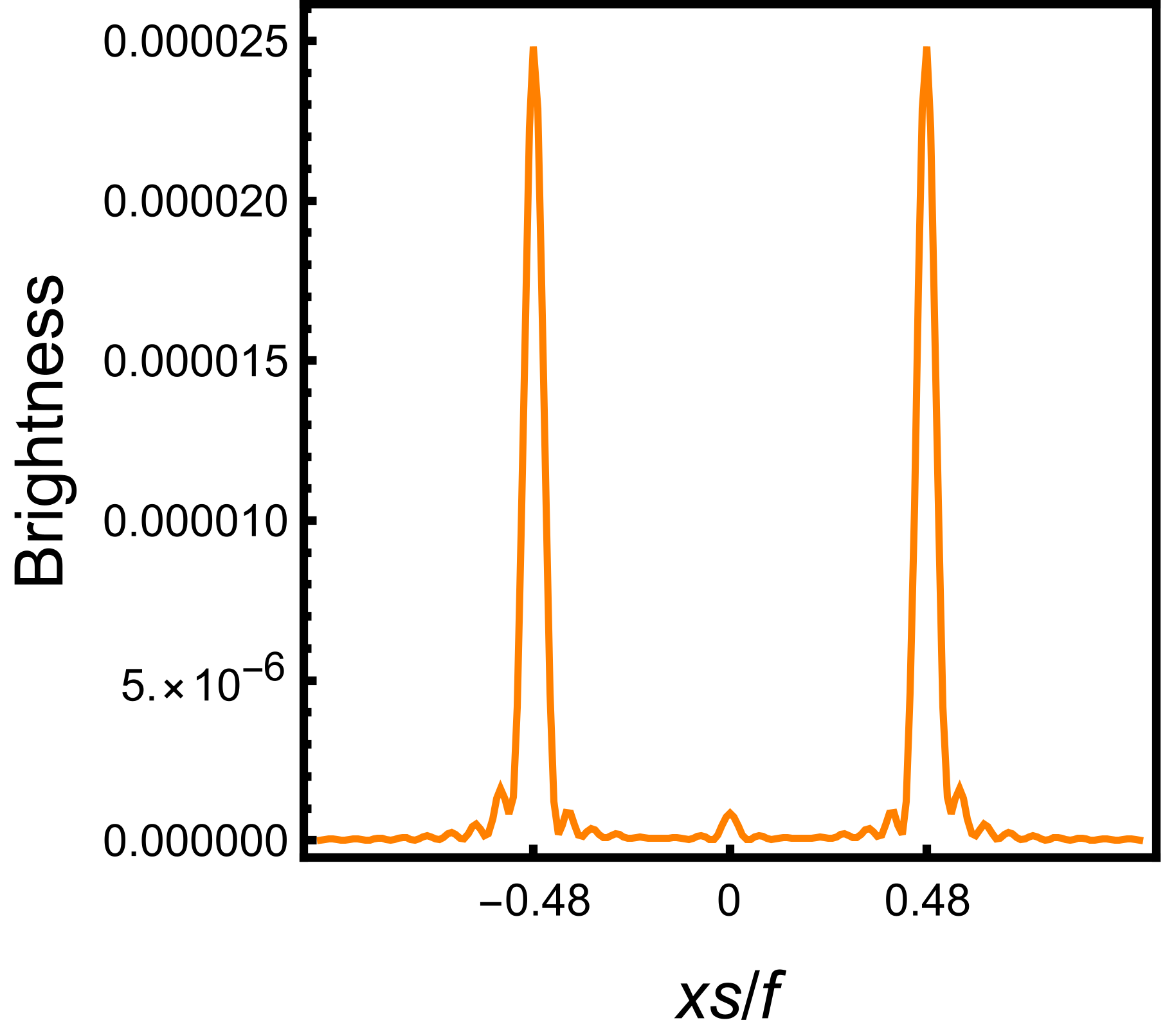}
    \caption{$\mu=0.1$}
  \end{subfigure}
  \hfill
  \begin{subfigure}[b]{0.48\columnwidth}
    \centering
    \includegraphics[width=\textwidth,height=0.8\textwidth]{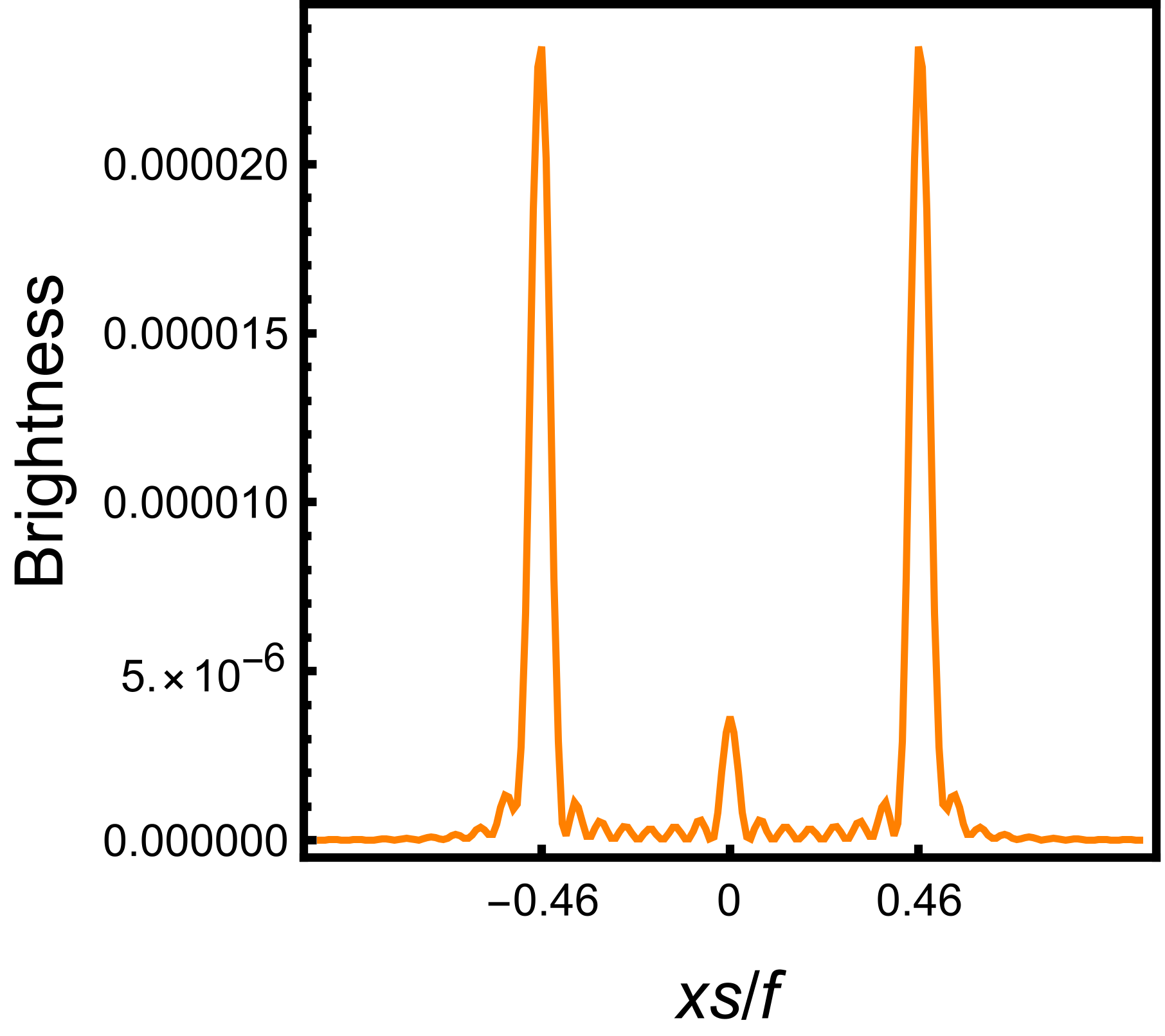}
    \caption{$\mu=0.3$}
  \end{subfigure}
\begin{subfigure}[b]{0.48\columnwidth}
    \centering
    \includegraphics[width=\textwidth,height=0.8\textwidth]{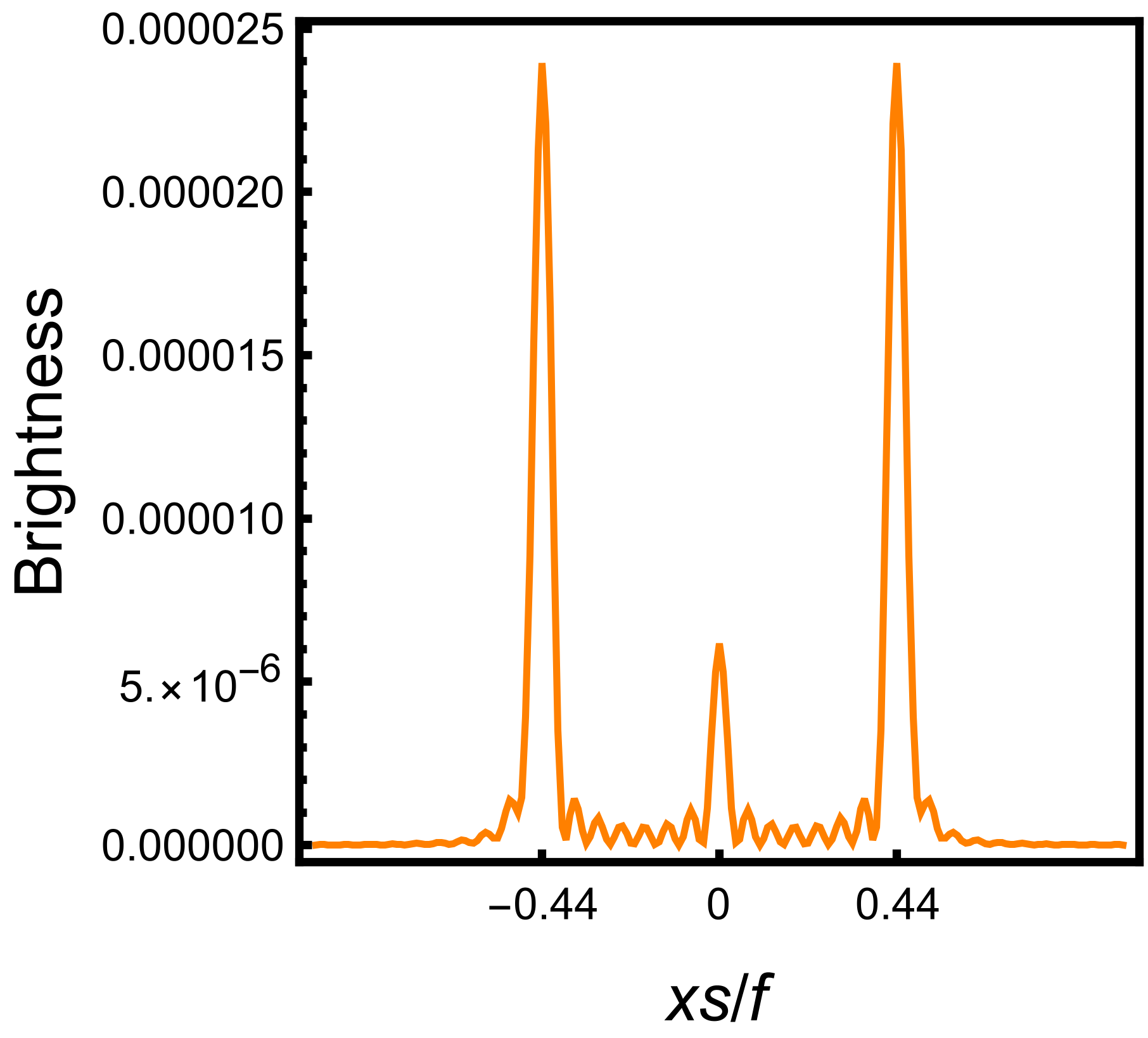}
    \caption{$\mu=0.5$}
  \end{subfigure}
  \hfill
  \begin{subfigure}[b]{0.48\columnwidth}
    \centering
    \includegraphics[height=0.8\textwidth,width=\textwidth]{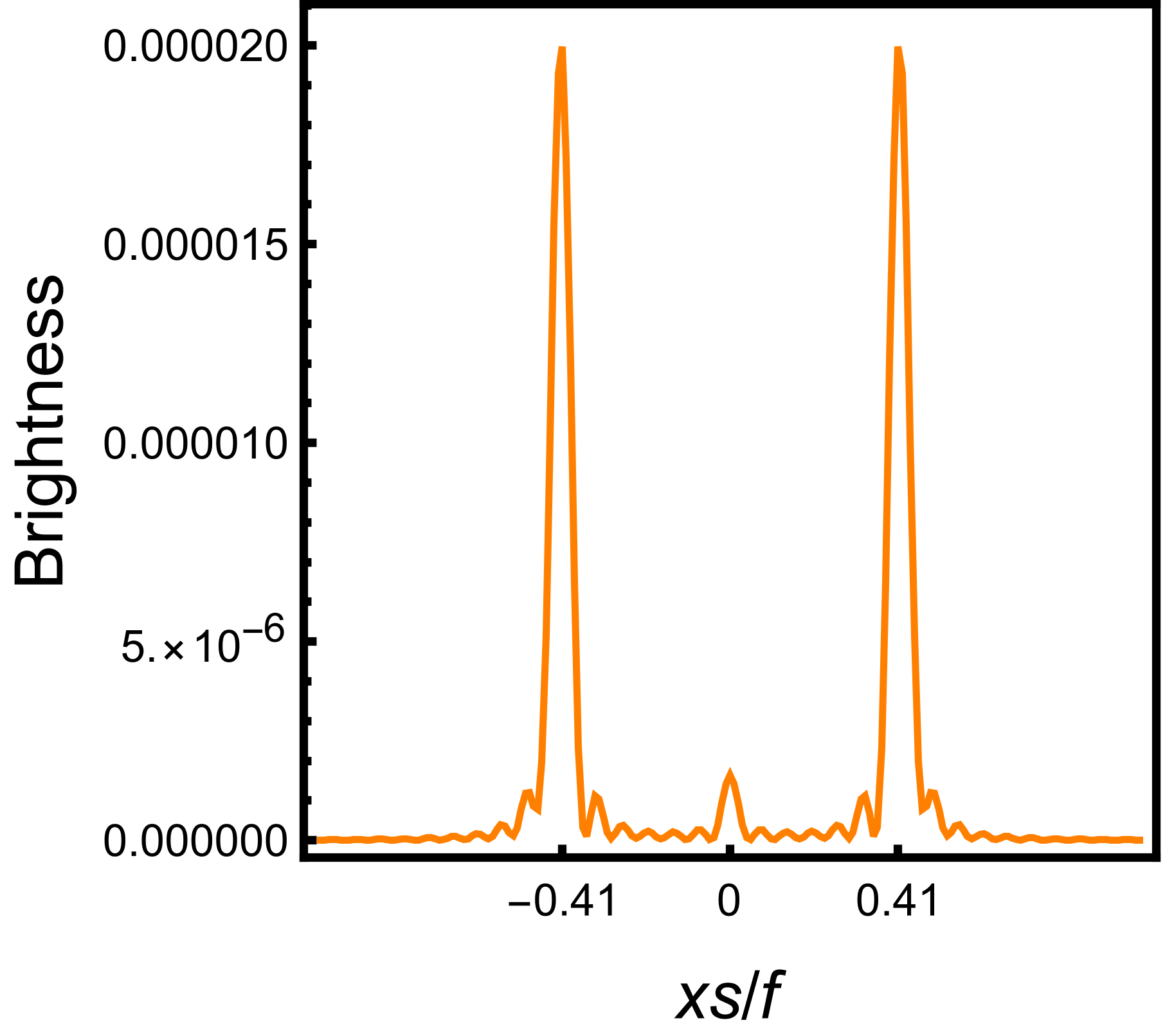}
    \caption{$\mu=0.7$}
  \end{subfigure}
  \caption{Effect of $\mu$ on the brigheness, where $\theta_{obs}=0$, $a=0.1$, $c=0.1$, $\omega=-\frac{2}{3}$, $e=0.5$, $y_{h}=5$, $\Omega=90$.}
  \label{20}%
\end{figure}

\section{Images from the viewpoint of geometric optics }
In this section, we discuss Einstein rings from the perspective of geometric optics.  Our primary focus is on studying the incidence angle of photons, based on the spherically symmetric spacetime background described in Eq.(\ref{1}) and Eq.(\ref{2}). We define conserved quantities within the metric framework as $\omega^{*} =F(r)\partial t/\partial \lambda $ and $\tilde{L} =r^{2} \partial \varphi /\partial t $, where $\omega^{*}$ represents the energy of the photon, $\tilde{L}$ denotes the angular momentum of the photon and $\lambda$ is the affine parameter. Due to the spherical symmetry of spacetime, it is only consider the case where photon orbits are located at the equator i.e., $\theta \equiv \pi/2 $. The four-velocity $\upsilon ^{\zeta } =(\mathrm{d}/\mathrm{d}\lambda )^{\zeta } $ satisfies \begin{eqnarray}
-F(r)(\mathrm{d}t/ \mathrm{d}\lambda)^{2} +F(r)^{-1} (\mathrm{d}r/ \mathrm{d}\lambda)^{2}\nonumber\\+r^{2} \sin \theta
(\mathrm{d}\varphi / \mathrm{d}\lambda)^{2} =0,
\end{eqnarray} and \begin{equation}
\dot{r}^{2}=\omega^{*}-\tilde{L} z(r).
\end{equation} where $z(r)=F(r)/r^{2}$ and $\dot{r} =\partial r /\partial \lambda$. The incident angle $n^{\zeta } \equiv \partial / \partial r^{\zeta }$, with boundary $\theta _{in}$ as normal vector, is defined as\cite{Liu:2022cev,Zeng:2023tjb} \begin{equation}
\cos \theta _{in}=\frac{g_{\alpha \beta } \upsilon ^{\alpha } n^{\beta } }{\left | \upsilon  \right | \left | n \right | }\mid _{r=\infty }
=\sqrt{\frac{\dot{r}^{2} /F }{\dot{r}^{2} /F+\tilde{L} /r^{2} } } \mid _{r=\infty },
\end{equation} which simplifies to \begin{equation}
\sin\theta _{in}^{2} =1-\cos \theta _{in}^{2} =\frac{\tilde{L} ^{2}z(r) }{{\dot{r}^{2}+\tilde{L} ^{2}z(r)} }\mid _{r=\infty }=\frac{\tilde{L} ^{2}}{(\omega^{*})^{2} }.
\end{equation}

When the two endpoints of a geodesic are aligned with the center of the BH, due to axial symmetry\cite{Hashimoto:2018okj}, an observer will perceive a ring-shaped image with a radius corresponding to the incident angle $\theta _{in}$. As a photon reaches the location of the photon sphere, it neither escapes nor falls into the BH, but instead begins to orbit around it. At this point, the angular momentum is denoted by $L$, and the equation of orbital motion for the photon at the photon ring is determined by the following conditions\cite{Hashimoto:2019jmw,Hashimoto:2018okj,Liu:2022cev} \begin{equation}
\dot{r}=0,\mathrm{d}z /\mathrm{d}r=0.
\end{equation} In the case, $\sin\theta _{in}=\mathit{L} /{\omega }$ is given, and from Figure \ref{19} $\sin \theta _{R}$ is observed to satisfy the relationship \begin{equation}
\sin \theta _{R} =r_{R} /f.
\end{equation}

The incident angle of the photon and the angle of the photon ring both describe the observable angle of the photon ring from an observer's perspective, and they should essentially be equal. That is \begin{equation}
r_{R} /f=L/\omega^{*}.
\label{27}
\end{equation}

\begin{figure}
	\centering 
\includegraphics[width=0.4\textwidth, angle=0]{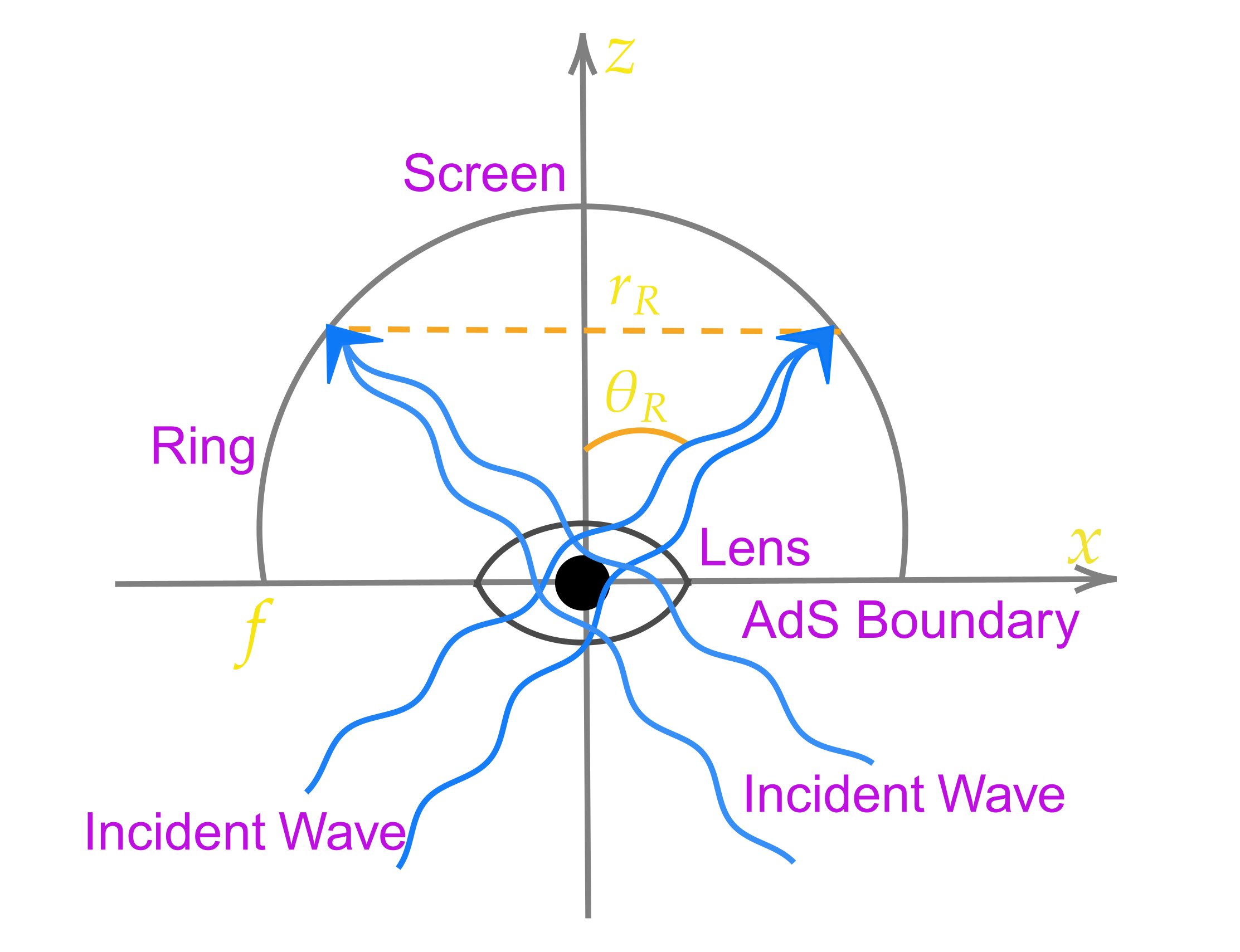}
	\caption{Relation between ring radius $r_{R}$ and ring angle $\theta _{R}$.} 
	\label{19}%
\end{figure}

\begin{figure}[htbp]
  \centering
  \begin{subfigure}[b]{0.48\columnwidth}
    \centering
    \includegraphics[width=\textwidth,height=0.9\textwidth]{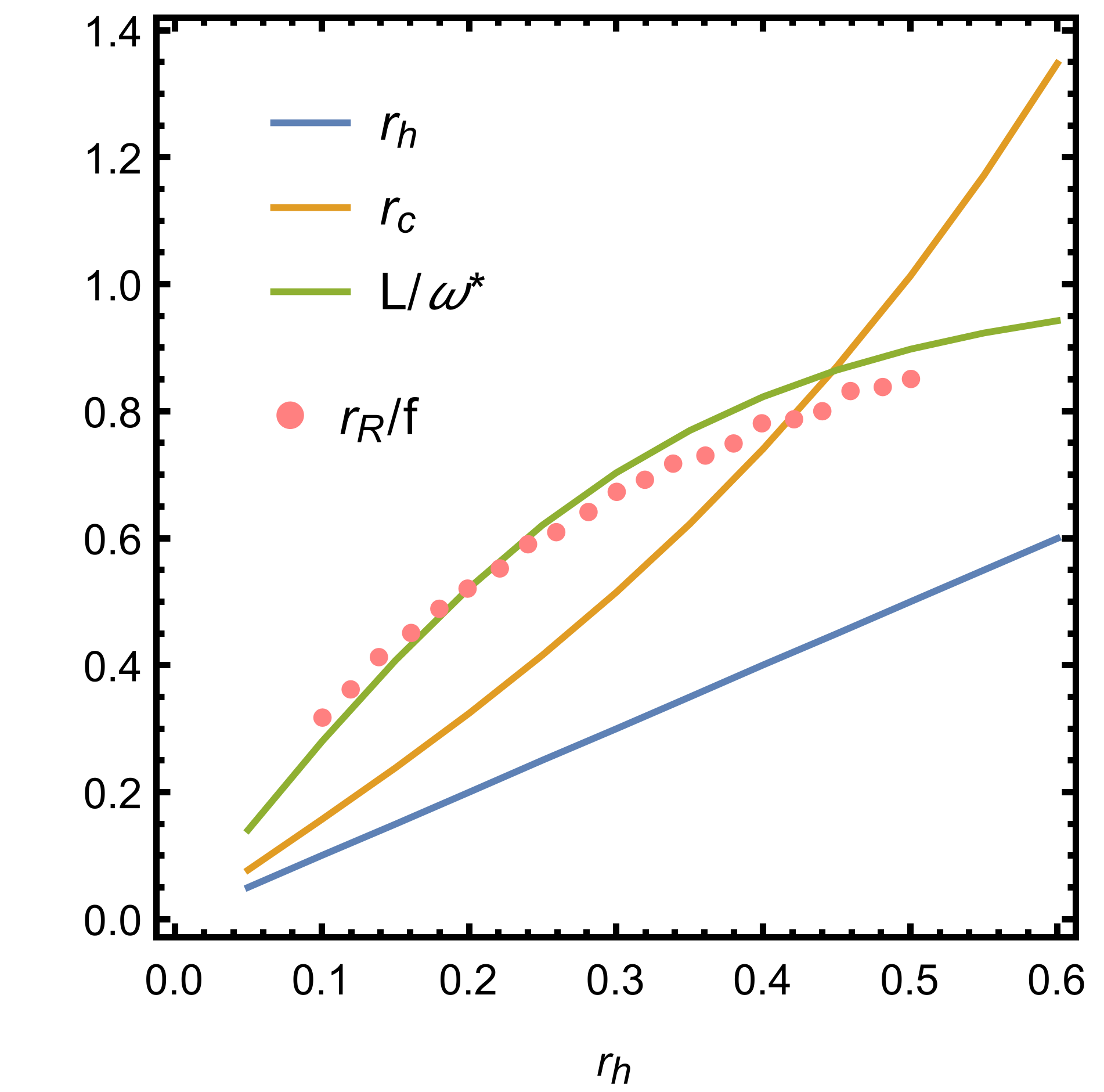}
    \caption{$a=0.01$}
  \end{subfigure}
  \hfill
  \begin{subfigure}[b]{0.48\columnwidth}
    \centering
    \includegraphics[width=\textwidth,height=0.9\textwidth]{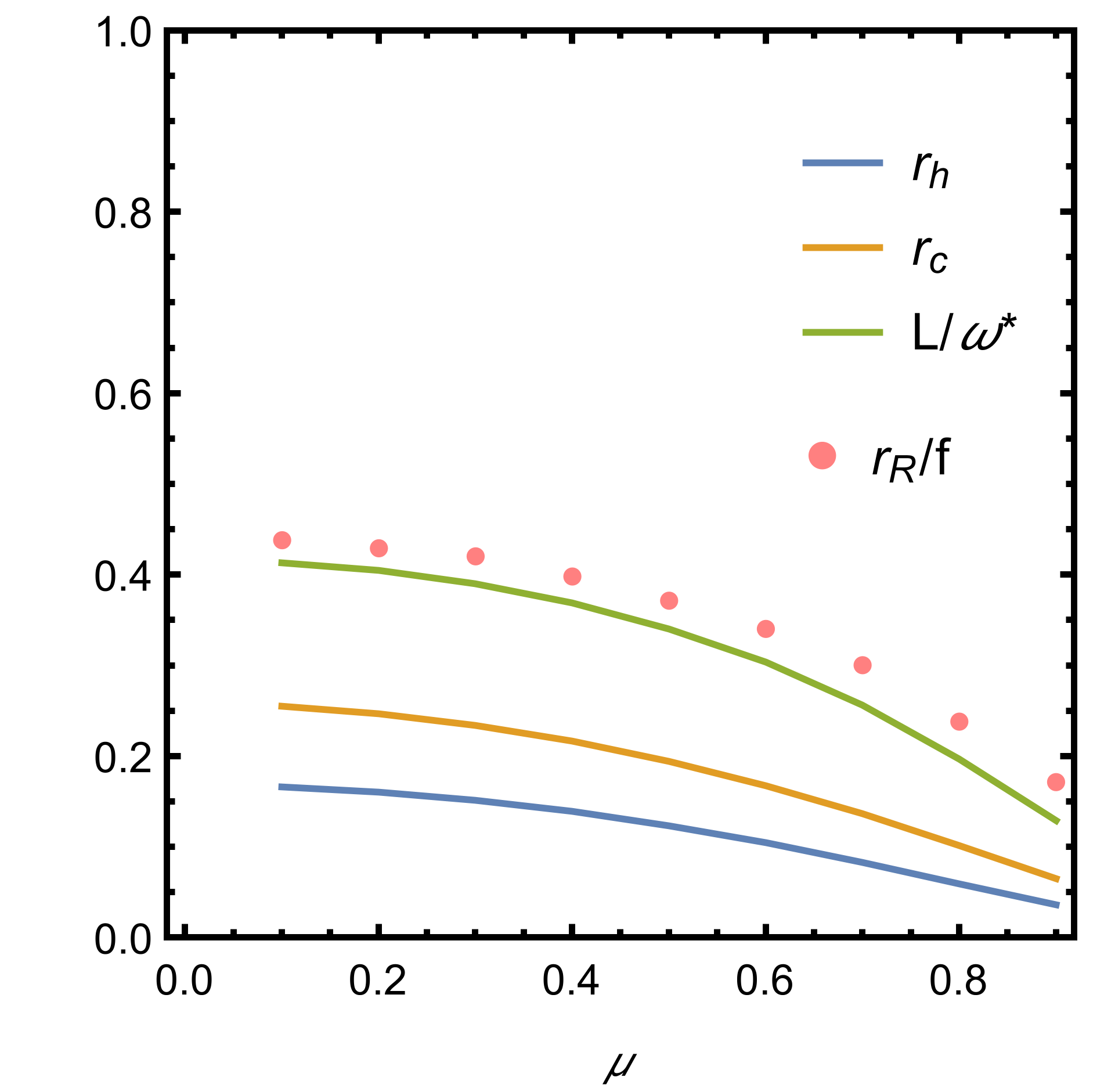}
    \caption{$a=0.01$}
  \end{subfigure}
\begin{subfigure}[b]{0.48\columnwidth}
    \centering
    \includegraphics[width=\textwidth,height=0.9\textwidth]{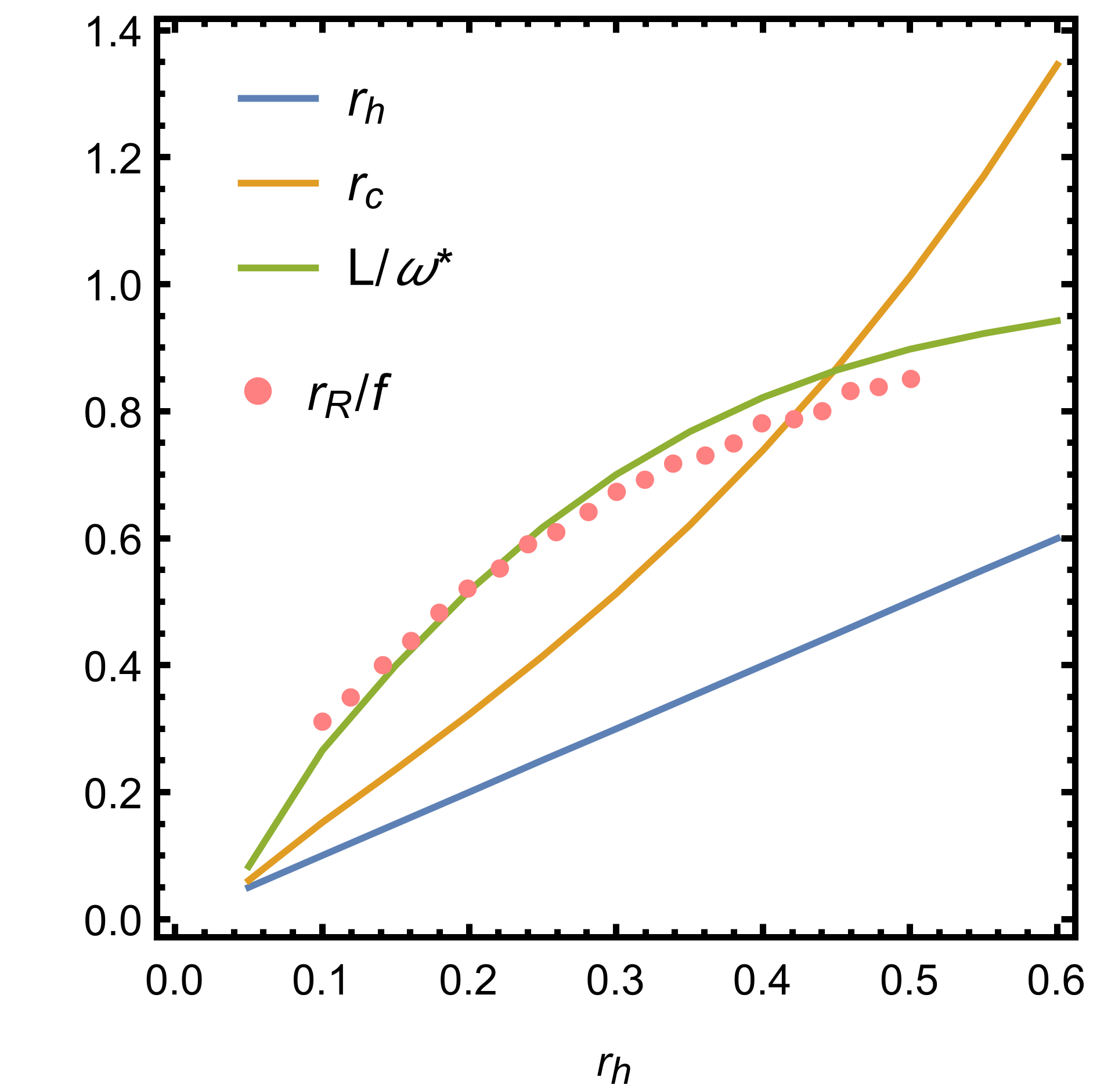}
    \caption{$a=0.05$}
  \end{subfigure}
  \hfill
  \begin{subfigure}[b]{0.48\columnwidth}
    \centering
    \includegraphics[height=0.9\textwidth,width=\textwidth]{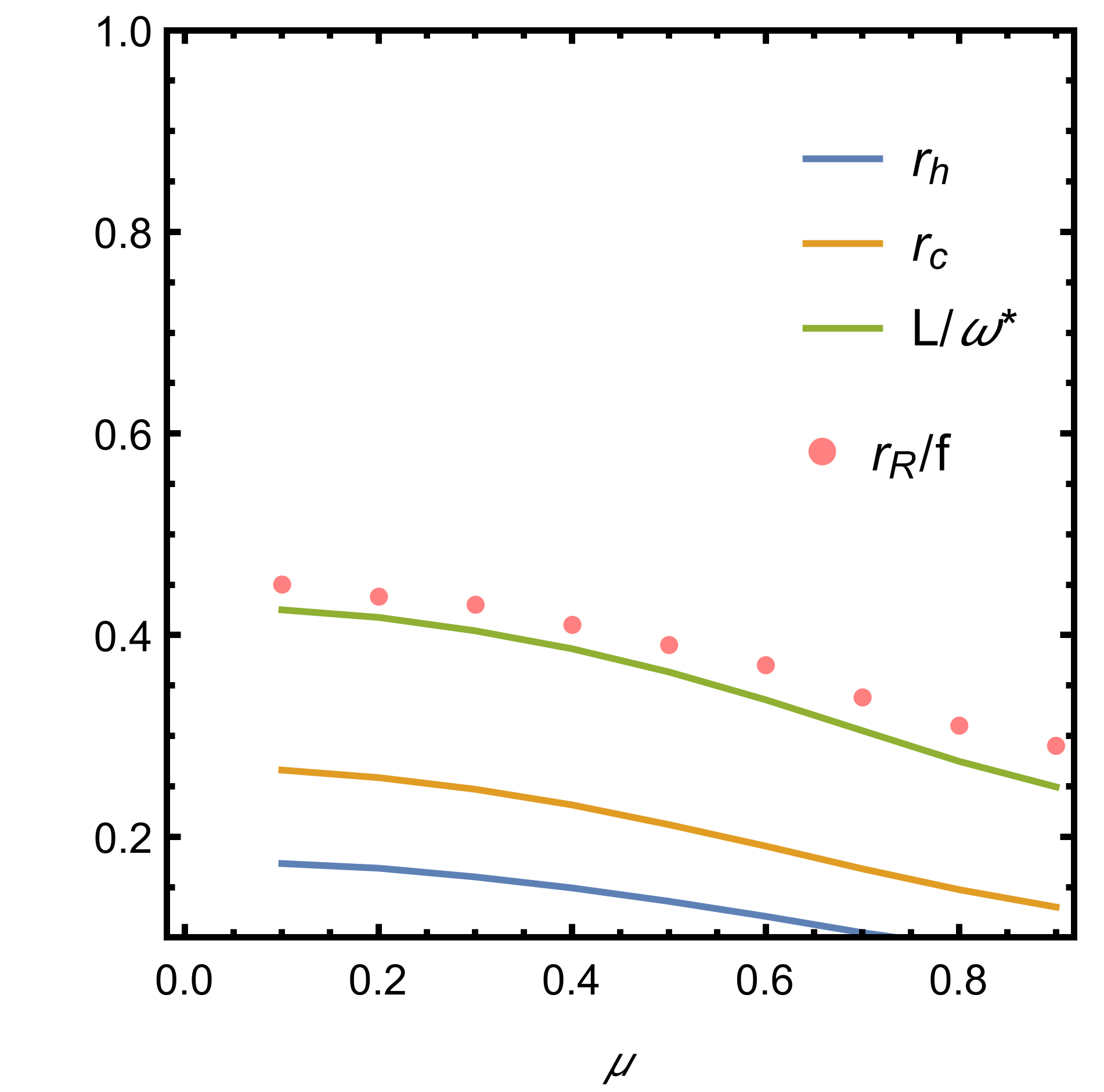}
    \caption{$a=0.05$}
  \end{subfigure}
  \begin{subfigure}[b]{0.48\columnwidth}
    \centering
    \includegraphics[width=\textwidth,height=0.9\textwidth]{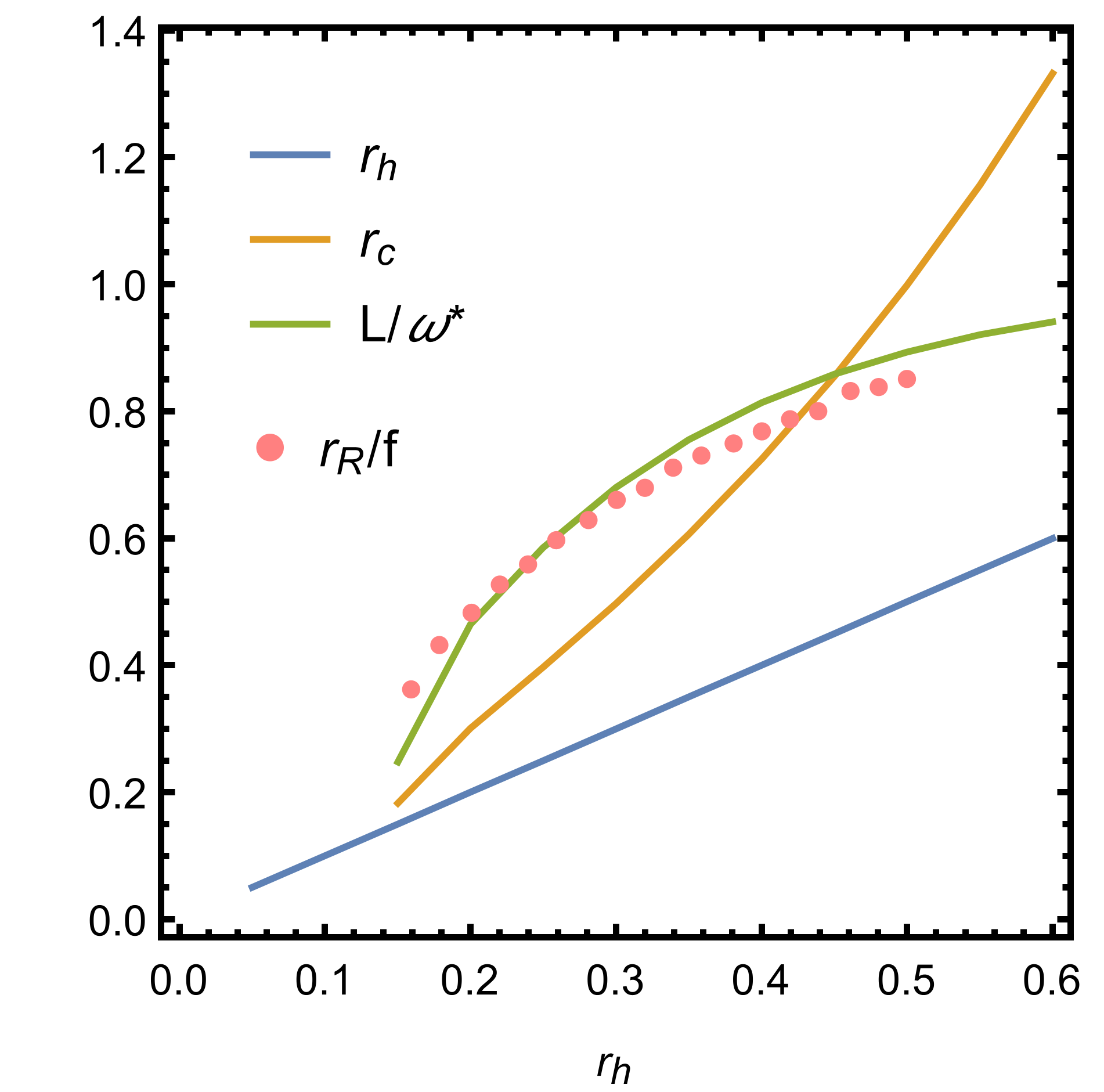}
    \caption{$a=0.15$}
  \end{subfigure}
  \hfill
  \begin{subfigure}[b]{0.48\columnwidth}
    \centering
    \includegraphics[height=0.9\textwidth,width=\textwidth]{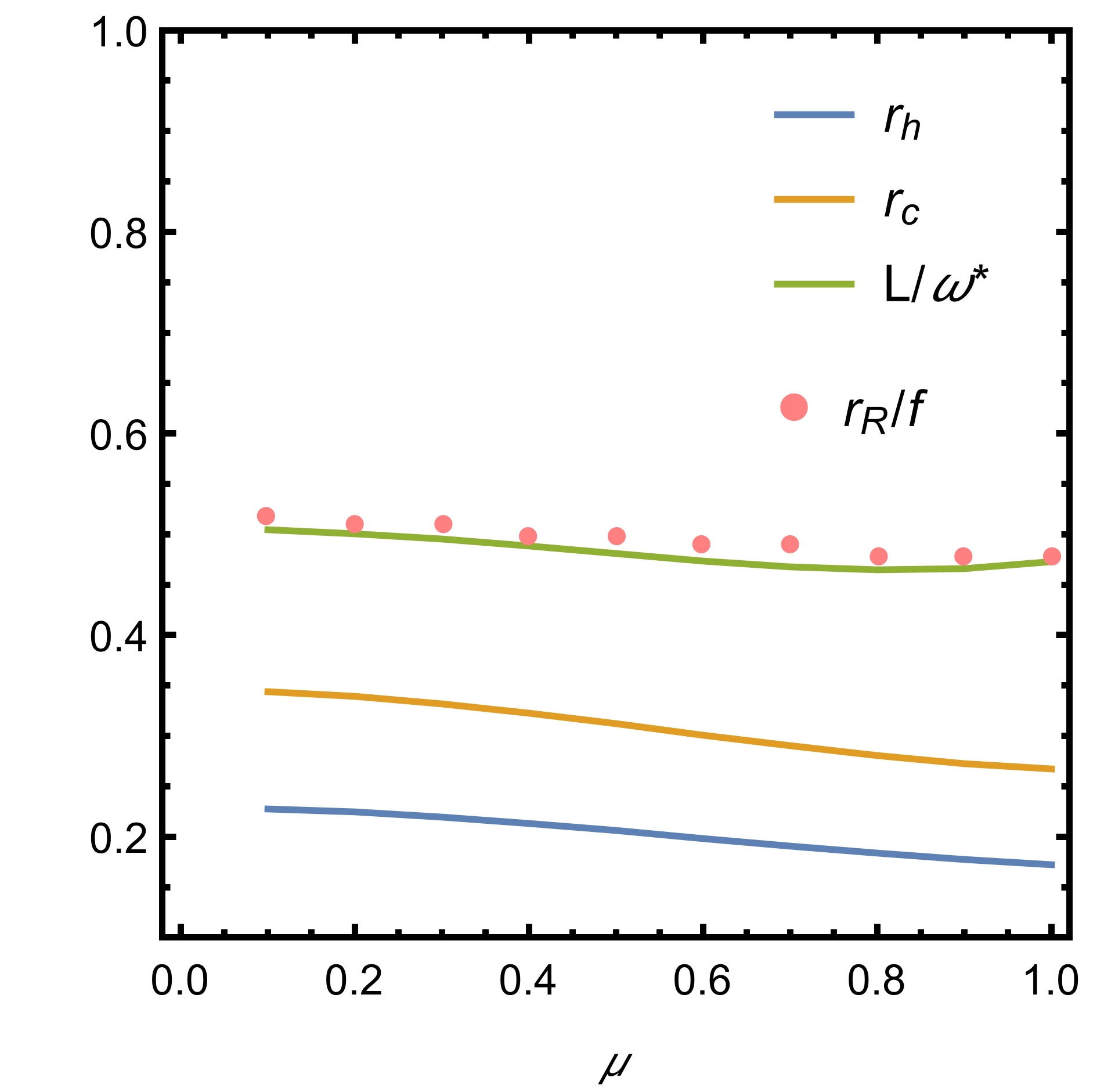}
    \caption{$a=0.15$}
  \end{subfigure}
  
  \caption{(Left Column) Comparison of Einstein ring radii between geometric optics and wave optics for different values of parameter $a$ when $\mu=0.5$, $c=0.1$, $\Omega=-\frac{2}{3}$, $e=0.5$, $\omega=90$. Here, the discrete pink points represent the observed angles of the Einstein rings derived from wave optics, while the green curve illustrates the incident angles of photons obtained based on geometric optics. (Right Column) Comparison of Einstein ring radii between geometric optics and wave optics for different values of parameter $a$ when $T=0.5$, $c=0.1$, $\Omega=-\frac{2}{3}$, $e=0.5$, $\omega=90$.}
  \label{22}%
\end{figure}

\begin{figure}[htbp]
  \centering
  \begin{subfigure}[b]{0.48\columnwidth}
    \centering
    \includegraphics[width=\textwidth,height=0.9\textwidth]{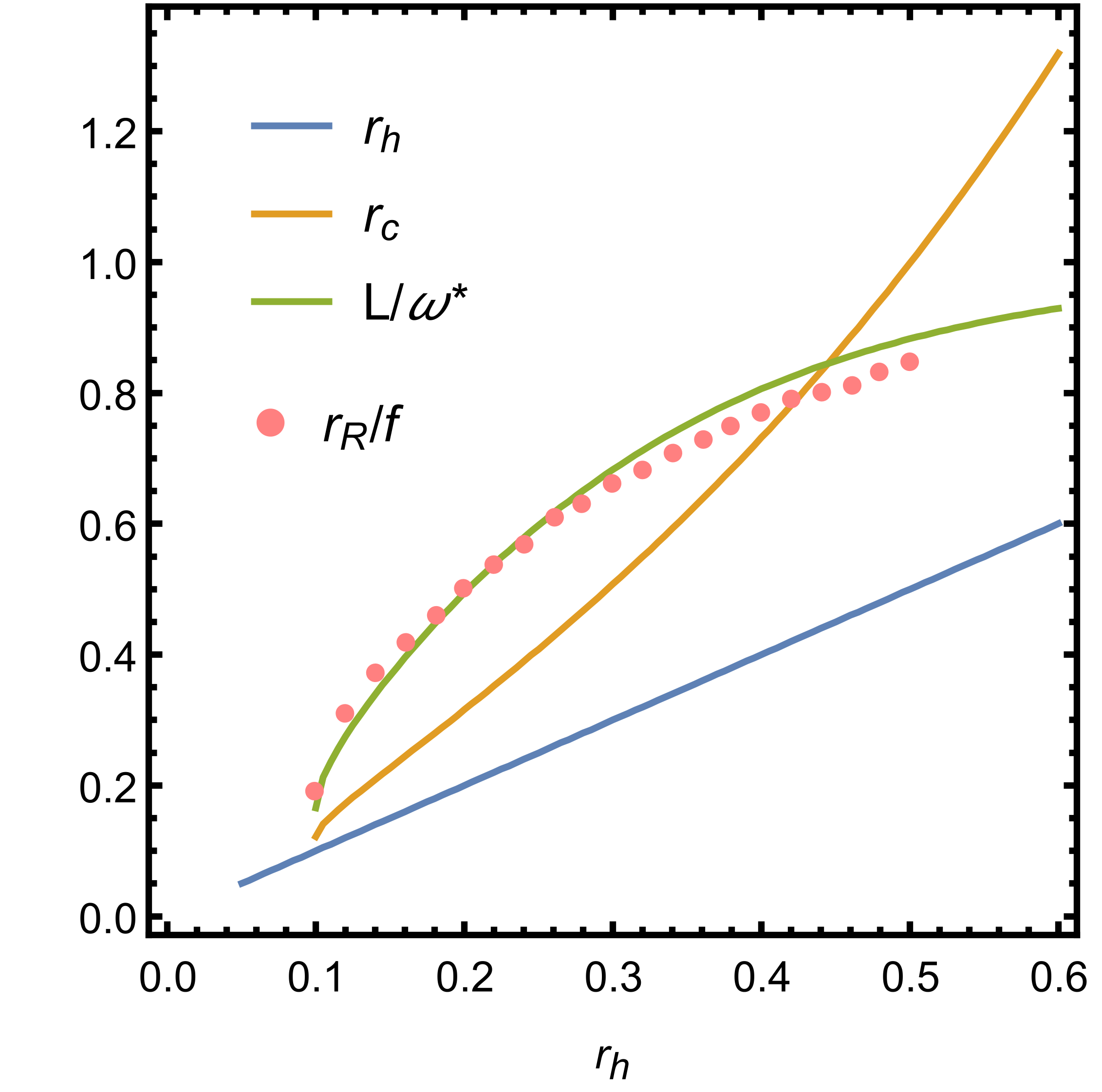}
    \caption{$c=0.05$}
  \end{subfigure}
  \hfill
  \begin{subfigure}[b]{0.48\columnwidth}
    \centering
    \includegraphics[width=\textwidth,height=0.9\textwidth]{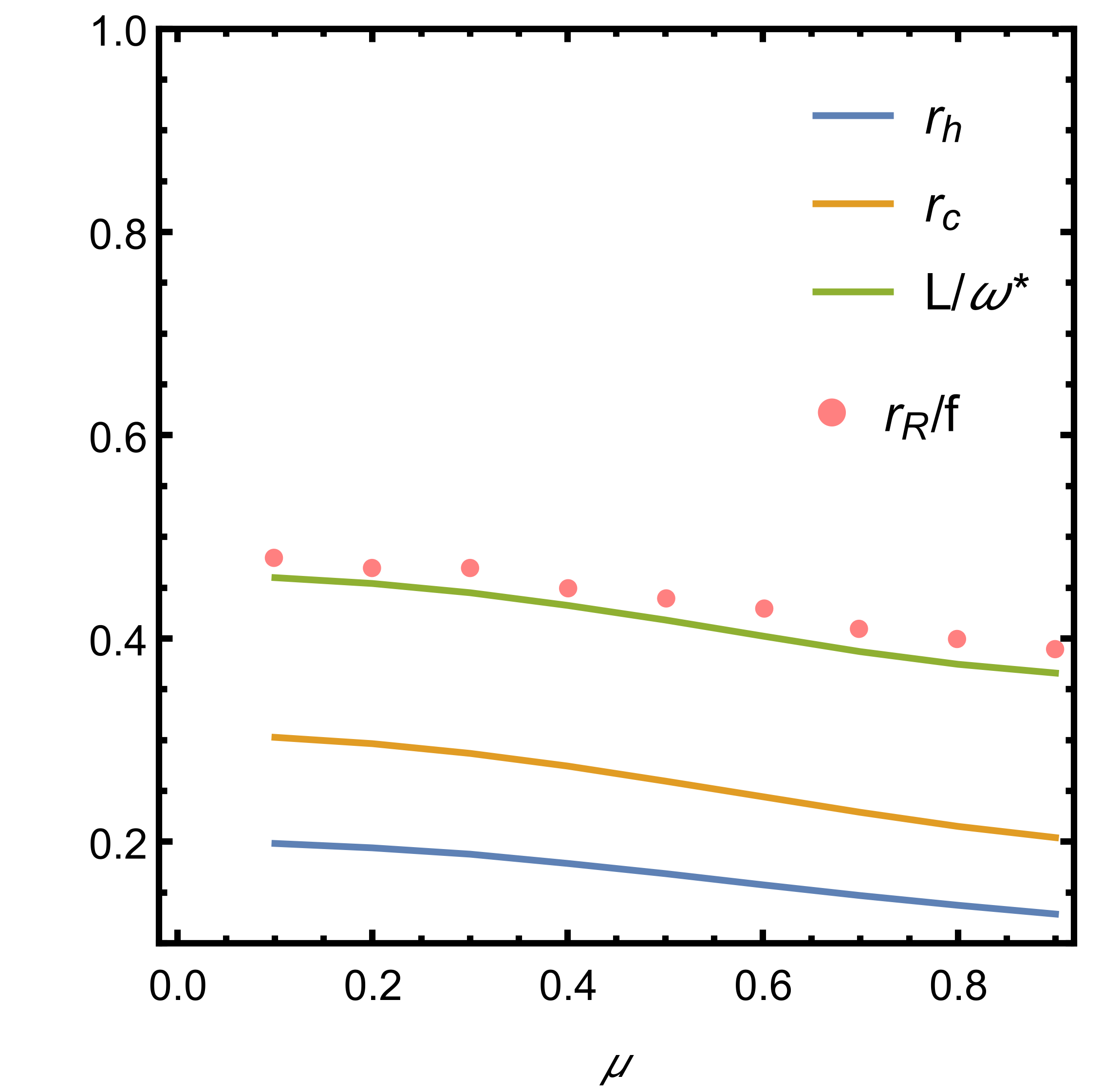}
    \caption{$c=0.05$}
  \end{subfigure}
\begin{subfigure}[b]{0.48\columnwidth}
    \centering
    \includegraphics[width=\textwidth,height=0.9\textwidth]{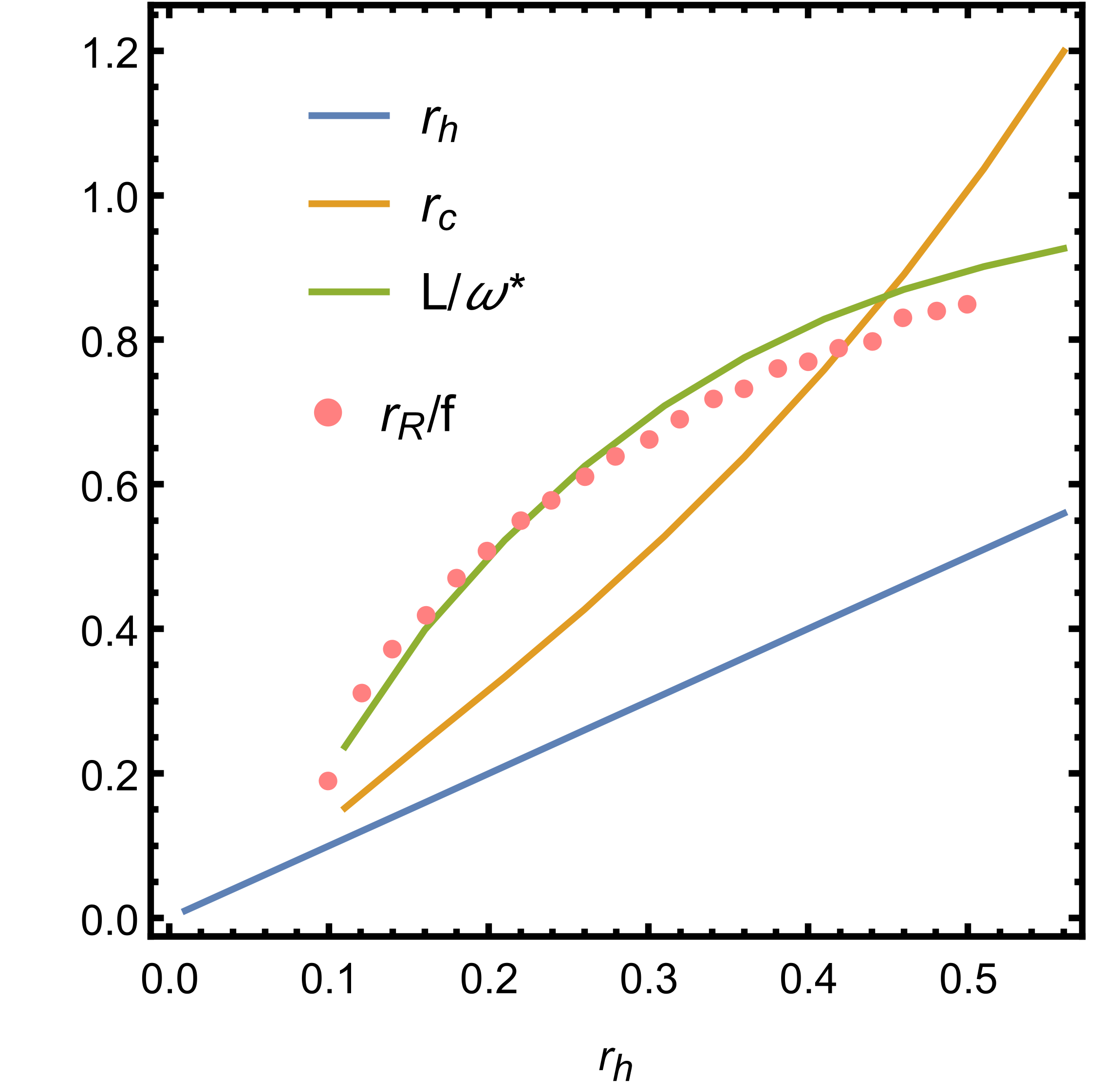}
    \caption{$c=0.1$}
  \end{subfigure}
  \hfill
  \begin{subfigure}[b]{0.48\columnwidth}
    \centering
    \includegraphics[height=0.9\textwidth,width=\textwidth]{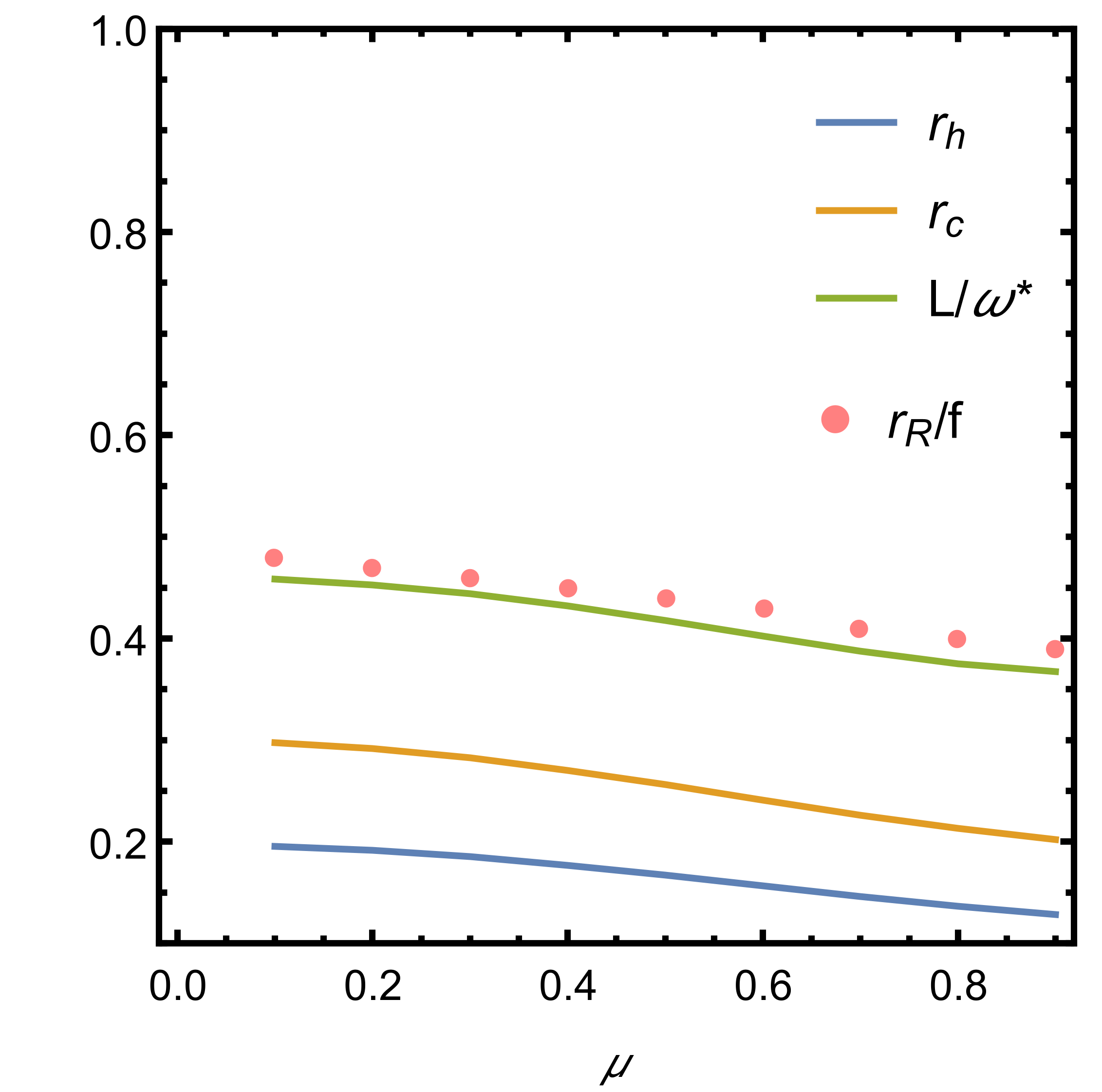}
    \caption{$c=0.1$}
  \end{subfigure}
  \begin{subfigure}[b]{0.48\columnwidth}
    \centering
    \includegraphics[width=\textwidth,height=0.9\textwidth]{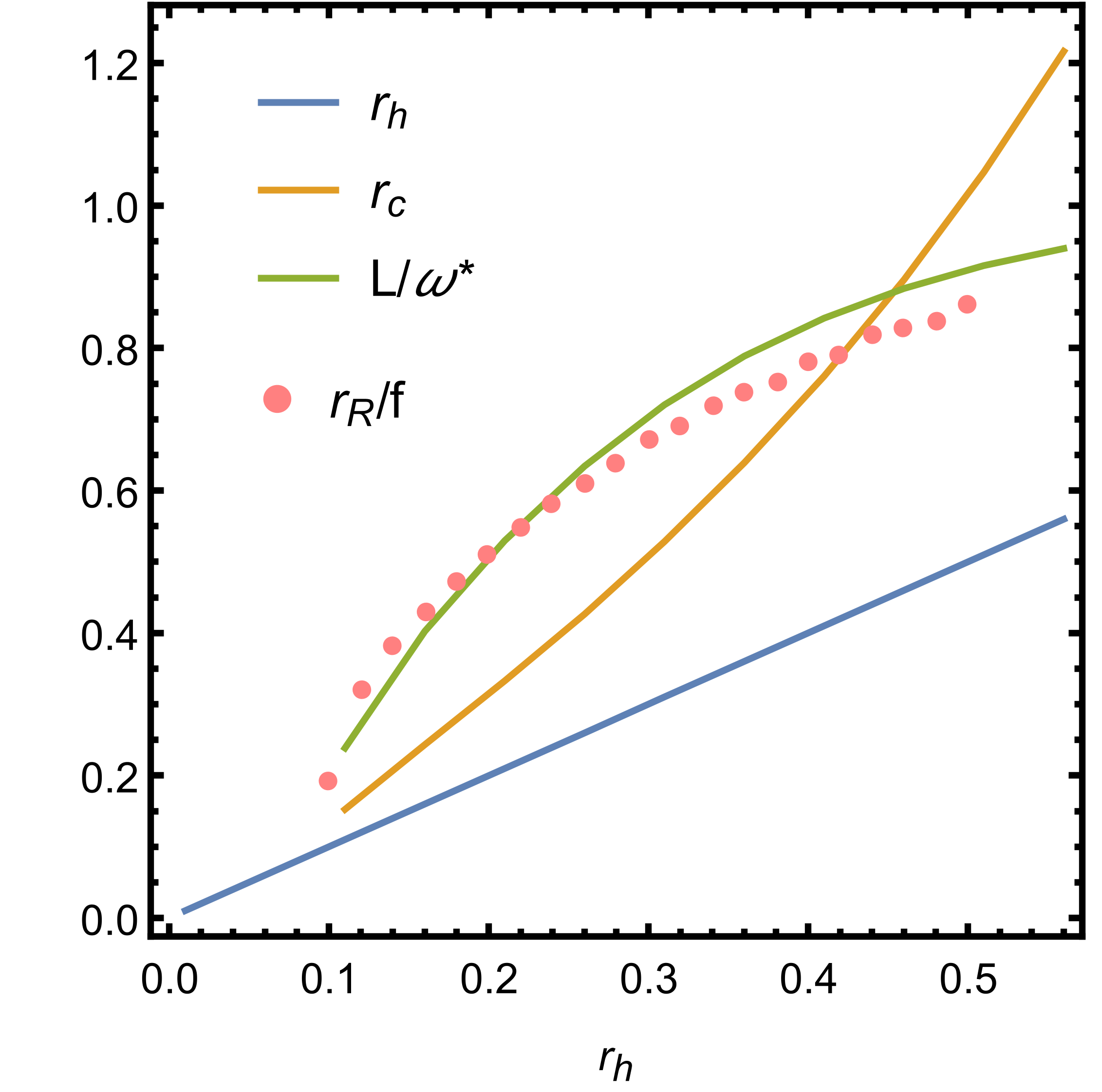}
    \caption{$c=0.15$}
  \end{subfigure}
  \hfill
  \begin{subfigure}[b]{0.48\columnwidth}
    \centering
    \includegraphics[height=0.9\textwidth,width=\textwidth]{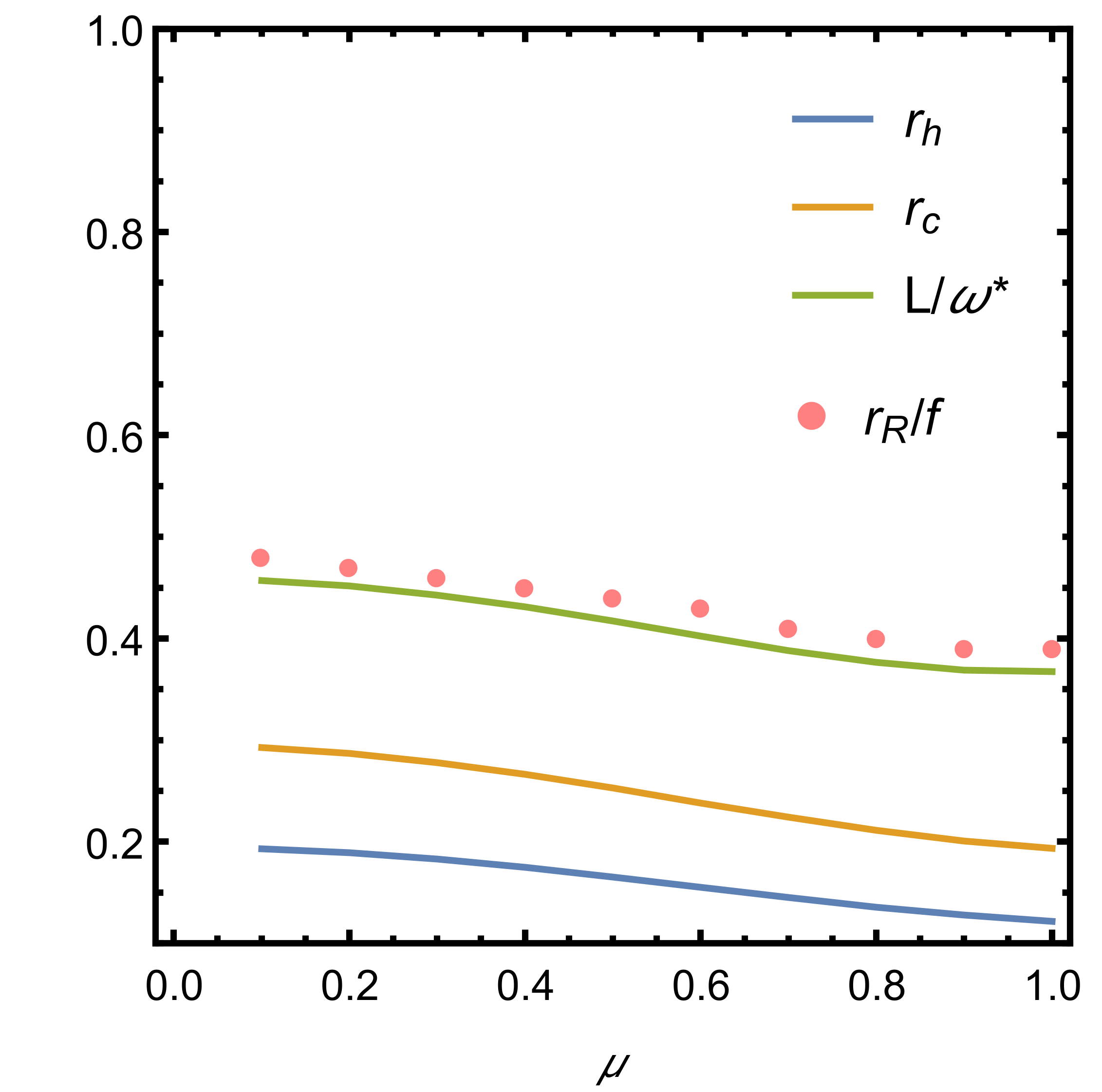}
    \caption{$c=0.15$}
  \end{subfigure}
  
  \caption{(Left Column) Comparison of Einstein ring radii between geometric optics and wave optics for different values of parameter $c$ when $\mu=0.5$, $a=0.1$, $\Omega=-\frac{2}{3}$, $e=0.5$, $\omega=90$. Here, the discrete pink points represent the observed angles of the Einstein rings derived from wave optics, while the green curve illustrates the incident angles of photons obtained based on geometric optics. (Right Column) Comparison of Einstein ring radii between geometric optics and wave optics for different values of parameter $c$ when $T=0.5$, $a=0.1$, $\Omega=-\frac{2}{3}$, $e=0.5$, $\omega=90$.}
  \label{23}%
\end{figure}

\begin{figure}[htbp]
  \centering
  \begin{subfigure}[b]{0.48\columnwidth}
    \centering
    \includegraphics[width=\textwidth,height=0.9\textwidth]{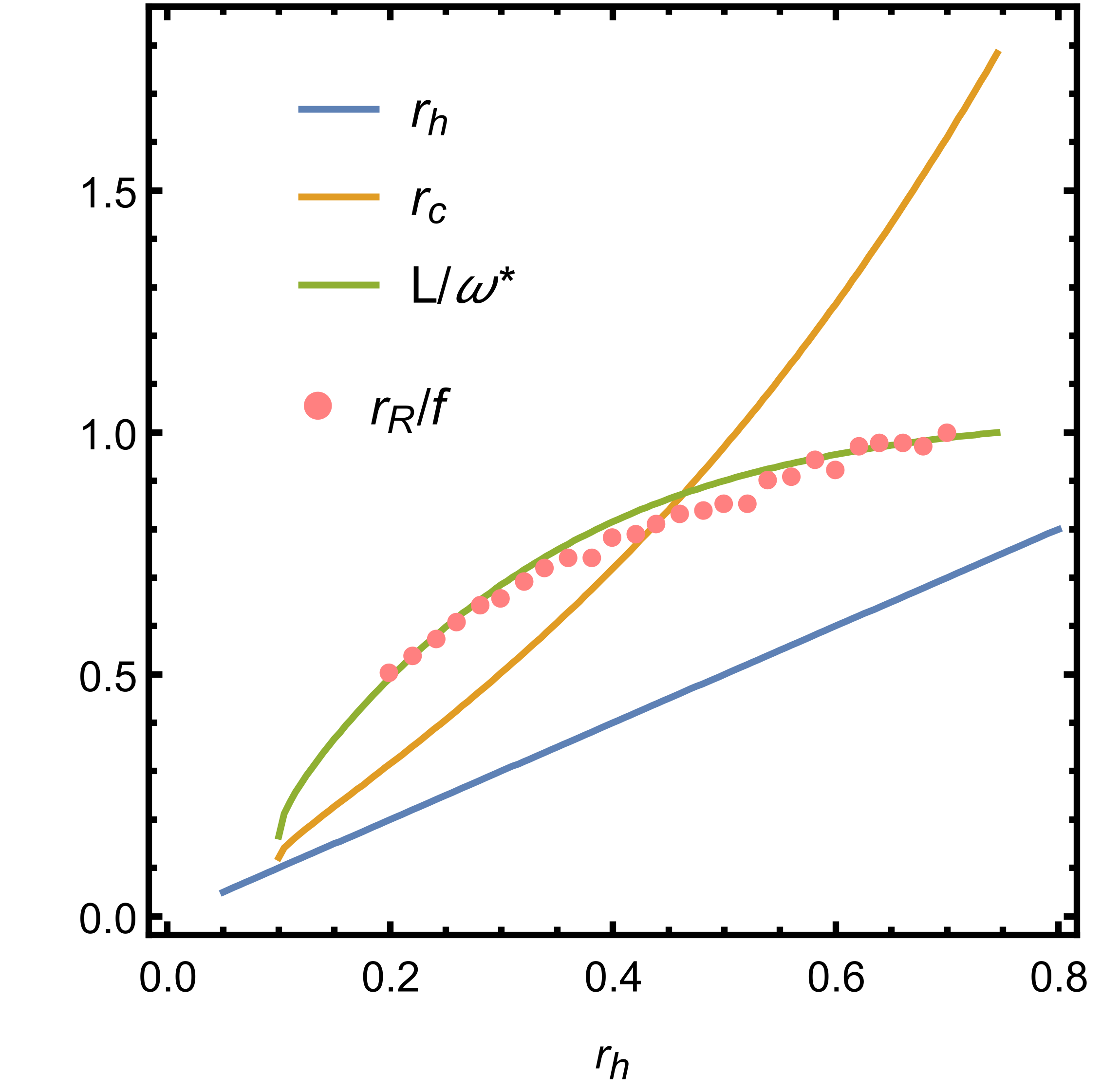}
    \caption{$\Omega=-1$}
  \end{subfigure}
  \hfill
  \begin{subfigure}[b]{0.48\columnwidth}
    \centering
    \includegraphics[width=\textwidth,height=0.9\textwidth]{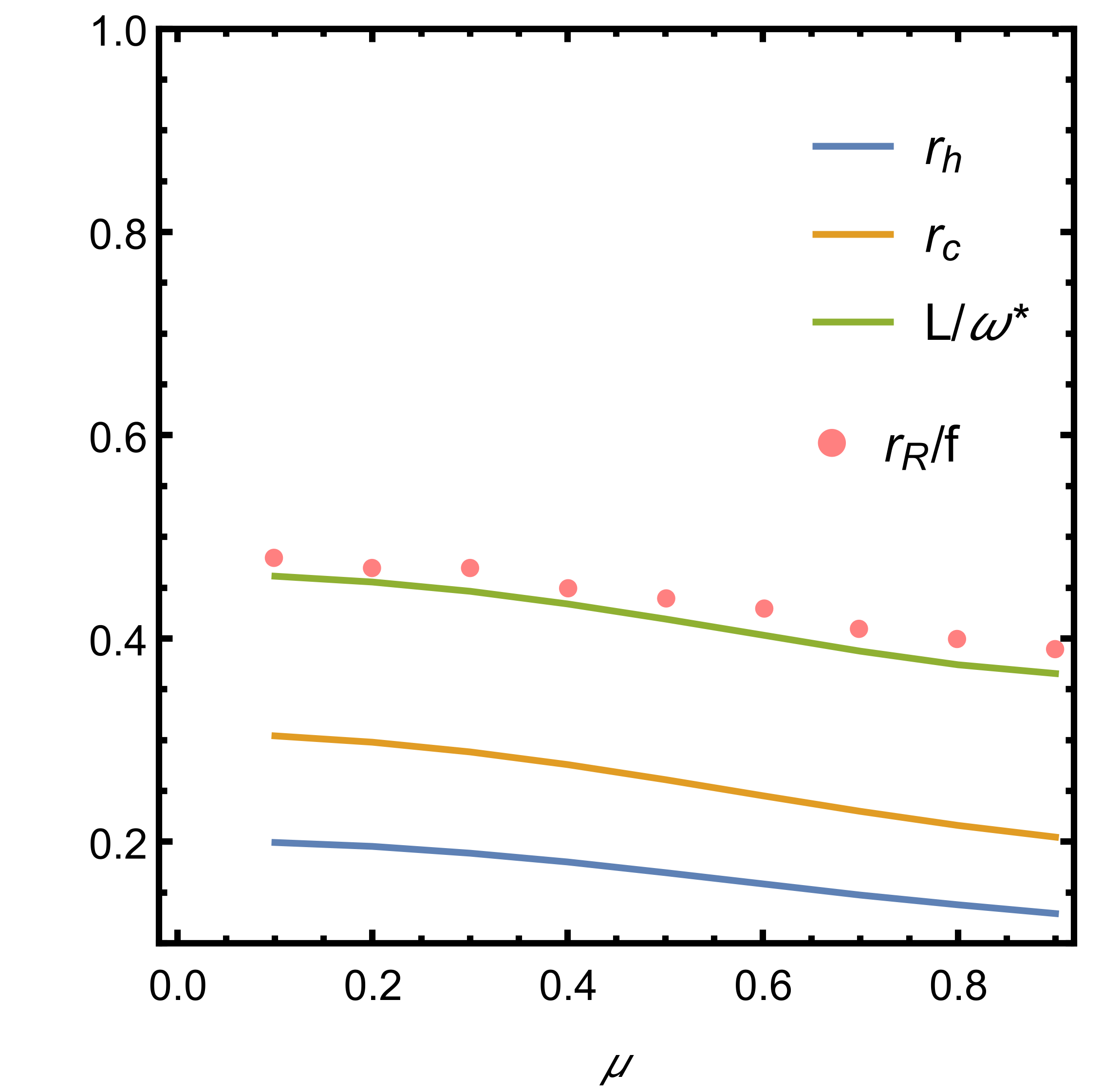}
    \caption{$\Omega=-1$}
  \end{subfigure}
\begin{subfigure}[b]{0.48\columnwidth}
    \centering
    \includegraphics[width=\textwidth,height=0.9\textwidth]{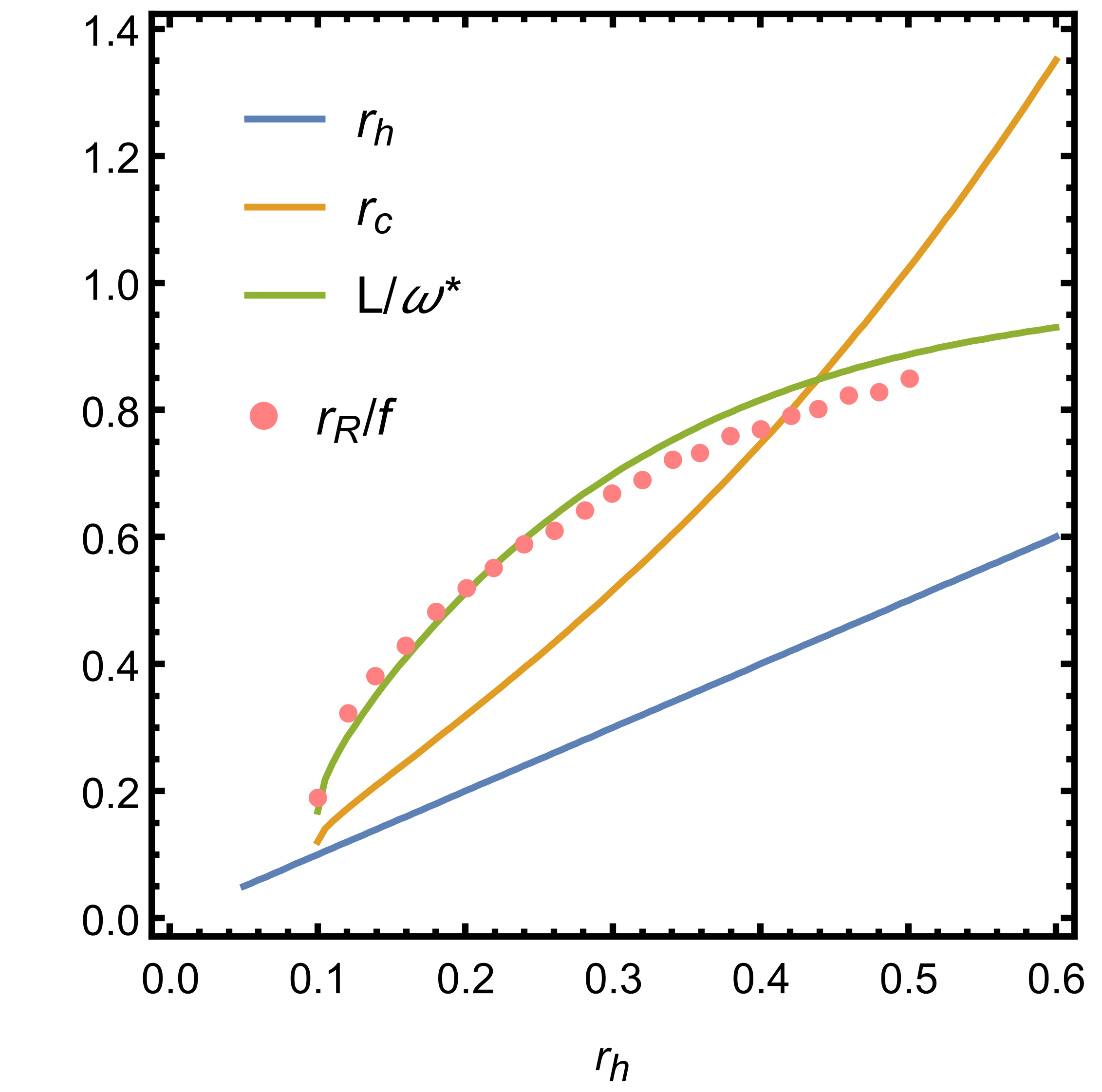}
    \caption{$\Omega=-1/3$}
  \end{subfigure}
  \hfill
  \begin{subfigure}[b]{0.48\columnwidth}
    \centering
    \includegraphics[height=0.9\textwidth,width=\textwidth]{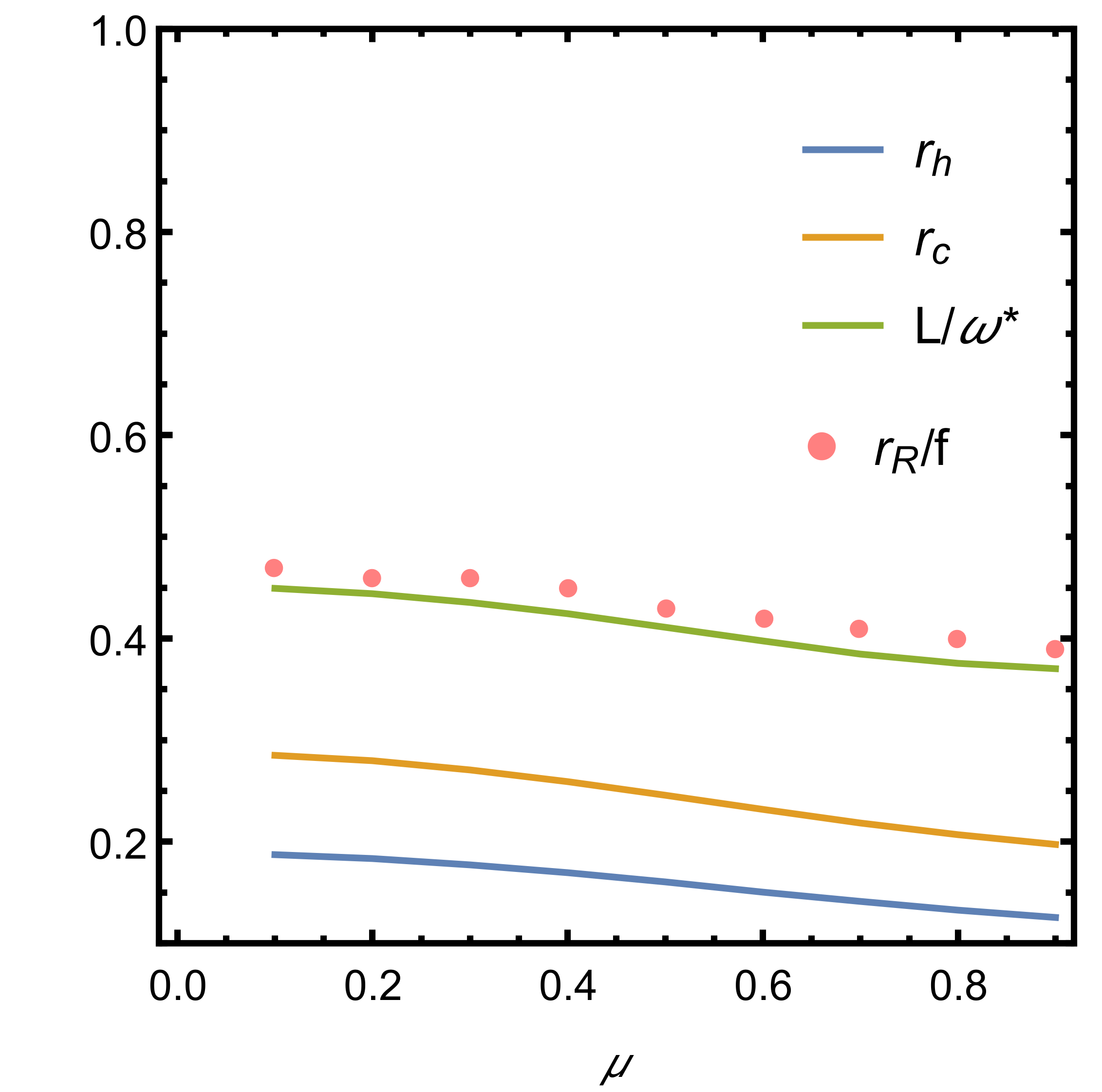}
    \caption{$\Omega=-1/3$}
  \end{subfigure}
  \begin{subfigure}[b]{0.48\columnwidth}
    \centering
    \includegraphics[width=\textwidth,height=0.9\textwidth]{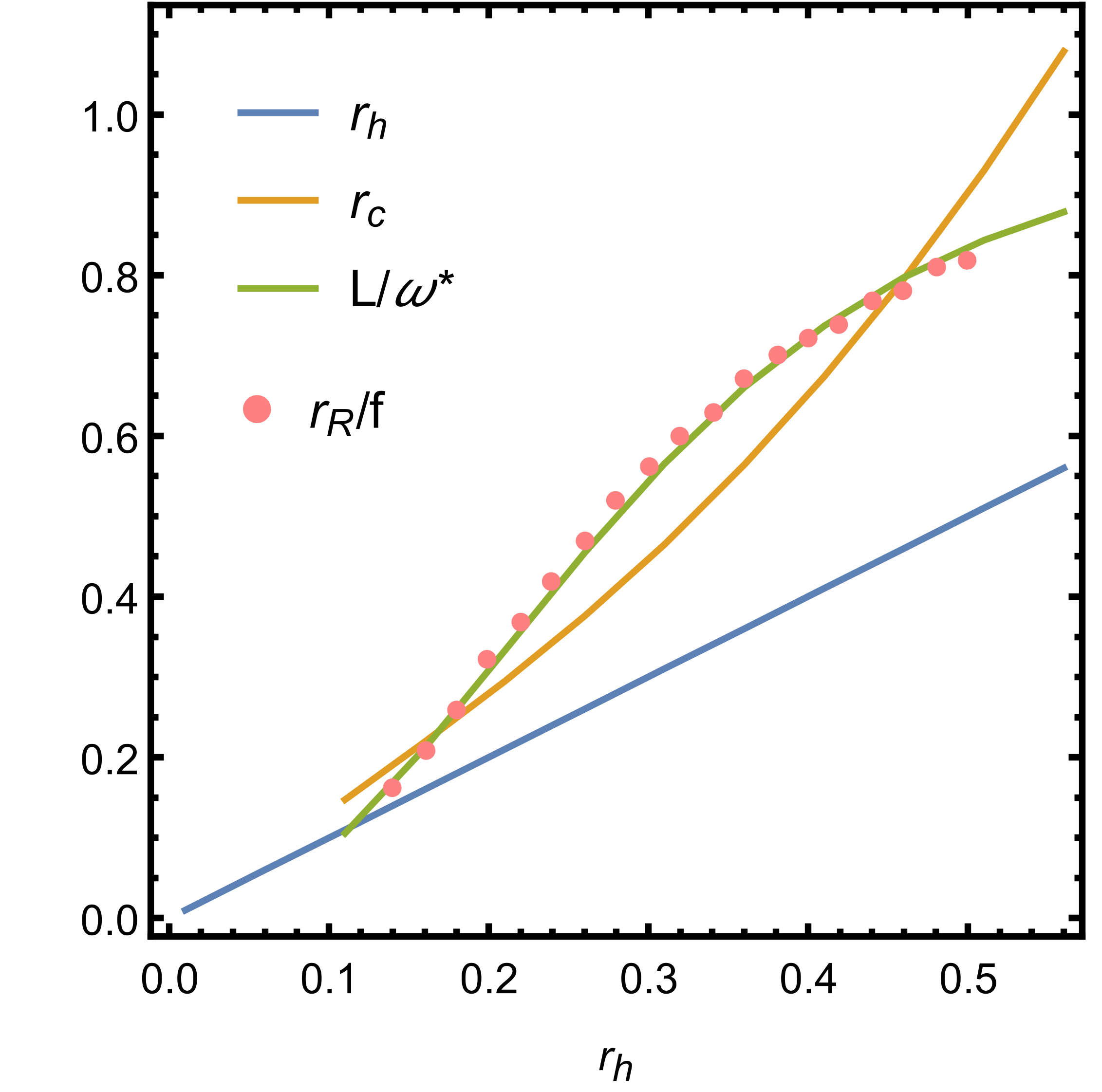}
    \caption{$\Omega=1/3$}
  \end{subfigure}
  \hfill
  \begin{subfigure}[b]{0.48\columnwidth}
    \centering
    \includegraphics[height=0.9\textwidth,width=\textwidth]{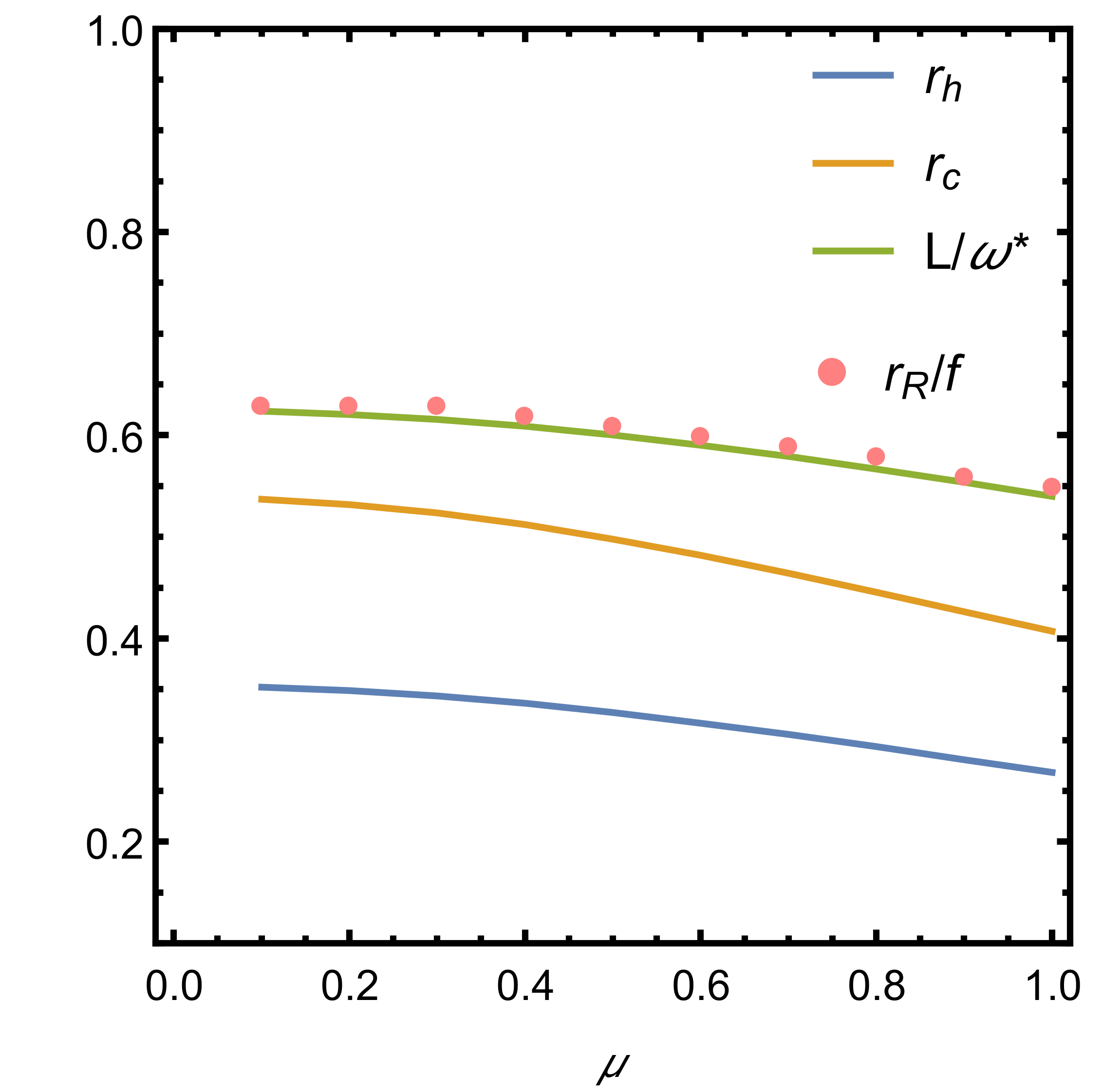}
    \caption{$\Omega=1/3$}
  \end{subfigure}
  
  \caption{(Left Column) Comparison of Einstein ring radii between geometric optics and wave optics for different values of parameter $\Omega$ when $\mu=0.5$, $a=0.1$, $c=0.1$, $e=0.5$, $\omega=90$. Here, the discrete pink points represent the observed angles of the Einstein rings derived from wave optics, while the green curve illustrates the incident angles of photons obtained based on geometric optics. (Right Column) Comparison of Einstein ring radii between geometric optics and wave optics for different values of parameter $\Omega$ when $T=0.5$, $a=0.1$, $c=0.1$, $e=0.5$, $\omega=90$.}
  \label{24}%
\end{figure}

As the physical parameters vary, we employ numerical methods to validate the results derived from Eq.(\ref{27}). Figure \ref{22} through \ref{24} respectively demonstrate the impact of varying parameters $a$, $c$ and $\Omega$ on the radius of the Einstein ring, with fixed chemical potential $\mu=0.5$ and temperature $T=0.5$. The results indicate that the angle of the Einstein ring observed through wave optics is in close agreement with the incident angle of the photon ring calculated using geometric optics. This conclusion holds regardless of the specific values of parameters $a$, $c$ and $\Omega$, although they may influence the fitting accuracy. Essentially, wave optics provides a feasible approach for the holographic reconstruction of the Einstein ring.

\section{Summary and conclusions}
Based on the AdS/CFT correspondence, an in-depth study of the Einstein ring phenomenon produced by Quantum Corrected AdS-Reissner-Nordström BHs in Kiselev Spacetime was conducted using wave optics methods. By analyzing the asymptotic behavior of the wave function in the boundary region, the response at the North Pole from a Gaussian wave source located at the South Pole was successfully obtained. The findings indicate that the response function corresponds to the diffraction pattern resulting from the scattering of the wave source by the BH. The amplitude of the response function decreases with increasing values of the parameter $c$, wave source frequency $\omega$, charge $e$, and the chemical potential $\mu$, respectively. In contrast, the amplitude increases with increasing values of the parameter $a$, $\Omega$ and temperature $T$.

To further investigate the Einstein ring phenomenon, we introduced an optical system with a convex lens and studied the impact of relevant physical parameters on the Einstein ring. The results show that the ring radius decreases with increasing values of parameters $a$, $\Omega$, temperature $T$, and chemical potential $\mu$ increase. Conversely, the ring radius increases with the increase of parameter $c$. Additionally, the ring radius becomes clearer as the wave source frequency $\omega$ increases.

Furthermore, within the framework of geometric optics, an investigation into the incident angle of photons was conducted. Theoretical analysis and numerical simulations revealed that the results obtained from geometric optics are consistent with those derived from wave optics. This observation holds true regardless of the values of parameters $a$, $c$, $\Omega$ chemical potential $\mu$, and temperature $T$, although these factors do influence the precision of the fitting.

\section*{Acknowledgements}
This work is supported by the National Natural Science Foundation of China (Grants No.
11675140, No. 11705005, and No. 12375043), and Innovation and Development Joint Foundation
of Chongqing Natural Science Foundation (Grant No. CSTB2022NSCQ-LZX0021) and Basic
Research Project of Science and Technology Committee of Chongqing (Grant No. CSTB2023NSCQMSX0324).  

\appendix




\end{document}